# Points defects produced by irradiation: influence of boundary conditions on their biased elimination

J.L. Bocquet [1]

Centre Borelli, ENS-Paris-Saclay 4 Avenue des Sciences, 91190 Gif-sur-Yvette - France

**Introduction**

The discovery of swelling of steels in power nuclear reactors launched a huge amount of research trying to explain the root of this damage. The consensus of the community hit rapidly the main reason. The elastic interaction energy of point defects (PD) with PD sinks, mainly dislocations, is not identical for interstitials 'I' and vacancies 'V'; the relaxation volume of I is indeed much larger than that of V (+2.0 atomic volume versus -0.22 atomic volume, respectively). Due to this *elastic bias* (EB), the interstitials I are more strongly attracted than vacancies by the long ranged elastic field; the dislocation will thus drain and absorb more I than V and give rise to an *absorption bias* (AB), characterized by $J_i > J_v$. The remaining highly supersaturated V, which are less mobile, condense eventually on the spot while building 2D and 3D clusters. The latter expand the material, make it swell and deteriorate its mechanical properties.

Many studies were devoted to establish the link between the EB and the AB [1-9], using mean field approaches (a lossy medium with an absorption term uniformly sprayed over the volume [1]) or an atomic scale modelling of individual sinks [2-9] in cells with various boundary conditions (zero flux condition or continuity of concentration through the boundary).

In the cell method a single sink is embedded in a volume surrounded by an outer surface of radius $r_{ext}$ without any other competitor; $r_{ext}$ is set to match the corresponding sink density and the defect concentration along this surface is the average concentration far from the sink in the medium <C>. The sink is surrounded by an inner surface of radius $r_{int}$ with $r_{int} << r_{ext}$; this inner surface is the one through which the incoming defects will be counted to measure the elimination flux. After solving a Laplace diffusion equation (no source term together with a fixed <C> at the outer surface) for each defect separately, an elastic bias (EB) is calculated after evaluation of the flux crossing the internal surface at $r_{int}$ with and without interaction $J_{d,rint}^{with}$ and $J_{d,rint}^{zero}$ respectively. An *elastic bias* (EB) is then defined by $X_d = 1 - J_{d,rint}^{with} / J_{d,rint}^{zero}$. The elimination of the defects in the whole system can then be partitioned between the different types of sinks [5]. Many papers debated about the last pending question, namely, the average defect concentration <C> to be chosen far from the sinks [2-5]. The *absorption bias* proposed by this approach is then defined for each type of sink by $AB = 1 - X_V / X_I$. However, the diffusion field used to calculate the defect fluxes is the field around one type of sink considered as alone and defects are introduced separately without recombining with each other.

---

[1] Corresponding author: jean-louis.bocquet@ens-paris-saclay.fr

A revisited version of the cell method was proposed more recently [10-12]. A single primary sink with its inner and outer surfaces was considered together with a zero flux condition on the outer surface. The two types of defects I and V are introduced simultaneously through a homogeneous source term $G$; the recombination events I + V were taken into account through a homogeneously spread recombination rate [13]. Solving numerically a Poisson equation for each type of sink (dislocation, cavity, grain boundary) yields the two elimination fluxes for I and V, namely $J_{I,rint}^{with}$, $J_{V,rint}^{with}$ which are measured simultaneously on the system under the same boundary conditions. The absorption bias was then defined by: $AB = 1 - J_{V,rint}^{with} / J_{I,rint}^{with}$.

It was found that this bias was always equal to zero whatever the creation rate $G$ and the elastic interaction of each defect with the primary sink. Indeed, the production and the recombination are symmetric processes and the zero-flux boundary condition is impermeable: thus, the incoming fluxes of I and V through the inner surface cannot be different from each other, whatever their unequal interactions with this primary sink. This is the reason why the authors introduced in their balance equation another type of sink (called secondary sink) through a uniformly spread absorption term, the latter reflecting the intensity of the absorption by this other competing sink and changing the surroundings into a lossy medium. The authors could show that the partition of the eliminated defects between the various types of sink was evaluated consistently: a cavity embedded in a lossy medium containing dislocations as a secondary sink yielded the same partition of eliminated defects as a dislocation embedded in a lossy medium containing cavities as a secondary sink.

Recently, the transport of PD in the elastic field of dislocations has received considerable improvements. Ab initio packages are widely used to calculate in detail the migration barriers of the defects, while taking a full account of their anisotropy: at the saddle point for the vacancy, at the stable and saddle point for the dumbbell interstitial with the help of their dipolar tensors in the two configurations [14-19]. Recent contributions studied sink networks made of parallel dislocations and highlighted interesting details about the shape of the flow lines in the neighborhood of the dislocations for each defect [18]. But despite this improvement in the quantitative treatment, the very definition of the absorption bias remains partially unsatisfying because the result depends always on the boundary conditions; the calculations are mainly done for PD which are separately introduced into the cell and which do not interact (when solving a diffusion equation or when using an Object-Kinetic-Monte-Carlo code), which remains far from experiments where the PD are created simultaneously and can recombine with each other.

From the reviews of the previous authors quoted above who compared at length the various ways of evaluating the AB, it is concluded that:

*the absorption bias AB calculated in a model system with a single primary sink is probably of limited relevance, since the root of the absorption bias (AB) lies in the simultaneous presence of sinks having different elastic properties (EB);

*a complete balance sheet of all the defects produced and eliminated must be done, including recombination, which ensures the simultaneous evaluation of absorption fluxes for the two defects, in the same system and under the same boundary conditions.

These are the reasons why the model system presented below contains two primary sinks with different elastic properties. It is restricted to be 1D in order to obtain rapidly concentration profiles for the two defects I and V which are created throughout the medium via a homogeneous source term.

Without claiming to describe an actual material under irradiation, the purpose is to explore the link between the elastic bias (EB) and the observed absorption bias (AB) in a model system and to determine the sensitivity of this link to the boundary conditions. We detail hereafter: the model system in section I, the elastic interaction of PD with sinks in section II, the irradiation contribution in section III, the diffusion model for the two defects in section IV, the procedure yielding the numerical results together with the control of the solution in section V, the results in section VI and the final discussion in section VII.

At the end of the main contribution, appendices A to C often paired with the sections enumerated above, gather together useful details.

**I Geometry of the model system**

The 1-D system is a matter slab made of N parallel (100) crystalline planes of a BCC lattice, which are perpendicular to Ox axis and numbered from 1 to N (Fig. 1). The inter-plane distance is a/2 where a is the lattice parameter. The system is bounded by to additive planes (numbered '0' on the left and 'N+1' on the right) playing the role of PD source-and-sink: their role consists in pinning down the PD concentrations to fixed values which depend on the temperature and on the elastic interaction associated with each sink.

The concentrations are assumed to be uniform in each atomic plane 'i'. Each site of plane 'i' (called afterwards site 'i' for short) has 4 neighbors on plane 'i-1' and 4 neighbors on plane 'i+1'. The sink-and-source planes are assumed to have the same crystalline structure as the bulk planes. It is assumed that a defect jumping on plane '0' (L sink) or 'N+1' (R sink) is immediately and irreversibly absorbed by the sink. The concentration $C_{d,i}$ of each defect on plane 'i' associated with the jump frequencies between adjacent planes yield the PD fluxes at the two boundaries, namely $J_{d,L}$ and $J_{d,R}$ on the left and right boundary respectively. With our notations and plane numbering, these outward fluxes correspond to $J_{d,L} = J_{d,1/2}$ and $J_{d,R} = J_{d,N+1/2}$.

During a sustained irradiation, the fluxes arriving on the left sink L are expected to be negative, while those on the right sink R are expected to be positive. The two absorption biases are historically expressed as:

$$AB_L = (|J_{I,L}| - |J_{V,L}|)/|J_{I,L}| \text{ together with } AB_R = (|J_{I,R}| - |J_{V,R}|)/|J_{I,R}|. \quad (1)$$

This historical choice yields very high values of the bias whenever the interstitial flux is small. Alternative and more symmetrical definitions could be defined to describe this imbalance.

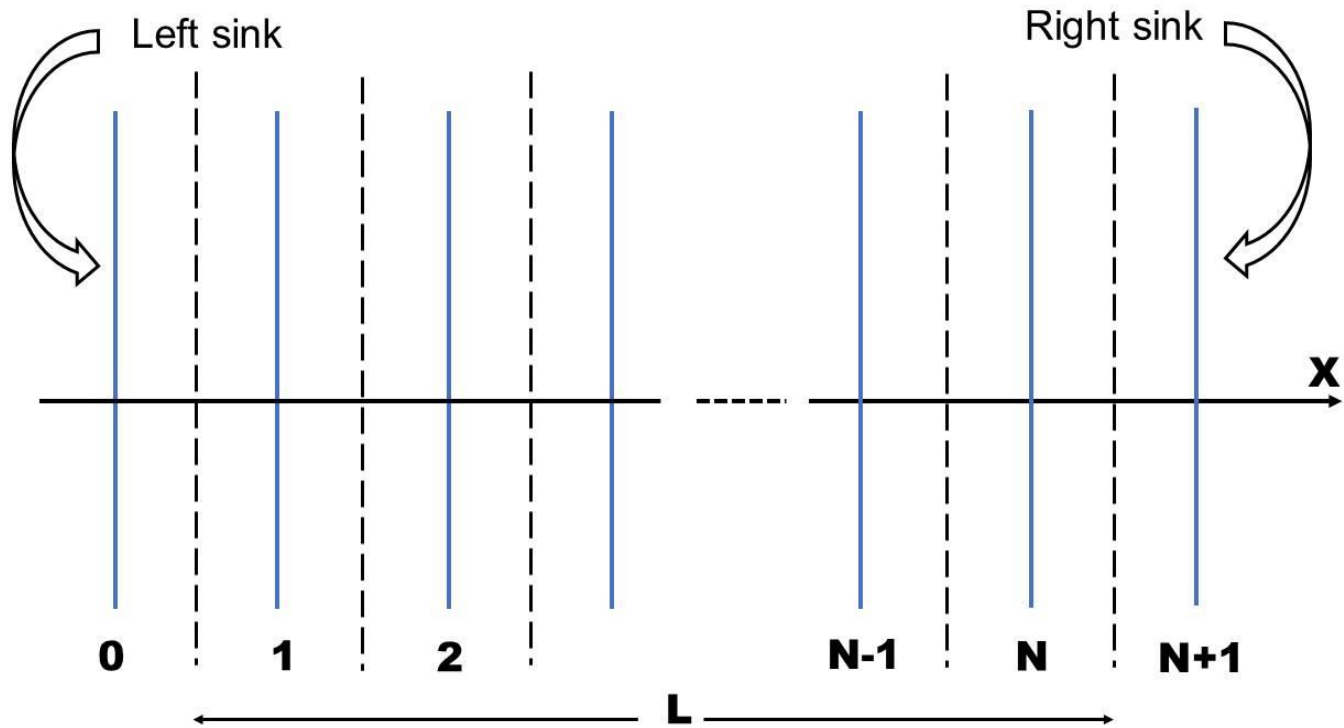

Figure 1.

Definition of the model system

The production of PD takes place via two channels in parallel, namely by thermal emission from the surface of the planar sink-and-sources and by the creation under irradiation. The elimination of PD takes place via three channels in parallel: namely, by recombination occurring immediately after creation (or aborted creation), by recombination occurring after diffusion (or diffusive recombination) and by absorption on the two sinks. Then the obvious constraint must be satisfied:

$$\left|J_{I,R}\right|+\left|J_{I,L}\right| \;=\; \left|J_{V,L}\right|+\left|J_{V,R}\right| \;=\; \Sigma_G^{eff} - \Sigma_{Rec}. \tag{2}$$

where $\Sigma_G^{eff}$, $\Sigma_{Rec}$ are the net creation rate in the system and the sum of mutual diffusive recombination events respectively. As a consequence, the imbalance of the fluxes (i.e. the numerator of each absorption bias) is equal in magnitude and takes opposite signs at the two sinks.

**II Elastic interaction of PD with planar sinks**

The chosen elastic interaction of the planar boundary L or R with the PD mimics that of a dislocation in α Fe while ignoring its angular dependence for sake of simplicity. Only the volume contribution is retained while dropping the effect of shears which are known to be of smaller magnitude. The expression combining the interactions between a defect 'd' located on plane 'i' with the two planar boundaries is given below:

$$E_{d,i} = kT \frac{\mu(1+\nu)\left|\Delta V_d^{rel}\right|}{3\pi(1-\nu)} \left( \frac{\{+,0,-\}b_L}{ia/2+\delta_L} + \frac{\{+,0,-\}b_R}{(N+1-i)a/2+\delta_R} \right) \quad (3)$$

where μ, ν are the shear modulus and Poisson ratio respectively, $b_L$ and $b_R$ the magnitude of the Bürgers vector on the left or on the right sink, $\Delta V_d^{rel}$ the relaxation volume of the defect 'd' measured in units of atomic volume. The '+' and '-' sign define the sign of the interaction of defect 'd' with the L and R sink : a minus sign '-' for an attraction denoted hereafter by the capital letter A and a plus sign '+' for a repulsion hereafter denoted by the capital letter R. The first term in the parenthesis accounts for the interaction of the defect with the left sink, which can be chosen > 0, = 0 or < 0; the second term accounts for the interaction of the defect with the right sink. The decrease of the shear modulus μ with temperature can be accounted for and amounts roughly to 18% between 300K and 1200K [20-21].

The above formula includes two distances $\delta_L$, $\delta_R$ which stand for the capture distance of the left (L) and of the right (R) sink respectively. They were set arbitrarily equal to $4b_L$, $4b_R$. Below this distance, the interaction does not follow the macroscopic elastic theory and the incoming defect is supposed to be immediately absorbed.

The signs of the interaction can be changed at will to account for different physical configurations:

- a wall of parallel edge dislocations with their tensile regions oriented toward the center of the slab will attract I and repel V; it is denoted by 'AR';
- a wall of parallel edge dislocations with their compressive regions oriented toward the center of the slab will attract V and repel I; it is denoted by 'RA';
- a cavity surface will be assumed to have no elastic interaction; it is denoted by '00'.

Each system is thus referenced by a quadruplet made of 'A', 'R' together with '0', naming first the L sink and its interactions with I and V in this order, and then the R sink and its interactions with I and V. For instance, AA | AA represents a L sink and a R sink attracting I and V, while AR | AA represents a L sink attracting I but repelling V together with a R sink attracting the two. 00 | AR mimics a neutral wall like a cavity on the left facing a dislocation wall on the right.

A further numerical value 'x' is added to the left part of the symbol only, if the strength of the elastic interaction at the left side is not identical to the strength of the interaction at the right side: for instance, AR2 | AA indicates that $b_L = 2\ b_R$. The absence of any numerical value means that the interactions are produced by Bürgers vectors of equal magnitude, i.e. a/2.

All the configurations explored hereafter use always the same right sink R with a Bürgers vector equal to a/2.

**III Irradiation model**

The irradiation is modelled by a uniform source term G (expressed in dpa s$^{-1}$) which creates isolated Frenkel pairs (FP). Only two values were probed, namely 10$^{-2}$ and 10$^{-4}$ dpa s$^{-1}$, with emphasis on the first one which yields more pronounced effects.

*the spatial correlation between the two partners of the FP upon creation is ignored, as well as their elastic interaction before recombining;

*notwithstanding the fact that only the component of the fluxes along the x-axis is considered, the recombination volume $V_{Rec}$ is however modelled as a true 3D object. The number $N_{Vrec}$ of atomic volumes it contains was set equal to 113: it encompasses up to 8 neighbor shells and is roughly spherical [22-23]. But it can be arbitrarily set equal to any other value (a few trials with a smaller value will be presented below). We give in Appendix A all the ingredients used to describe its geometry together with the quantities which introduce the recombination events into the balance equations.

**IV Jump frequencies and flux formulation**

The net defect flux from plane 'i' to plane 'i+1', is measured between the two planes and is defined by

$$J_{d,i+1/2} = \frac{a}{2}\left(4C_{d,i}W_{d,i\to i+1} - 4C_{d,i+1}W_{d,i+1\to i}\right) \tag{4}$$

where 'd' stands for the type of defect (I or V), $C_{d,i}$, $C_{d,i+1}$ are the defect concentrations expressed in atomic fractions on site 'i' and 'i+1', $W_{d,i\to i+1}$, $W_{d,i+1\to i}$ are the defect jump frequencies between one atomic site belonging to plane 'i' and one atomic site belonging to plane 'i+1'.

The diffusion model consists in defining the energy barriers to be crossed by each PD to reach a neighbor plane. The energies at the stable positions $E_{d,i}$ and $E_{d,i+1}$ are determined by the elastic interaction; the energy at the saddle position $E_{d,i+1/2}^{sadd}$ is defined through the quantity $\Delta_{d,i+1/2}^{sadd}$ depicted on Fig. 2 according to:

$$E_{d,i+1/2}^{sadd} = \left(E_{d,i} + E_{d,i+1}\right)/2 + \Delta_{d,i+1/2}^{sadd}. \tag{5}$$

The above description is not restrictive. The quantities $\Delta^{sadd}_{d,i+1/2}$ can be chosen with more or less sophistication. In the present case, for sake of simplicity, they will be equated to a fixed value, namely the migration barrier for each defect in the bulk without any elastic interaction $\Delta^{sadd}_{d,i+1/2} = E^{m}_{d,bulk}$. They can however be replaced by more precise values which are obtained with empirical potentials, or better, with ab initio calculations, while taking account of the shear components of the sink strain field. This improvement has been brought by numerous authors in a recent past [14-19].

We explain in Appendix B the reason why a discrete formulation of the flux must be adopted instead of the continuous expression deduced from the Nernst-Einstein approximation. Indeed, in the immediate neighborhood of the absorbing surface, where the elastic interaction hits its maximum, the drift component of the flux is no longer simply proportional to the elastic driving force. Ignoring this point may change the relative weights of the drift and gradient terms in the total flux to such an extent that it changes the sign of the flux. This modification of the terms entering the evaluation of fluxes, although never implemented in practice, was formerly suggested by previous authors [24-25].

Several barrier evaluations are compared in Appendix C. The interaction energy of the dumbbell defect with the strain field of a dislocation depends on the orientation of its dissociation axis. Including this feature would require a distinct diffusion equation for each orientation of the dumbbell, which is beyond the scope of the present contribution since our model remains basically one-dimensional. Fortunately, the major part of the interaction stems from the volumetric component; it is shown that the brute evaluation given in Eq. 3 is not too far from the most sophisticated ones, if conveniently modified. Thus, it can be reasonably expected that the qualitative conclusions of our calculations will not be drastically impacted by the crudeness of the description.

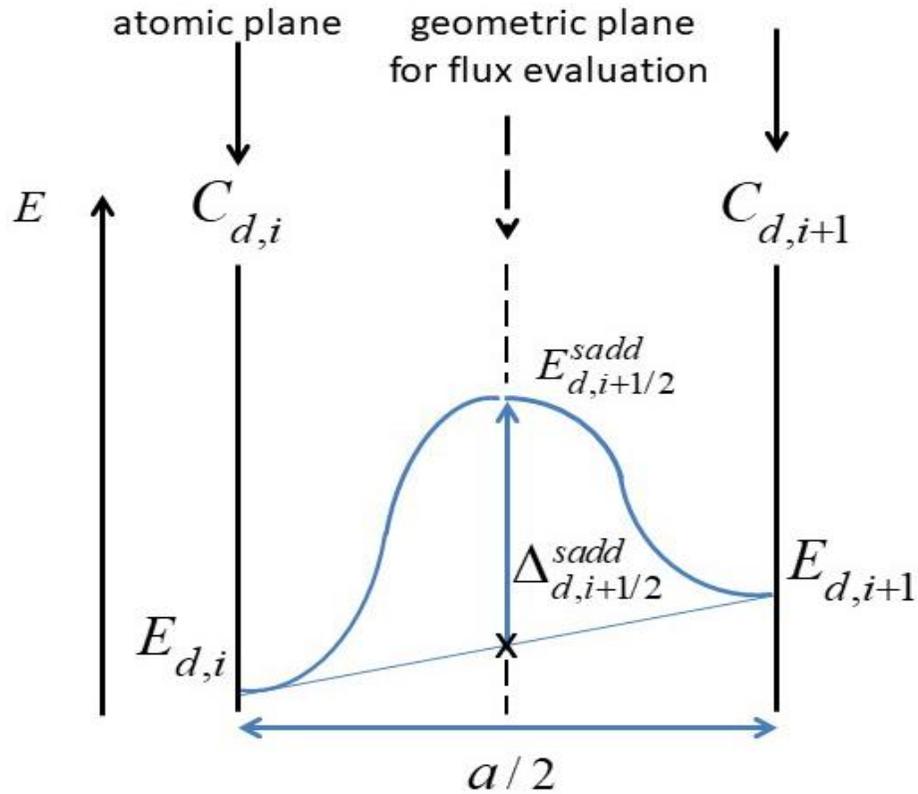

*Figure 2.*

*General model for jump barriers in the presence of a driving force*

**V Rates equations at stationary state and the resulting equations systems**

The starting global balance equation is written for a site belonging to plane 'i' and contains a diffusion term, a recombination term and a defect production term. The boundary conditions connect the last atomic plane of the matter slab ('1' and 'N') with the L and R sink-source plane ('0' and 'N+1') respectively.

$$\frac{\partial C_{d,i}}{\partial t} = -div(J_d)\big|_i + G_i^{eff} - Rec_i \tag{6}$$

where 'd' stands for the defect type, $G_i^{eff}$ is the effective source term taking into account the loss of defect due to aborted creation (see below Eq. 7) and $Rec_i$ the loss due to diffusive recombination.

This global equation is translated at the atomic level, resting only on the jump frequencies between adjacent planes which were defined earlier and on the site occupation probabilities. The matter balance per unit time is calculated for each site 'i', taking account of all the events which change its occupancy: departures and arrivals of the defect on the site due to diffusion, production by irradiation flux together with the elimination by a newly created anti-defect, diffusive recombination on the spot.

The Laplacian term is then discretized for each site 'i' according to:

$$-div(J_d)\big|_i = \left(J_{d,i-1/2} - J_{d,i+1/2}\right)/(a/2) = 4C_{d,i-1}W_{d,i-1\to i} - 4C_{d,i}(W_{d,i\to i-1} + W_{d,i\to i+1}) + 4C_{d,i+1}W_{d,i+1\to i}$$

where the multiplying factor 4 is the number of neighbors on an adjacent plane. The recombination term $Rec_i$ is no longer calculated through the homogeneous approach [13]: the latter becomes highly questionable in the vicinity of sinks where concentration gradients are steep and where the jump frequencies, which vary rapidly from one site to the next one, are very different from the jump frequency in the bulk. This is the reason why it is calculated exactly by listing, one after the other, all the jumps producing a recombination event. The contribution of the recombination appears as a relative probability which decreases the net result of each jump: a jump from site 'i' to 'i+1' does transport the defect from 'i' to 'i+1', provided the arrival site does not belong to the recombination volume of a nearby anti-defect. The details were reported in Appendix A.

The effective source term $G_i^{eff}$ due to irradiation is expressed in units of dpa s$^{-1}$. It is not equal to the nominal creation rate $G$, because the creation process is self-limited: indeed, no new defect can be created on some site 'i' as soon as this site belongs to the recombination volume Vrec of a nearby pre-existing anti-defect. A first order correction replaces merely the nominal creation rate $G$ by an effective one defined by:

$$G_i^{eff} = G\left(1 - V_{Rec}(C_{V,i} + C_{I,i})\right). \tag{7}$$

This effective creation rate depends now on the local concentrations of PD and accounts for this "aborted creation event" which is a non-thermal process. It could be improved further to allow for some overlap of the recombination volumes: as a result, the same corrective factor appears, but now squared [22]. The limiting concentration of FP is equal to 1/Vrec in the two approximations and we retain only the first order (and linear) correction term for sake of simplicity. This self-limitation effect was observed during irradiation below 10K and was used in the past to measure experimentally the recombination volume. The inclusion of this aborted creation event during the calculations can be switched on and off at will with convenient flags; it will be indicated by $Rc1$ or $Rc0$, respectively.

Tables Ia and Ib below gather together the terms which enter the equations monitoring the population of I on each plane 'i'. Table Ia gathers together the contributions of the I jumps leaving site 'i'.

| | Departures from site 'i' |
|---|---|
| Jumps from site 'i' to a first neighbor 'i-1' or 'i+1' | $-C_{I,i}\left(4W_{I,i\to i+1} + 4W_{I,i\to i-1}\right)$ |
| Jumps of a vacancy reaching the recombination volume Vrec(i) of I sitting on 'i' | $-C_{I,i}\,V2I(i)$ |
| Creation of an anti-defect (V) on any site belonging to the recombination volume of I | $-C_{I,i}*G*N_{Vrec}$ |

*Table Ia. Terms contributing to a decrease of the population of I on site 'i*

Table Ib gathers together the contributions of the I jumps arriving on site 'i'.

| | Arrivals on site 'i' |
|---|---|
| Jumps from site 'i-1', provided 'i' does not belong to the recombination volume of a nearby V | $+ 4C_{I,i-1}\, W_{I,i-1 \to i}\left(1 - I2V^{+}(i)\right)$ |
| Jumps from site 'i+1', provided 'i' does not belong to the recombination volume of a nearby V | $+ 4C_{I,i+1}\, W_{I,i-1 \to i}\left(1 - I2V^{-}(i)\right)$ |
| Creation on site 'i' by irradiation flux | $+ G_i^{\,eff} = G\left(1 - V_{Rec}(C_{I,i} + C_{V,i})\right)$ |

*Table Ib. Terms contributing to an increase of the population of I on site 'i'*

The effect of recombination is thus spread among the coefficients of the three concentrations $C_{I,i-1}$, $C_{I,i}$, $C_{I,i+1}$. The expressions $V2I(i)$, $I2V^{+}(i)$, $I2V^{-}(i)$ were given in Appendix A.

The following remarks are worth to be mentioned:

*the recombination event takes place when I and V, initially separated by a distance larger than the recombination radius, jump in the direction of each other. It takes account of the jumps of the two partners (jumps of I approaching V and the reverse) but the corresponding jump frequencies are evaluated at different locations in the system (i.e. in the energy gradient of the elastic force). As a consequence, a given recombination event is not seen to take place at the same location in the system by each partner:

* "from the point of view of I": I (and thus V, too) vanishes on site 'i' (1rst and 2nd line of Table Ib);

\* "from the point of view of V": V (and thus I, too) vanishes at some distant site belonging to the recombination Vrec(i) of I.

In the same way, I disappears from site 'i' because a vacancy V arrived at some distant site of its recombination volume Vrec(i) (2nd line of Table Ia). This is the reason why the recombination events in the whole system, which rest on the knowledge of locally varying jump frequencies, cannot be summed up under the form of a unique term on each plane 'i' as it was in the continuous and homogeneous approach using bulk diffusivities [13];

\* the creation event does not account of the fact that I and V are created on different lattice sites;

\*the irradiation source $G_i^{eff}$ contains a first order self-limitation term describing the creation process in the presence of pre-existing defects (last line of Table Ib) and the correction uses the PD concentrations on plane 'i' only. This is also an approximation because it should include the recombination volumes of pre-existing defects centered on sites 'k' lying at some distance from 'i'; but this point was ignored in order to simplify the formulation and the numerical solution.

Including these recombination events due to diffusion in the calculations can be switched on or off at will with convenient flags; they will be mentioned by $Rd1$ or $Rd0$, respectively. Merging together the two pieces of information about the recombination events gives rise to a single symbol $Rcjd\ i \quad i,j=\{0,1\}$ which will also appear in the figures displayed afterwards.

Since only the stationary state is looked for, the time derivative is set to zero. The starting equation can be split into a system of quasi-linear equations, containing as many equations as the number of atomic planes for each PD. The terms which do not contain the concentration of the defect under consideration are shifted to the RHS of the equations.

For the interstitial species:

$$(1-\delta_{i1})\ ail(i)\ C_{I,i-1} + aid(i)\ C_{I,i} + (1-\delta_{iN})\ aiu(i)\ C_{I,i+1} = -G\left(1-N_{Vrec}C_{V,i}\right) - \delta_{i1}\ ail(1)\ C_{I,L}^{eq} - \delta_{iN}\ aiu(N)\ C_{I,R}^{eq}$$

$$i \in [1,N] \tag{8}$$

The Kronecker symbols $\delta_{i1}$, $\delta_{iN}$ are used to take account of the concentrations at the left (L) and right (R) boundaries, namely $C_{I,L}^{eq}$ and $C_{I,R}^{eq}$, which do not belong to the set of unknowns.

With the help of Table Ia and Ib, the coefficients of the lower, main and upper diagonals are easily defined by:

$$\begin{aligned}
ail(i) &= + 4\, W_{I,i-1\to i}\left(1 - I2V^{+}(i)\right) \\
aid(i) &= -(4W_{I,i\to i+1} + 4W_{I,i\to i-1}) - V2I(i) - G*N_{Vrec} \\
aiu(i) &= + 4\, W_{I,i-1\to i}\left(1 - I2V^{-}(i)\right)
\end{aligned} \qquad (9)$$

The major contribution comes from the jump frequencies; the recombination jumps appear under the form of a correction, which remains small at the higher temperatures and becomes more important at lower ones.

Similar expressions for the vacancy species are deduced by exchanging the symbols I and V and yield the system of equations below:

$$(1-\delta_{i1})\, avl(i)\, C_{V,i-1} + avd(i)\, C_{V,i} + (1-\delta_{iN})\, avu(i)\, C_{V,i+1} = -G\left(1 - N_{Vrec}C_{I,i}\right) - \delta_{i1}\, avl(1)C_{V,L}^{eq} - \delta_{iN}\, avu(N)C_{V,R}^{eq}$$

$$i \in [1, N] \qquad (10)$$

together with:

$$\begin{aligned}
avl(i) &= + 4\, W_{V,i-1\to i}\left(1 - V2I^{+}(i)\right) \\
avd(i) &= -(4W_{V,i\to i+1} + 4W_{I,i\to i-1}) - I2V(i) - G*N_{Vrec} \\
avu(i) &= + 4\, W_{V,i-1\to i}\left(1 - V2I^{-}(i)\right)
\end{aligned} \qquad (11)$$

Each system of equations is then alternately solved as if it were truly tri-diagonal. The solution of system (8) for I is first solved and used as an input for the solution of system (10) for V, and conversely at the next iteration. The nonlinear terms, which represent the effect of the recombination events, modify only slowly the coefficients of the three diagonals and a limited number of iterations is necessary, from several tens at higher temperatures up to several thousand around room temperature. Fortunately, solving a tri-diagonal system remains always an easy task and the back and forth procedure keeps its efficiency above room temperature. The convergence basin is large and any starting point is convenient.

The iterations are stopped when the following empirical criterion was met:

$$\sum_{i=1}^{N} \frac{|RHS(i) - LHS(i)|}{|RHS(i)|} < \varepsilon \qquad (12)$$

where RHS and LHS stand for the values of the right and left-hand-sides of each equation and $\varepsilon$ is set arbitrarily equal to $10^{-10}$. It was found that the solution could no longer be noticeably improved when performing more iterations. For the case of thermal equilibrium ($G_i^{eff} = 0$), the ratio in Eq. 12 above is replaced by the numerator.

The recombination events happening per unit time in the whole system are counted exactly through two independent procedures:

- "from the point of view of I" by summing the quantities V2I, I2V$^+$, I2V$^-$ which enter the coefficients of system (8):

$$\sum_{i=1}^{N} \left( V2I(i) + I2V^+(i) + I2V^-(i) \right); \qquad (13)$$

- "from the point of view of V" by summing the quantities I2V, V2I$^+$, V2I$^-$ which enter the coefficients of system (10):

$$\sum_{i=1}^{N} \left( I2V(i) + V2I^+(i) + V2I^-(i) \right). \qquad (14)$$

It is checked throughout the calculations that the two summations (13) and (14) yield strictly equal quantities (within round-off errors), since each pair of sites ('i', 'j') which is a candidate for a recombination event is scanned in the two systems of equations (but not in the same order), either through a jump of I approaching V or a jump of V approaching I.

The three processes working in parallel for defect elimination are

* the self-limited or aborted creation (AC) which is a non-thermal process,

* the diffusive recombination losses which imply thermally activated jumps

* the absorption fluxes on the sinks.

Their relative weights are expressed below as proportions of the total nominal defect production $N \times G$ and by construction always sum up exactly to unity for each defect.

The convergence is slowed down around room temperature where diffusion to sinks becomes less efficient and where the elimination of PD happens mainly by diffusive recombination or aborted creation. The terms connected to diffusive recombination in the coefficients are no longer a small correction and represent progressively an overwhelming part, which requires a larger number of iterations.

### VI Results

All the calculations presented below were done with jump barriers evaluated through the brute hydrostatic approximation (Eq. 3) or the modified interaction discussed in Appendix C. Results are displayed with the notation 'brute' and 'modif'. Voigt-and-Reuss' (V+R) averaged expressions of µ and ν (Equation C4 in Appendix C) are used throughout the calculations. The temperature dependence of the elastic constants is included or not at will with a convenient flag through the notations 'CijT' or 'Cij0' respectively.

The calculations can switch on or off the irradiation flux and/or the diffusive recombination events as explained above; thus concentration profiles will be obtained for all configurations with or without recombination. It is worth noticing that the non-thermal recombination event upon creation, due to the overlap of recombination volumes, is a channel for elimination which must always be kept: at room temperature, for large systems and large production rates, point defect concentrations were found to reach unrealistic values larger than unity. Indeed, the equations which are used above do not include any upper limit for the value of the solution.

The system configurations explored below will be of the following types: 00 | AR, AR2 | AR, RA | AR. The first system will be studied in more detail and considered afterwards as a reference to which the others will always be compared for highlighting the differences. The symmetric configuration AR | AR will be briefly commented first.

**VI-1 Symmetric configurations**

The system is made of two sinks with identical elastic properties and has a symmetry plane passing through the center of the slab. Along this plane, the elastic forces and the fluxes vanish and the concentrations of the two defects reach their maximum. It is the only configuration in which the zero-force surfaces (ZforceS) and the zero-flux surfaces (ZfluxS) coincide.

This configuration assigns an equal strength to the two sinks and no bias can be observed of course, as already noted by former authors [10-12]: the latter worked with a single dislocation in a cylinder with a zero-flux boundary condition, the latter playing the same role as the symmetry plane at the center of the slab in the present system:

*the fluxes of I and V at the sinks are always equal and do not depend on the elastic interaction;

*when recombination is ignored, they do not depend further on temperature and amount to $N\,G/2$ (Eq. 2)

*the dependence on temperature and on the type of interaction is observed only if recombination is allowed, but the two defect fluxes remain equal anyway.

At last in thermal condition (G = 0) calculations find that two non-vanishing inward fluxes coming from both ends (positive at L sink and negative at R sink) show up in order to compensate the consumption of defects produced by the recombination events in the bulk at that temperature. Even without irradiation, the system is no longer in equilibrium but rather in a stationary state. The point does not quantitatively matter in the present situation where the defect and anti-defect have very high formation energies; but it might have some influence in materials where I and V bring contributions to matter transport of comparable magnitude (like silicon).

### VI-2 Configuration 00|AR (N=101)

The distance between the two sinks corresponds to the distance between parallel dislocations in a 3D system with a density close to $10^{16}$ m$^{-2}$. The calculations are performed with the elastic constants at 0K (denoted by 'Cij0') unless otherwise indicated, together with the brute and modified interactions. The fluxes are measured as the number of atoms crossing the lateral facet of the system (area = $a^2/2$) during the unit time.

The L sink is elastically neutral whereas the R sink attracts I and repels V. The elastic forces produced by the two sinks never compensate each other and there are no ZforceS. The fluxes are evaluated in two different ways: the so-called 'lattice flux' is calculated with Eq. B3 of Appendix B, whereas the so-called 'NE flux' is evaluated with Eq. B5.

### IV-2-1 Influence of recombination on concentration profiles and absorption bias

On Fig. 3 are displayed the relative contributions of different elimination channels for the two species and for the brute interaction; as already mentioned, the contributions sum up to unity for each species at any temperature. The role of recombination events can be clearly detected up to 500K: the purple crosses indicate the amount of aborted creation whereas the red crosses indicate the amount of diffusive recombination. This threshold temperature depends on the size of the system (see below for N = 1001).

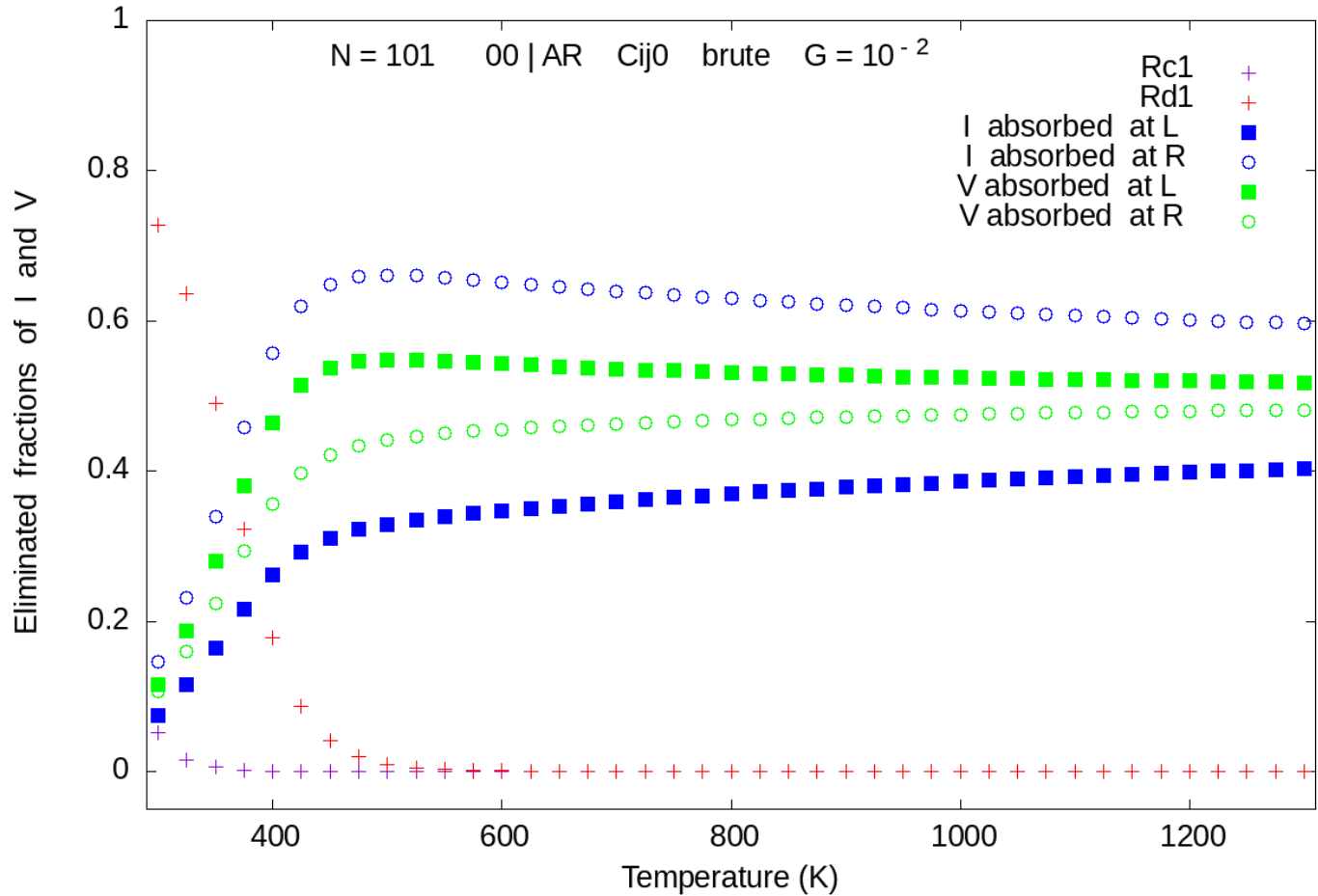

*Figure 3.*

*Eliminated fractions of I and V species as a function of temperature with the brute interaction and taking the recombination into account Rc1d1.*

Due to the sustained homogeneous production of PD, the concentrations in the slab are always larger than the thermal concentrations at the absorbing boundaries; as a consequence the ZfluxS always exist. On Fig. 4-5 are displayed the I and V concentration profiles at 300K and 800K calculated with the brute interaction, when evaluated with or without recombination (Vrec =

113). Aborted creation plays a weak role: when switching from Rc0 to Rc1 the change of the profiles is much smaller than the change brought by switching from Rd0 to Rd1.

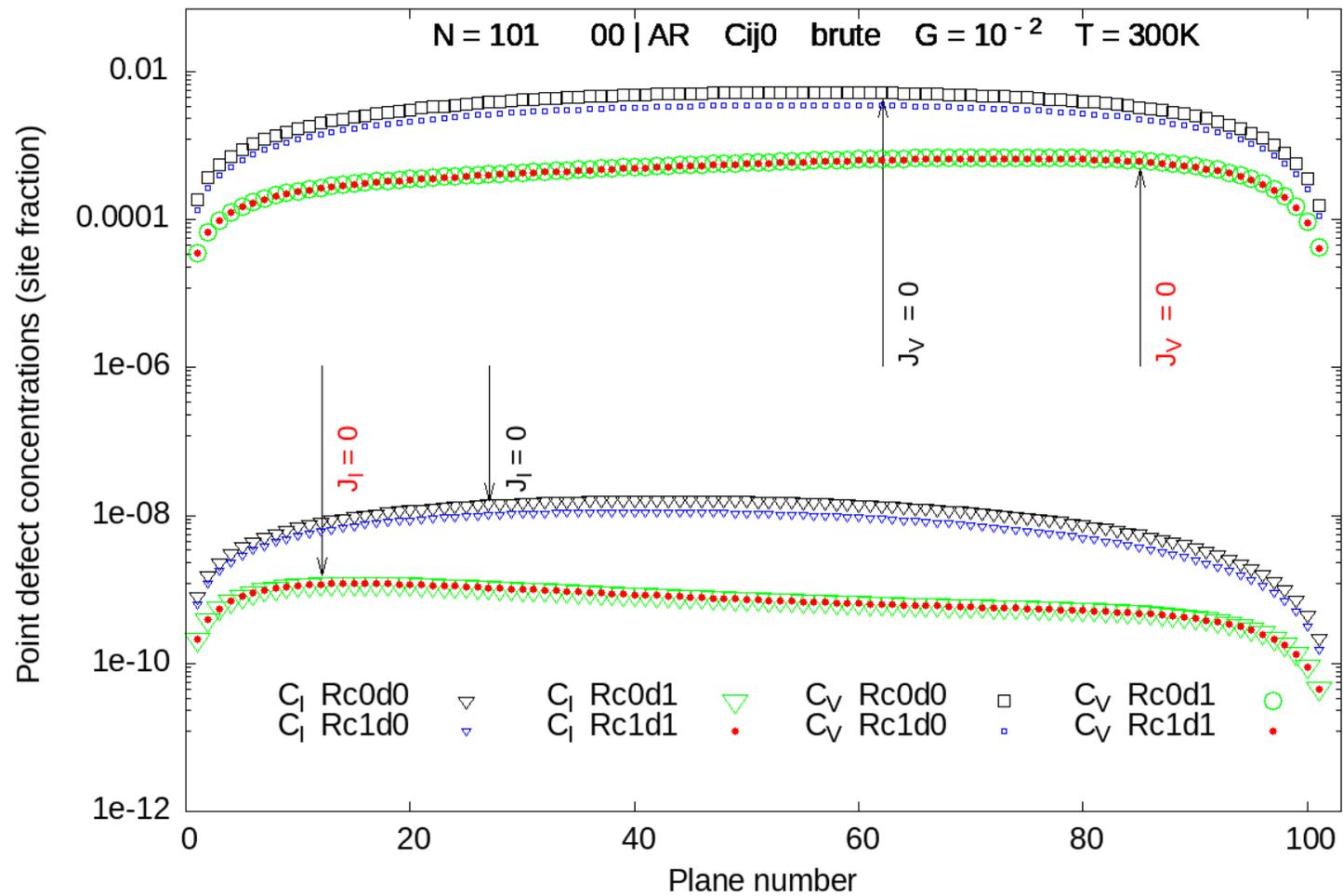

*Figure 4.*

*Concentration profiles for I and V at 300K, with or without recombination, together with the ZfluxS.*

Neglecting the recombination overestimates the concentrations by one order of magnitude at 300K; but the difference nearly vanishes above 800K. At last the ZfluxS have different locations for the two defects and their positions change with temperature.

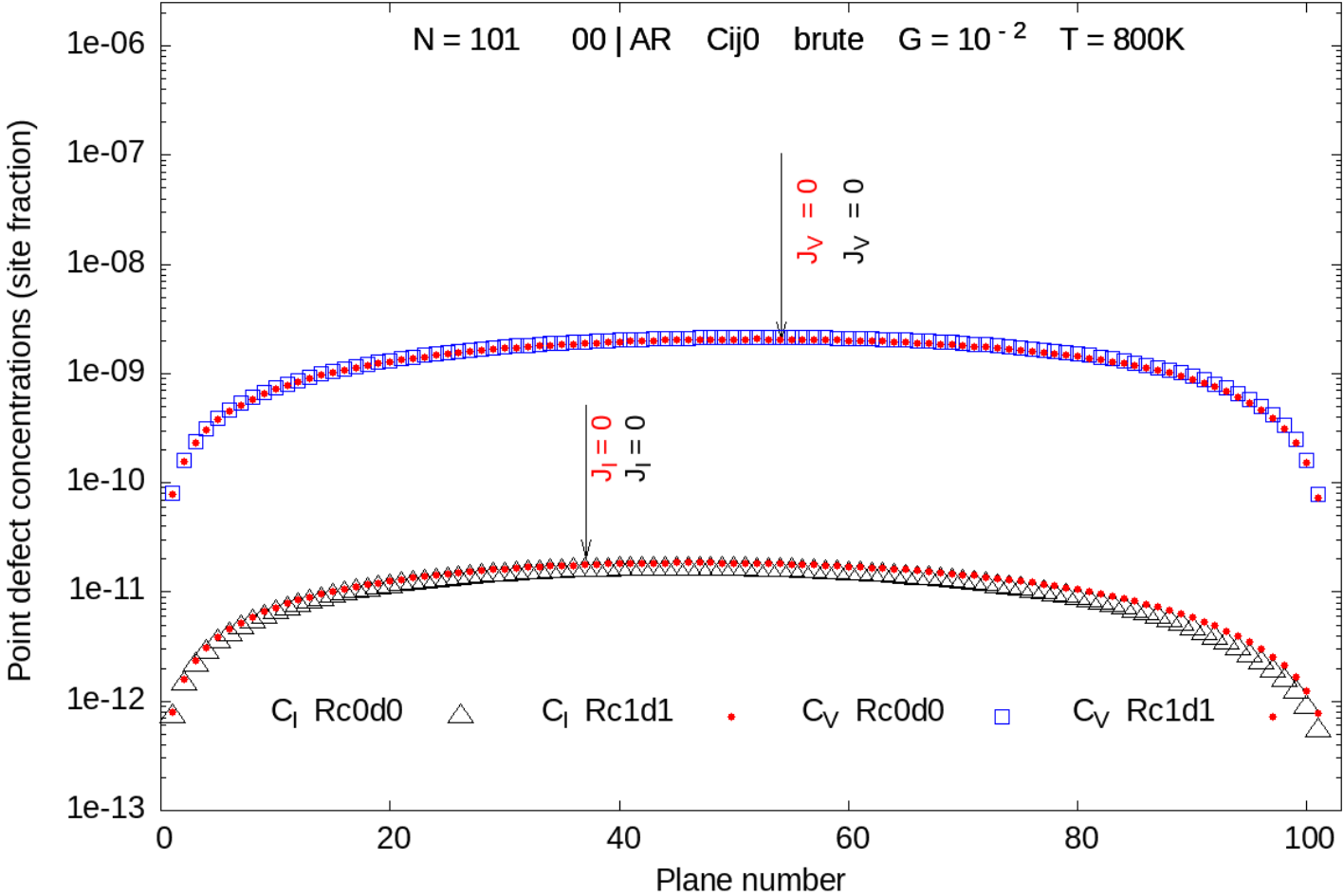

Figure 5.

Concentration profiles for I and V at 800K, with or without recombination, together with zero-flux planes.

The L and R absorption biases, when evaluated with expression (1), exhibit a general decreasing trend with increasing T as expected, but this decrease is not monotonous. The absolute value of the negative L bias and positive R bias start with an initial increase between 300K and 450K (Fig. 6-7).

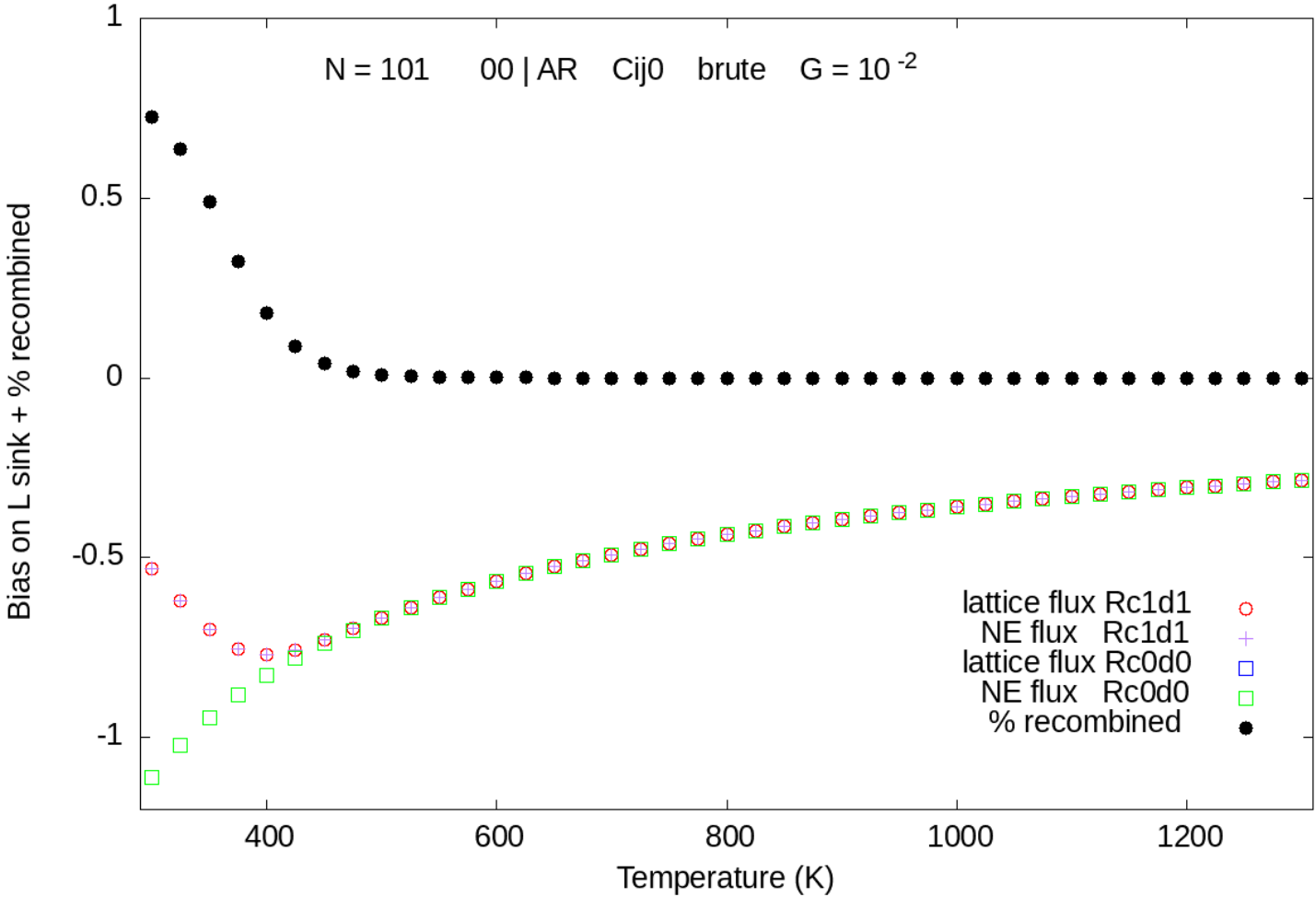

Figure 6.

*Absorption bias at the L sink with lattice and NE fluxes with or without recombination*

The sign change of the derivative is connected to the amount of recombination in the system. Indeed, more recombination decreases the two fluxes by an equal amount but change their ratio. For the L sink having no elastic force, the lattice and NE evaluations of the fluxes coincide nearly everywhere but at the lower temperatures. For the R sink developing a stronger interaction, the two evaluations depart markedly from one another over a large temperature domain.

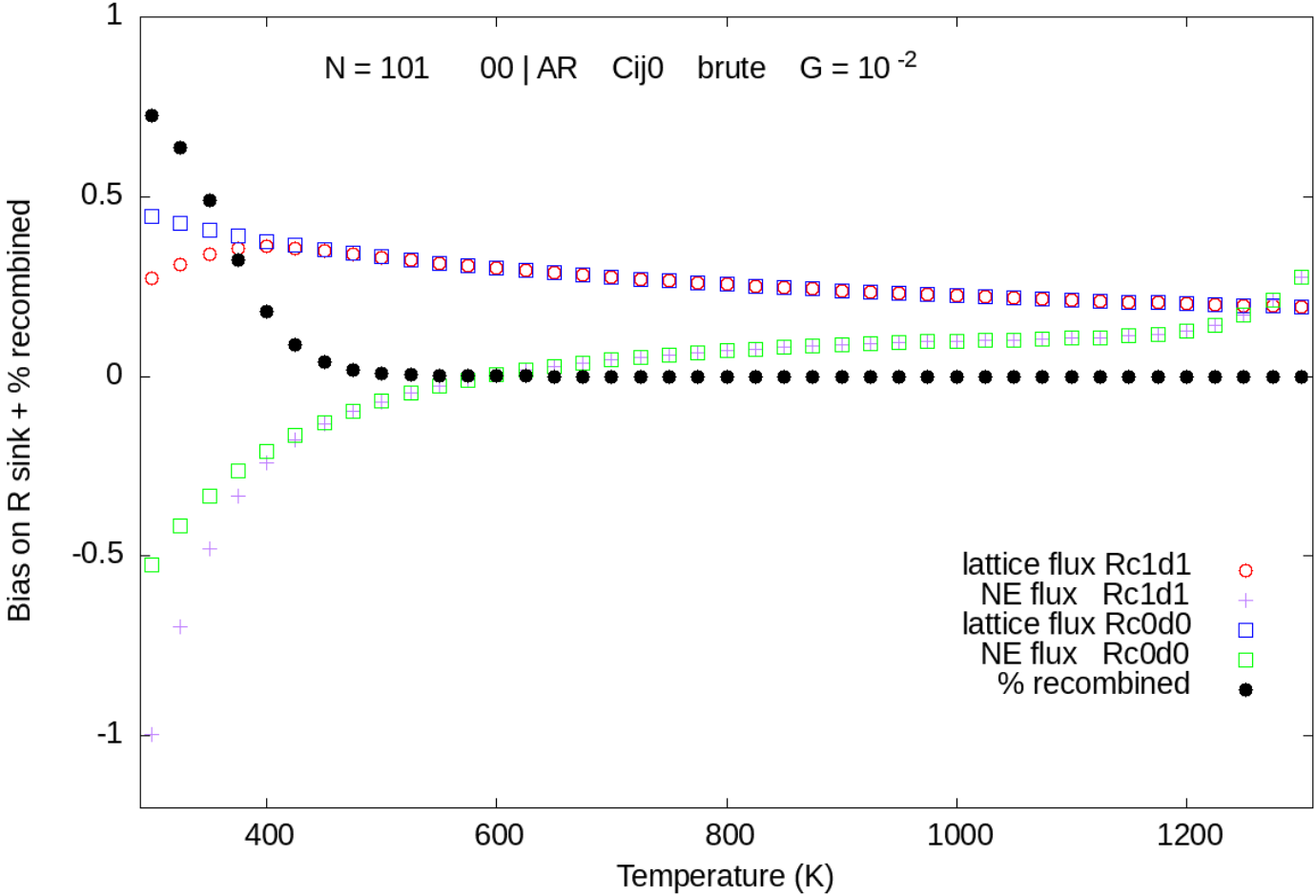

*Figure 7.*

*Absorption bias at the R sink with lattice and NE fluxes with or without recombination.*

### IV-2-2 Influence of the flux formulations on the gradient and drift terms

Figure 8 displays the two components of the interstitial flux (gradient term and drift term of Eq. B3 and B5) in the region close to the R sink where the interaction is noticeable. The correction on the gradient term shows up on the departure of the two blue curves; the correction between the two drift terms (NE and lattice) shows up between the yellow and green curves, respectively. The resulting fluxes (NE and lattice) are displayed with the orange and purple curves, respectively.

The corresponding curves for the vacancy, which experiences an elastic force roughly ten times smaller show that the departure between the two evaluations can be hardly perceived (Fig. 9), for the negative drift component (since V are repelled by the R sink) as well as for the positive gradient component.

On Figure 10 are displayed the lattice and NE evaluations of I and V total fluxes close to the R sink. The NE evaluation of the I flux and of the V flux (green and blue curve, respectively) exhibit a cross-over around plane n° 100, producing an absorption bias of the wrong sign (already detectable on Fig. 7), whereas the lattice expressions (purple for I and blue for V) yield an absorption bias AB with the correct sign.

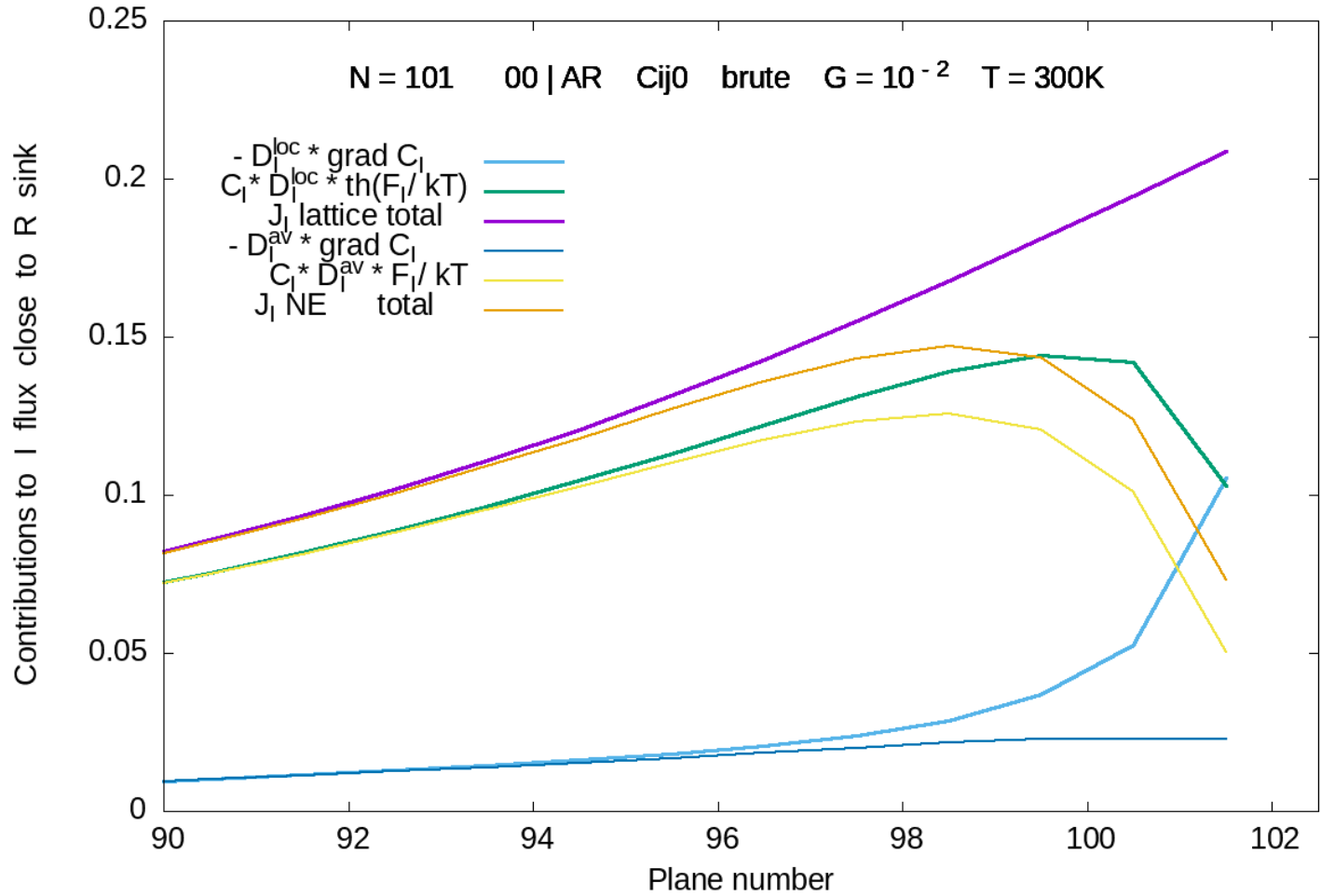

*Figure 8.*

*Spatial zoom on I flux components at the R sink with lattice and NE expressions and inclusion of full recombination Rc1d1.*

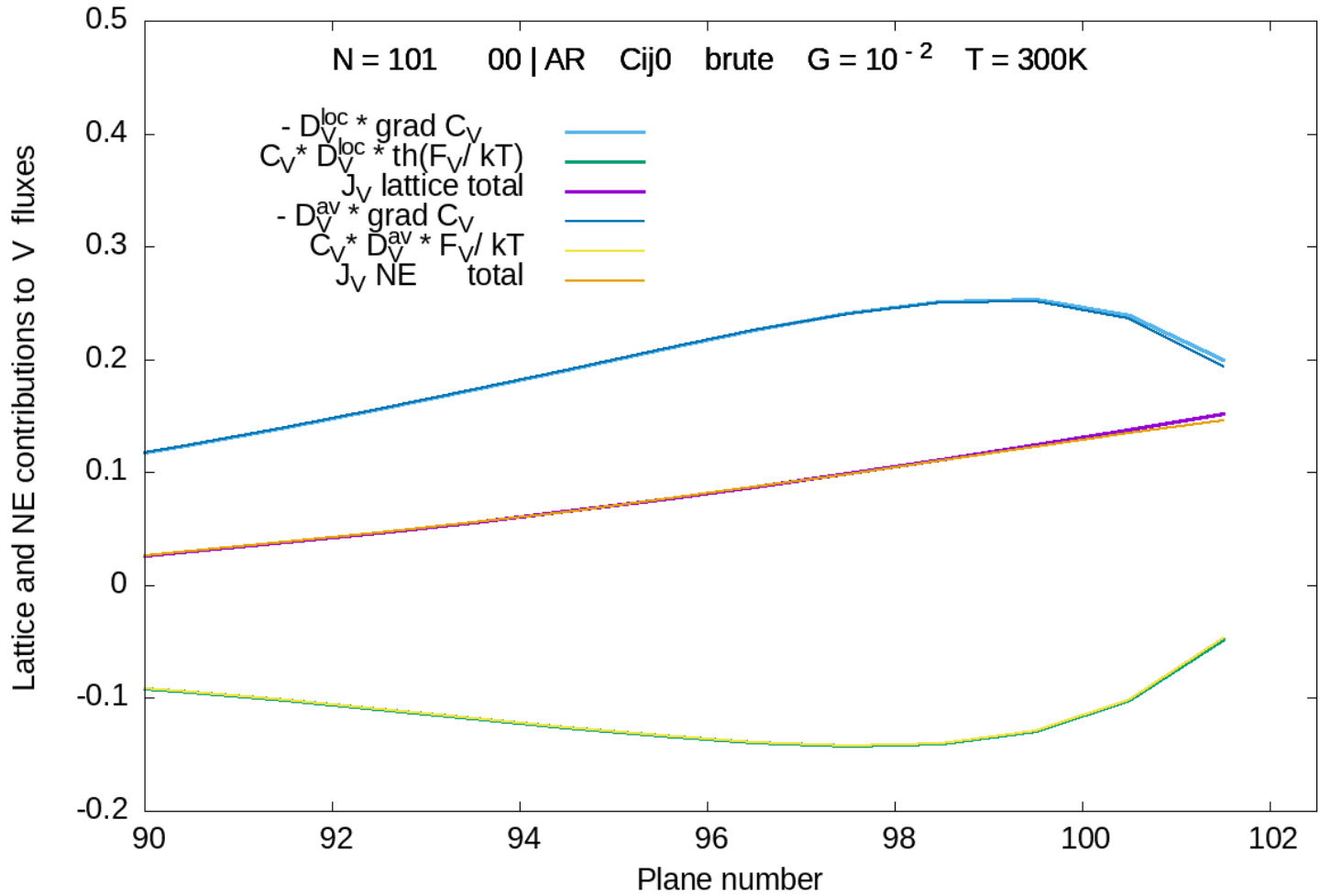

*Figure 9.*

*Spatial zoom of V flux components at the R sink with lattice and NE expressions and inclusion of full recombination Rc1d1.*

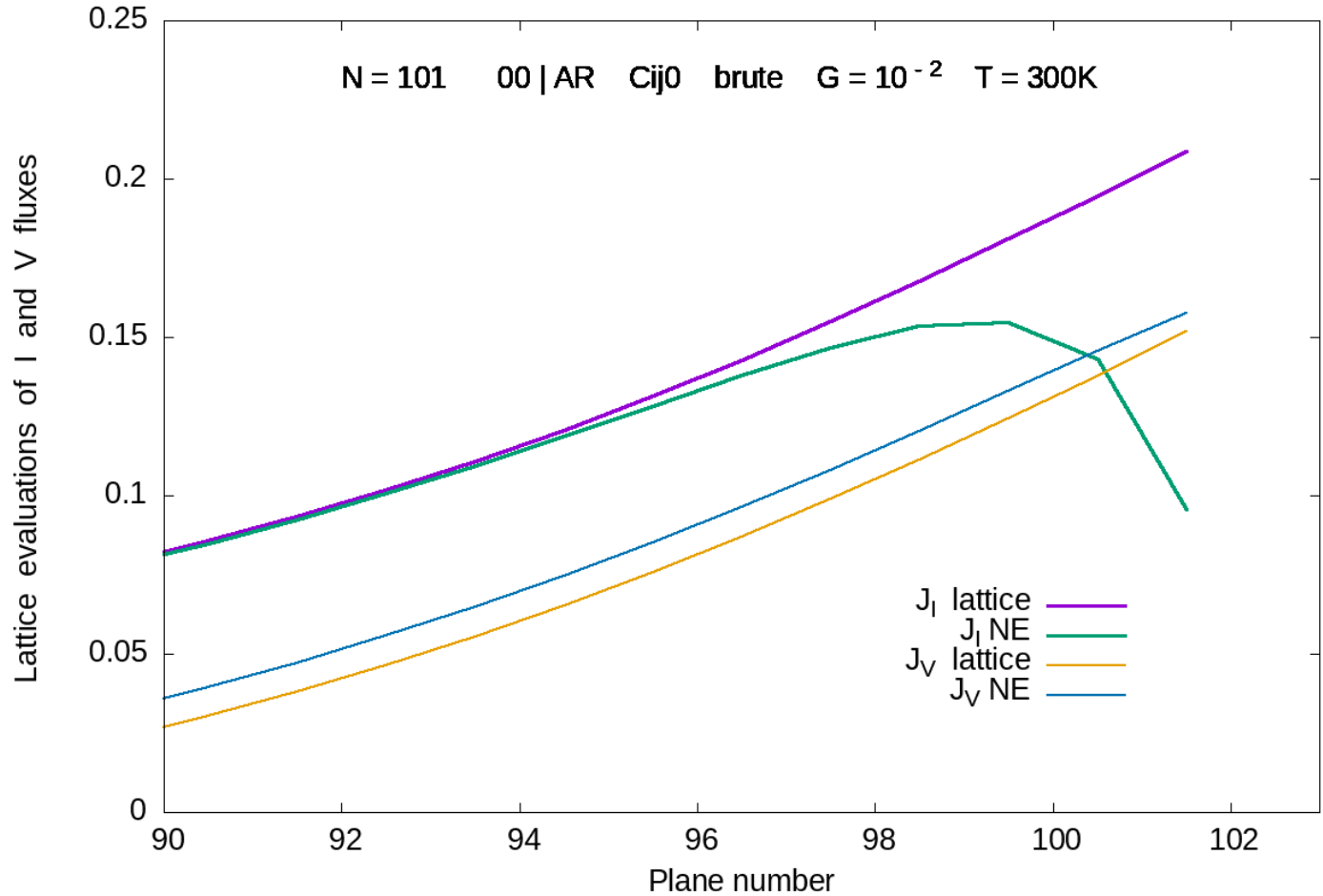

Figure 10.

*Spatial zoom on I and V fluxes at the R sink with lattice and NE expressions and inclusion of full recombination Rc1d1.*

### IV-2-3 Influence of the interaction on the absorption bias AB

The AB changes slightly with the detail of the interaction (brute interaction versus modified). Since our modified interaction is always softer than the brute one, the bias is slightly decreased (roughly by 20%) over the whole temperature range, as can be seen on Fig. 11.

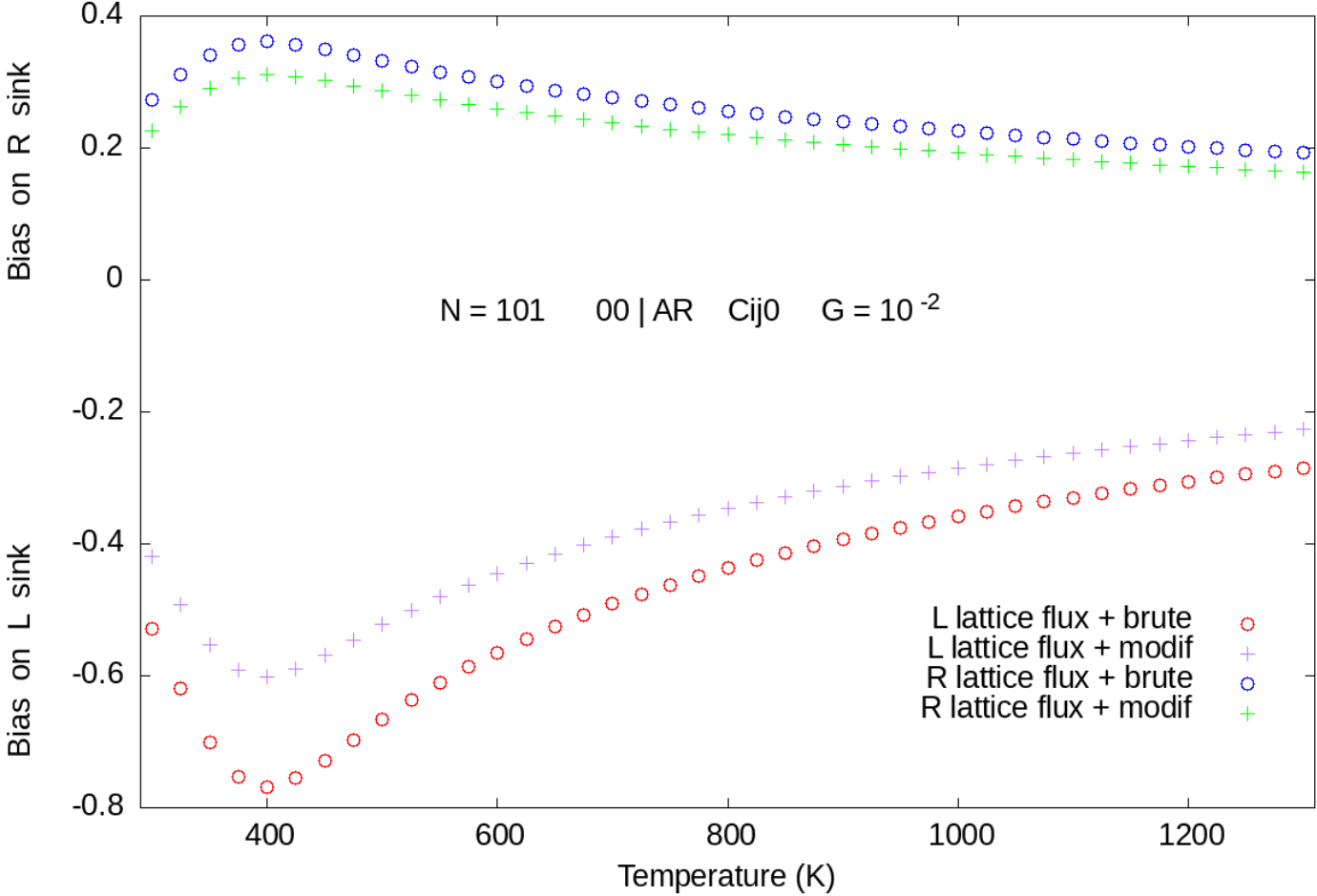

Figure 11.

Absorption bias at the L and R sink with brute and modified interaction together with elastic constants at 0K and inclusion of full recombination Rc1d1.

### IV-2-4 Influence of temperature dependent elastic constants

If the temperature dependence of the elastic constants is taken into account, the resulting softening of the interaction is detectable only at the higher temperatures: the decrease of the absorption bias is small (compare Fig. 7 above and 12 below).

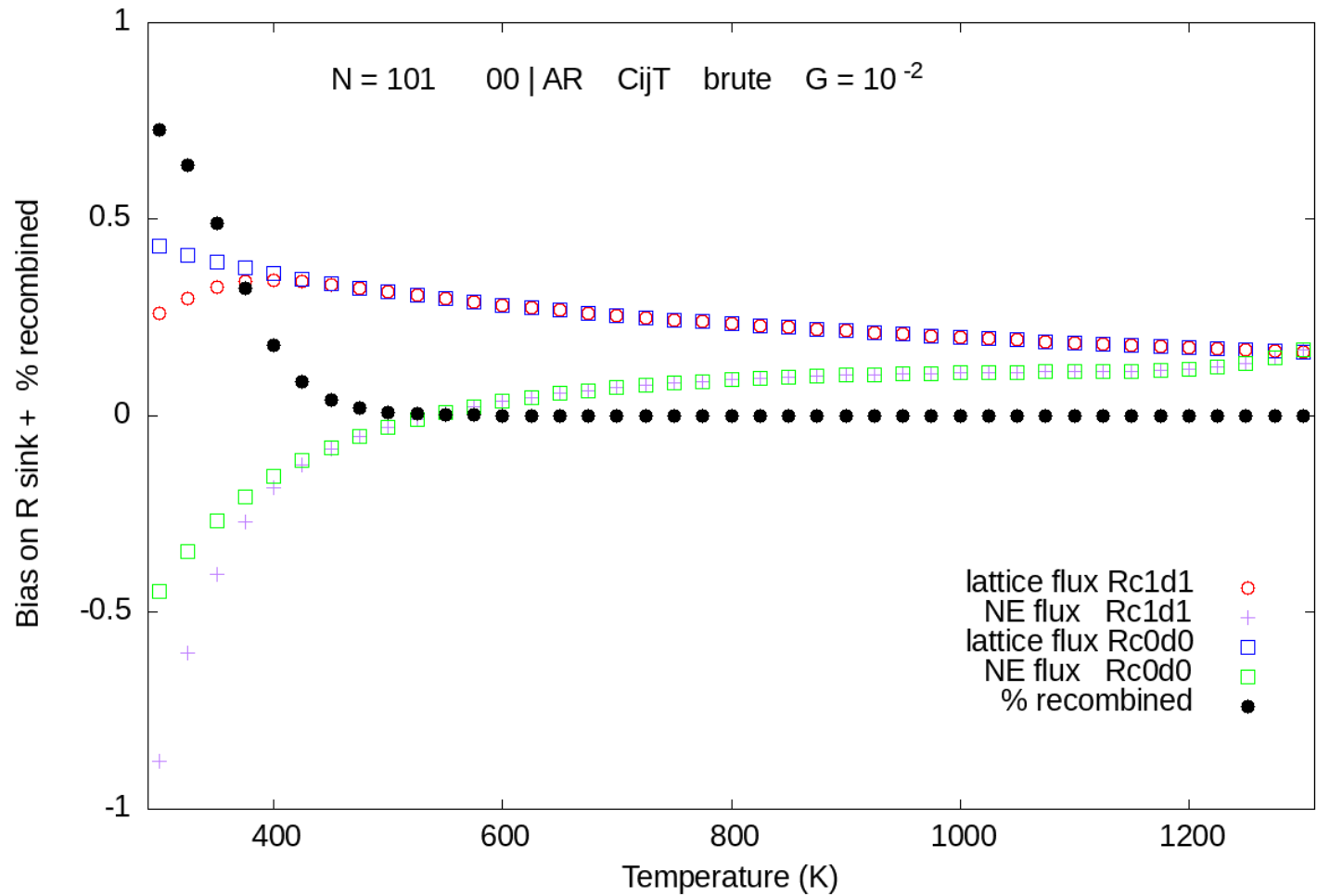

Figure 12.

Absorption bias at the R sink with lattice and NE fluxes with or without recombination.

The elastic constants are reduced by 18% (denoted with 'CijT') at the higher temperatures where the resulting softening of the interaction starts to be detectable. The absorption bias AB is slightly decreased (compare Fig. 11 above and 13 below).

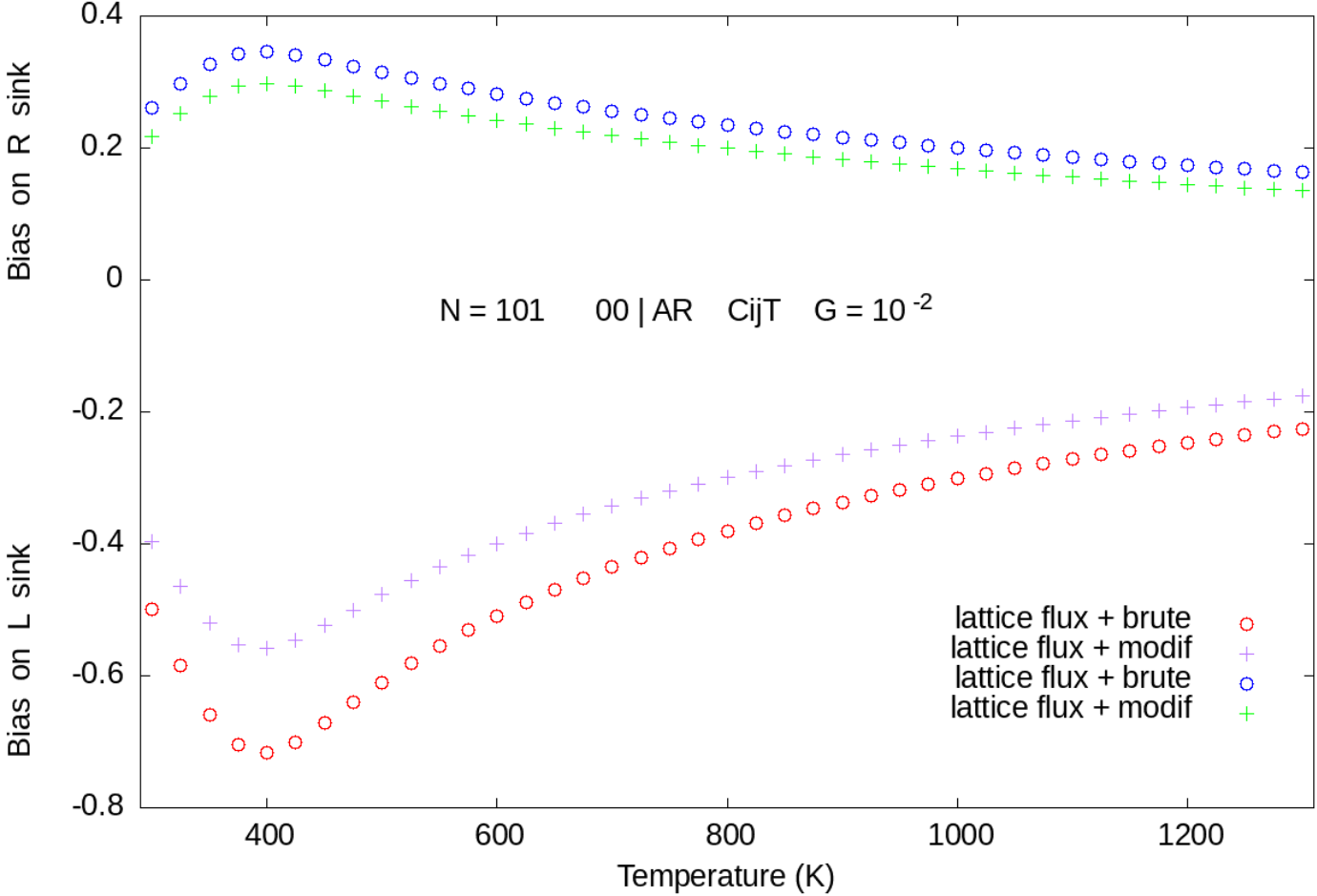

Figure 13.

*Absorption bias at the L and R sink with the brute and modified interaction together with temperature dependent elastic constants and inclusion of full recombination Rc1d1.*

### IV-2-5 Influence of the creation rate

When the creation rate is lowered down to $10^{-4}$ dpa/s, the above outlined trends still hold. The effects of recombination events are now detectable on a smaller temperature range and vanish around 350K instead of 450K. The fluxes are approximately divided by a factor of $10^2$, but the biases displayed on Fig. 14 are of comparable magnitude with those displayed on Fig. 11.

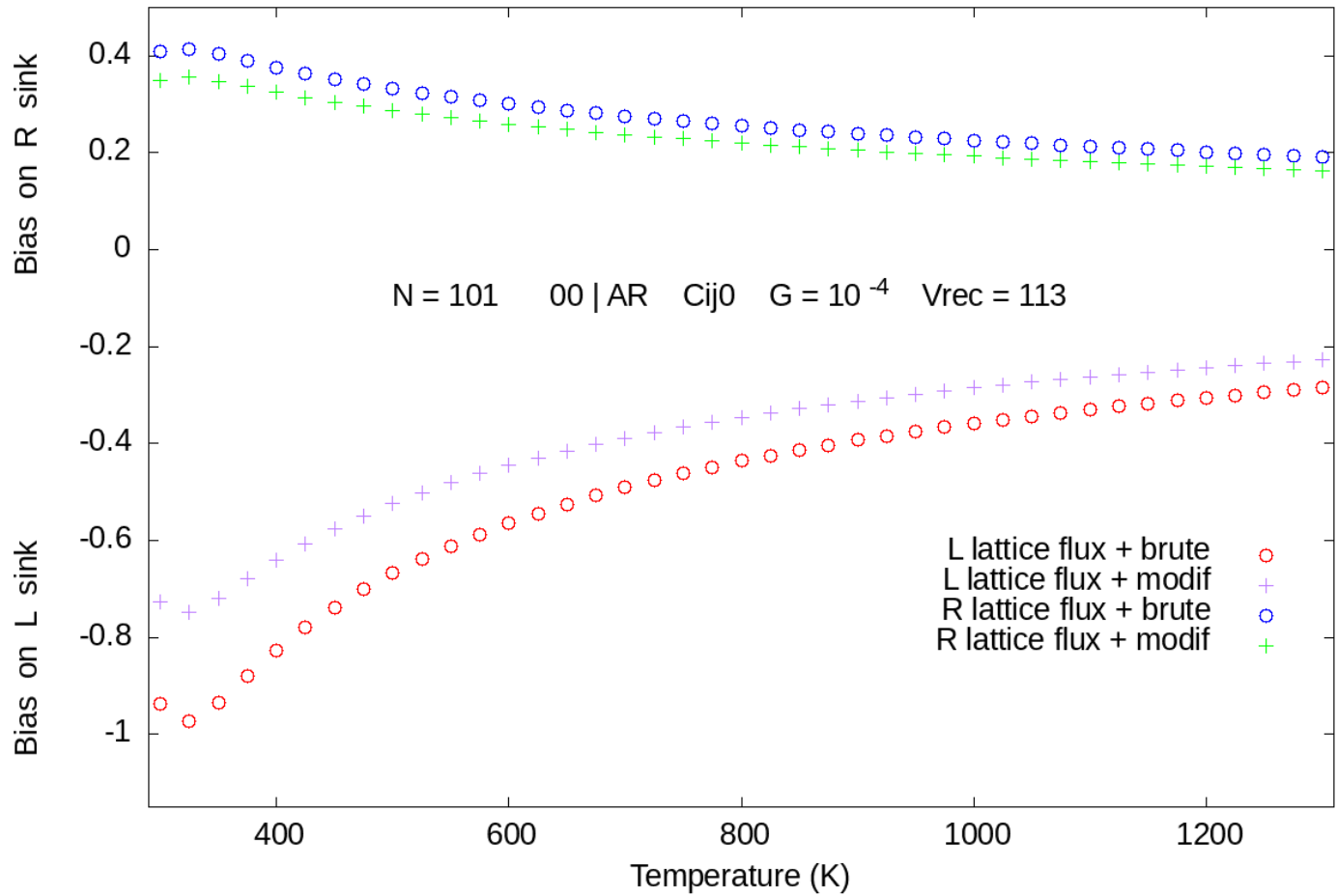

*Figure 14.*

*Absorption bias at the L and R sink with the brute and modified interaction with a reduced creation rate and inclusion of full recombination Rc1d1.*

### VI-2-6 Influence of system size (N=1001)

The system is now made of 1001 planes. The distance between the two sinks corresponds to the distance between parallel dislocations in a 3D system with a density equal to $10^{14}$ m$^{-2}$. The system experiences the creation of ten times more defects. However, detailed inspection of the data shows that the I and V fluxes at both ends are slightly smaller than the fluxes measured on the small system by at most 50%.

Fig. 15-16 display the concentration profiles for the vacancy V and for the interstitial I in the slab. The size of the system is such that the two sinks, in the absence of recombination, cannot eliminate defects at a sufficient rate to maintain the concentration level at an acceptable level, mainly for the vacancies: indeed, when recombination is not taken into account (Rc0d0), the maximum concentration for V is around 0.4 which is physically unreasonable. It is observed that including successively all the contributions of the recombination events lowers gradually the curve from the starting curve obtained with Rc0d0 down to Rc1d0, Rc0d1 and finally to Rc1d1. The contribution of the diffusive recombination (i.e. the change from Rd0 to Rd1) is as already noted before more important than the contribution of aborted creation (i.e. the change from Rc0 to Rc1).

The overshoot beyond unity suggests that the equation system is inadequate because nothing prevents the solution from exceeding unity; the differential equation should probably be reformulated with new variables $C_v/(1+C_v-C_i)$ and $C_i/(1+C_v-C_i)$

Fig.17 compares the two curves obtained with Rc1d1 but with different recombination volumes; the value of the latter apparently does not matter much since Vrec=9 is nearly as efficient as Vrec=113.

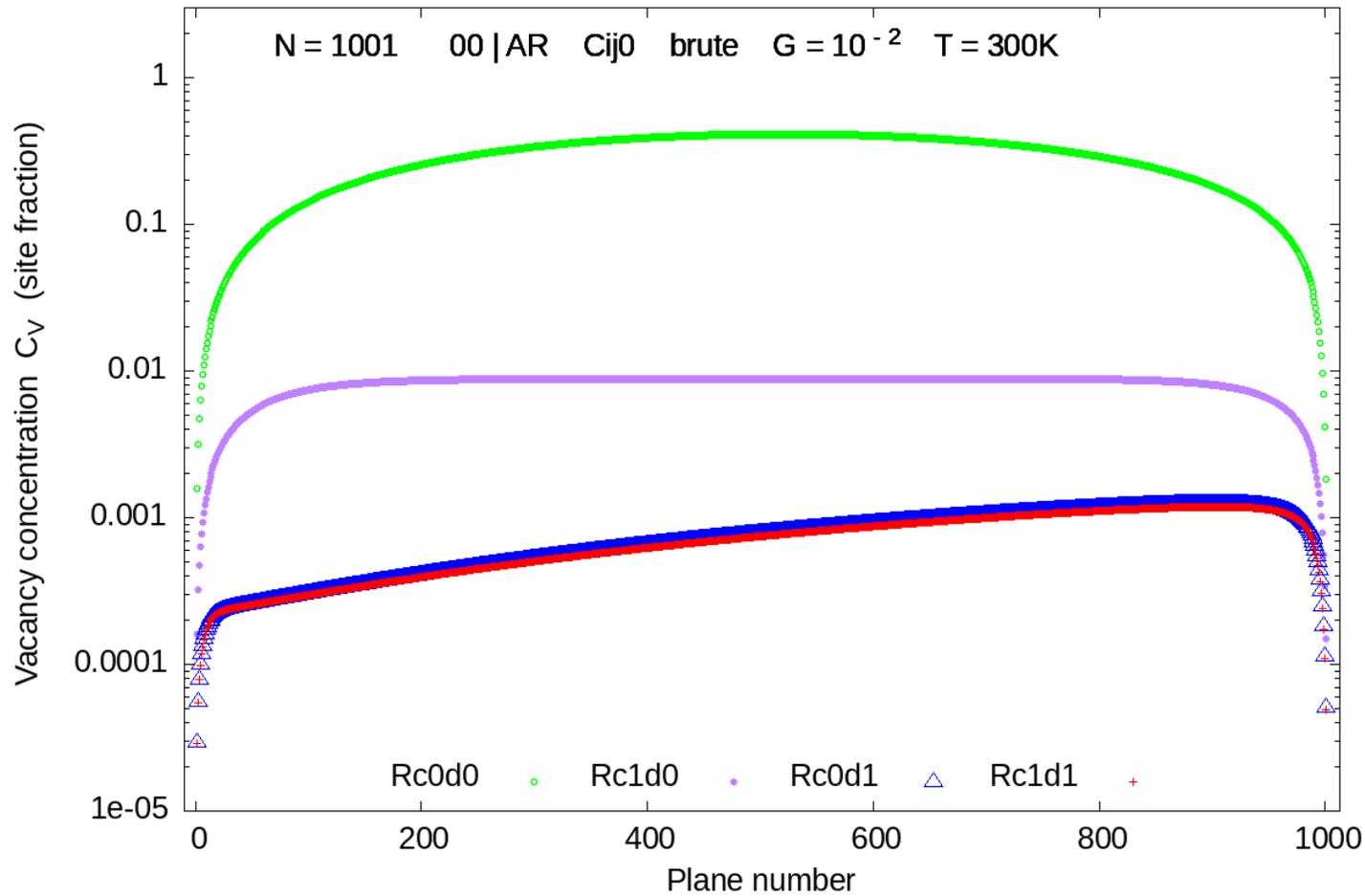

*Figure 15.*

*Concentration profiles for vacancies V at 300K, with or without recombination. Rc1 includes the recombination through aborted creation; Rd1 includes the recombination after diffusion jumps.*

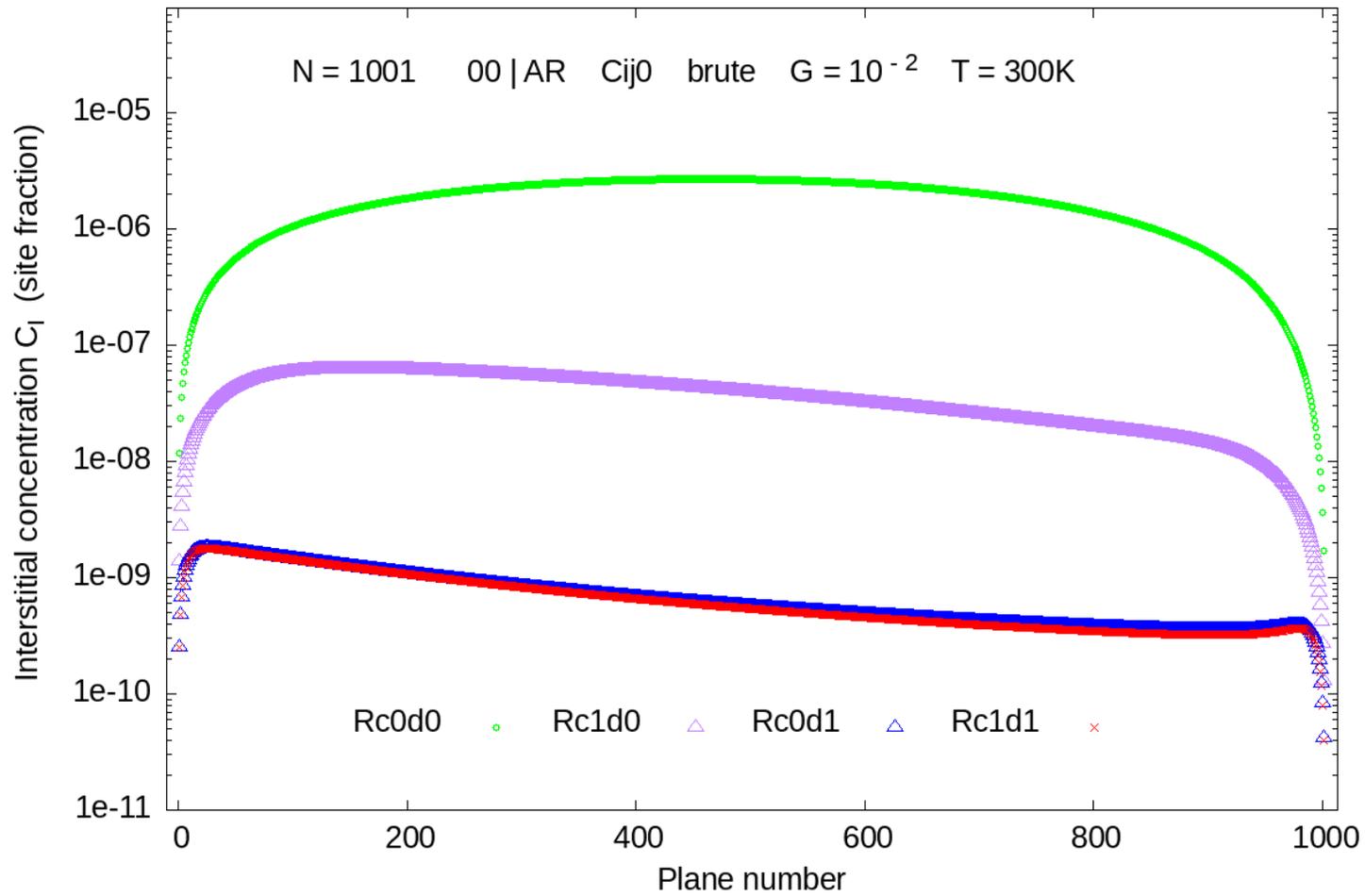

*Figure 16.*

*Concentration profiles for vacancies I at 300K, with or without recombination. Rc1 includes the recombination through aborted creation; Rd1 includes the recombination after diffusion jumps.*

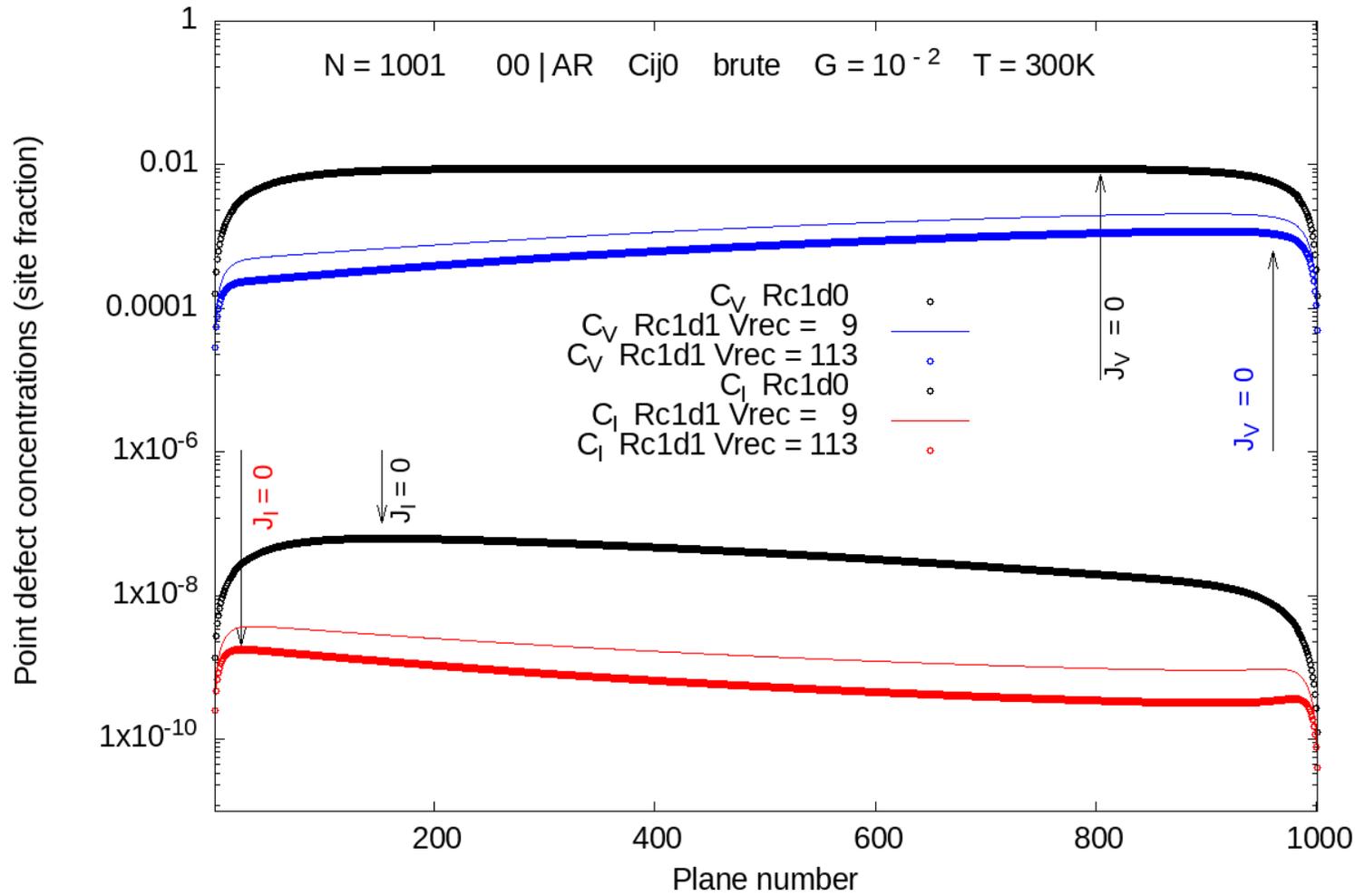

*Figure 17.*

*Concentration profiles for vacancies V and interstitials I at 300K, with full recombination Rc1d1 but two different recombination volumes.*

On Figure 18 are displayed the relative contributions of the various elimination channels. The recombination events play a detectable role up to 1000K, against 500K for the smaller system. The initial increase of diffusive recombination between 300K and 350K is due to the simultaneous decrease of the aborted creation process.

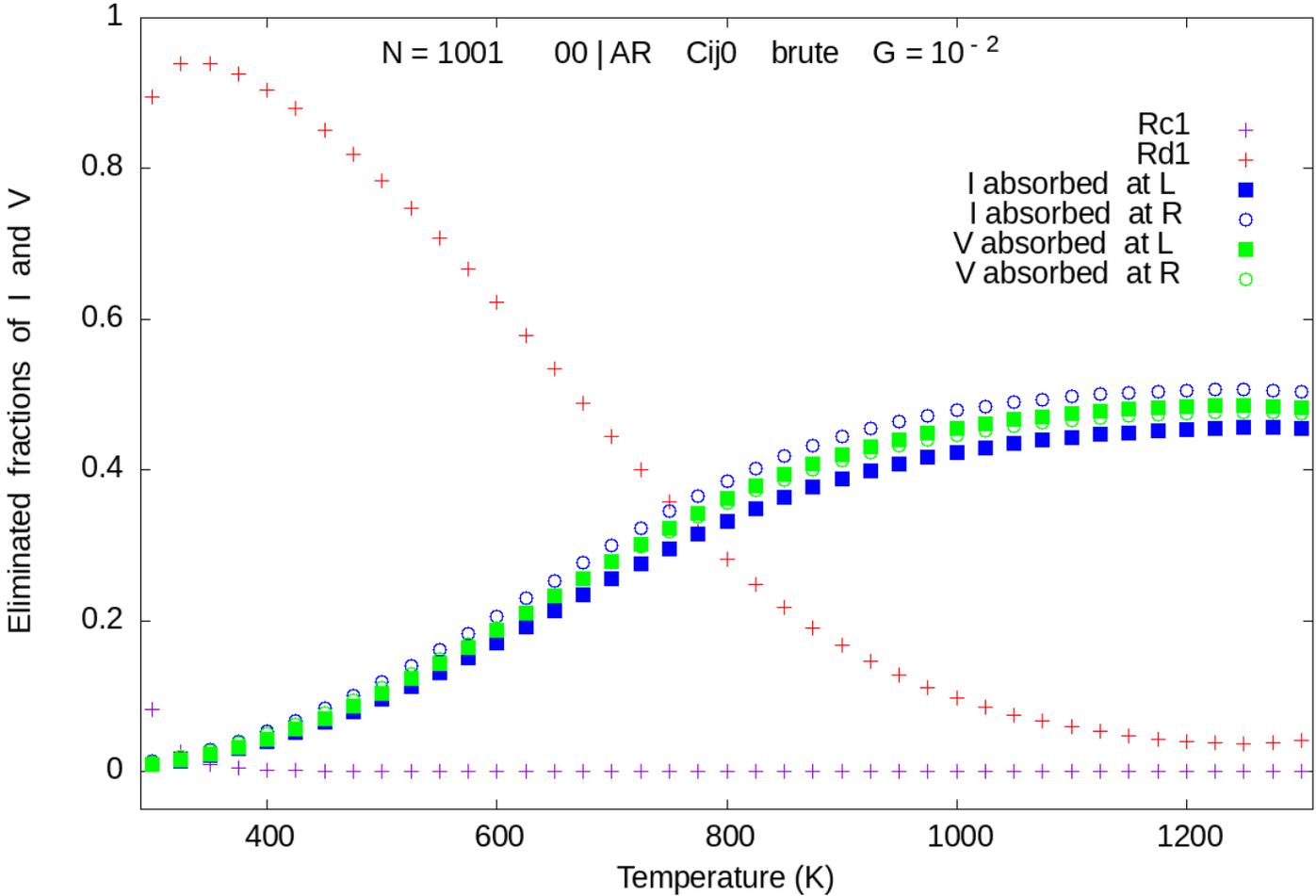

Figure 18.

Eliminated fractions of I and V species as a function of temperature with the brute interaction and taking full recombination into account Rc1d1

The L and R bias are definitely smaller for the larger system. At 300K the R bias is decreased by a factor of 8 down to 0.035 (Fig. 19) against 0.275 for the smaller system (Fig. 11).

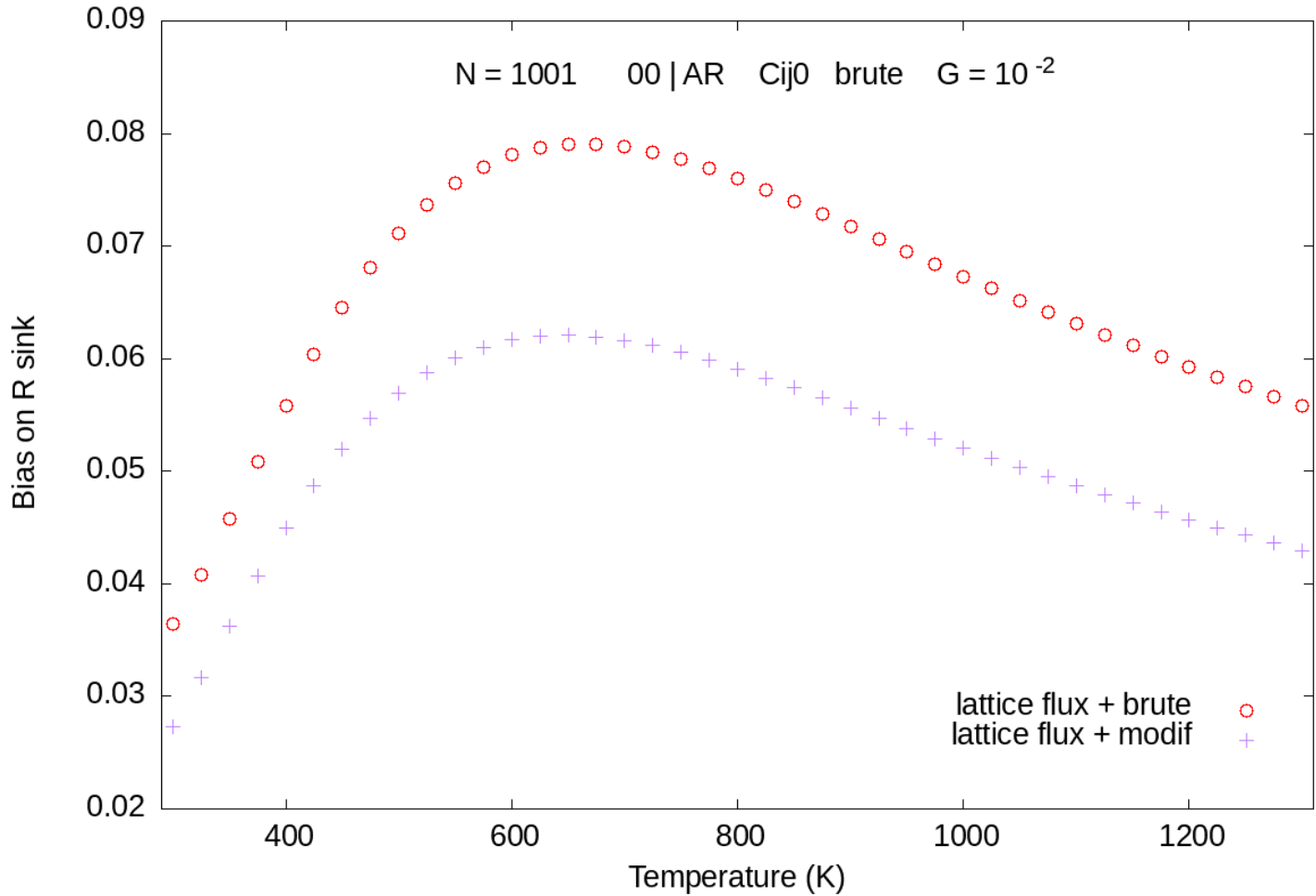

Figure 19.

*Absorption bias on R sink with the brute and modified interactions and the lattice expressions of the fluxes. Recombination fully included Rc1d1.*

**VI-4 Configuration AR2|AR1 (N=101)**

In the small system the L sink is now replaced by a wall with an interaction energy with PD of the same AR type (attracting the I defects and repelling the V ones) but twice larger. ZforceS now exist; they coincide with the ZfluxS only for the interstitials. They are displayed on Fig. 20.

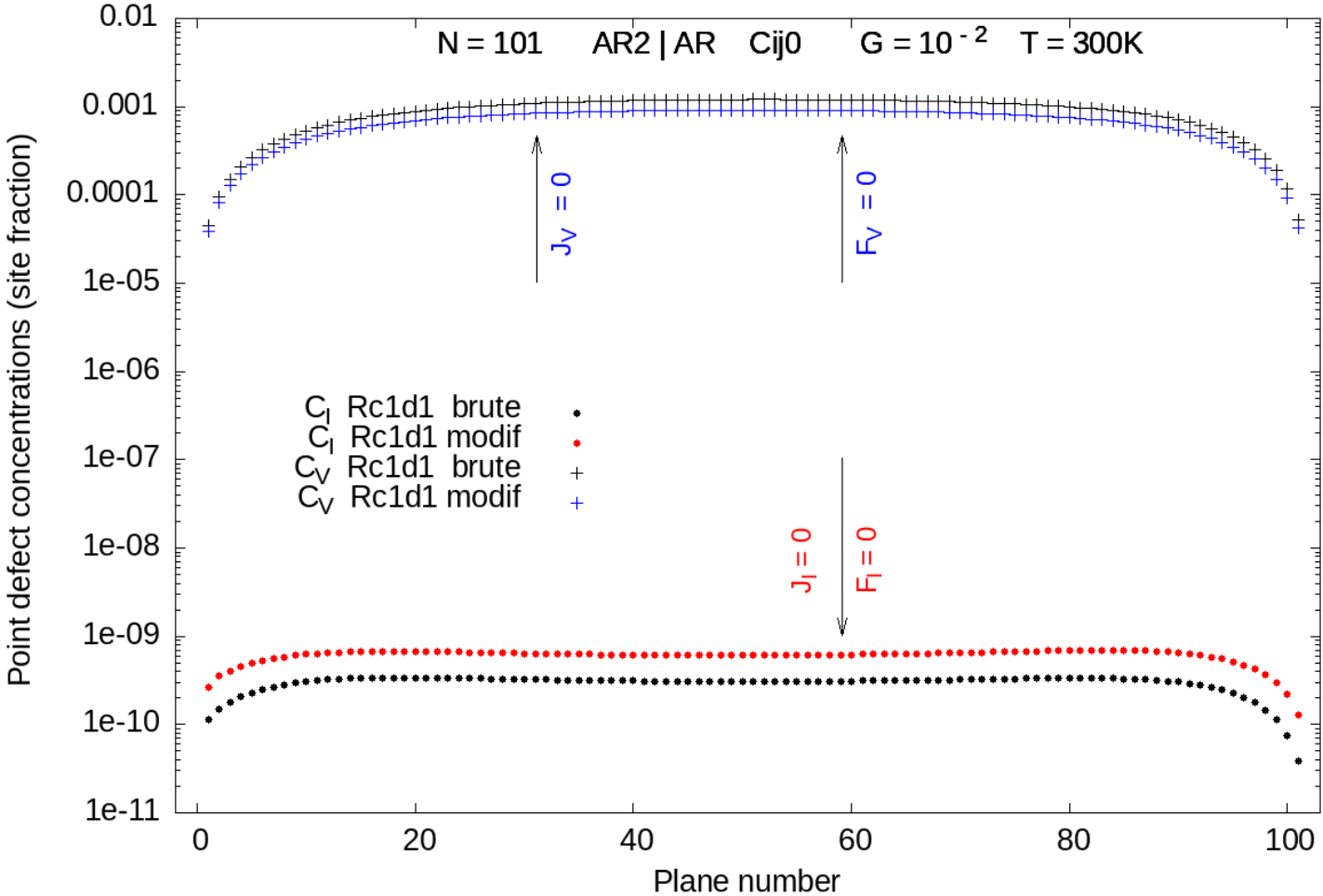

Figure 20.

Concentration profiles for I and V species with recombination fully taken into account Rc1d1.

ZForceS and ZfluxS are indicated for the two profiles.

The resulting absorption bias at the L and R sinks are displayed on Fig.21. The new feature is the fact that the R sink exhibits now a reversed negative absorption bias, while the L sink absorbs more interstitials than vacancies (compare with Fig. 11).

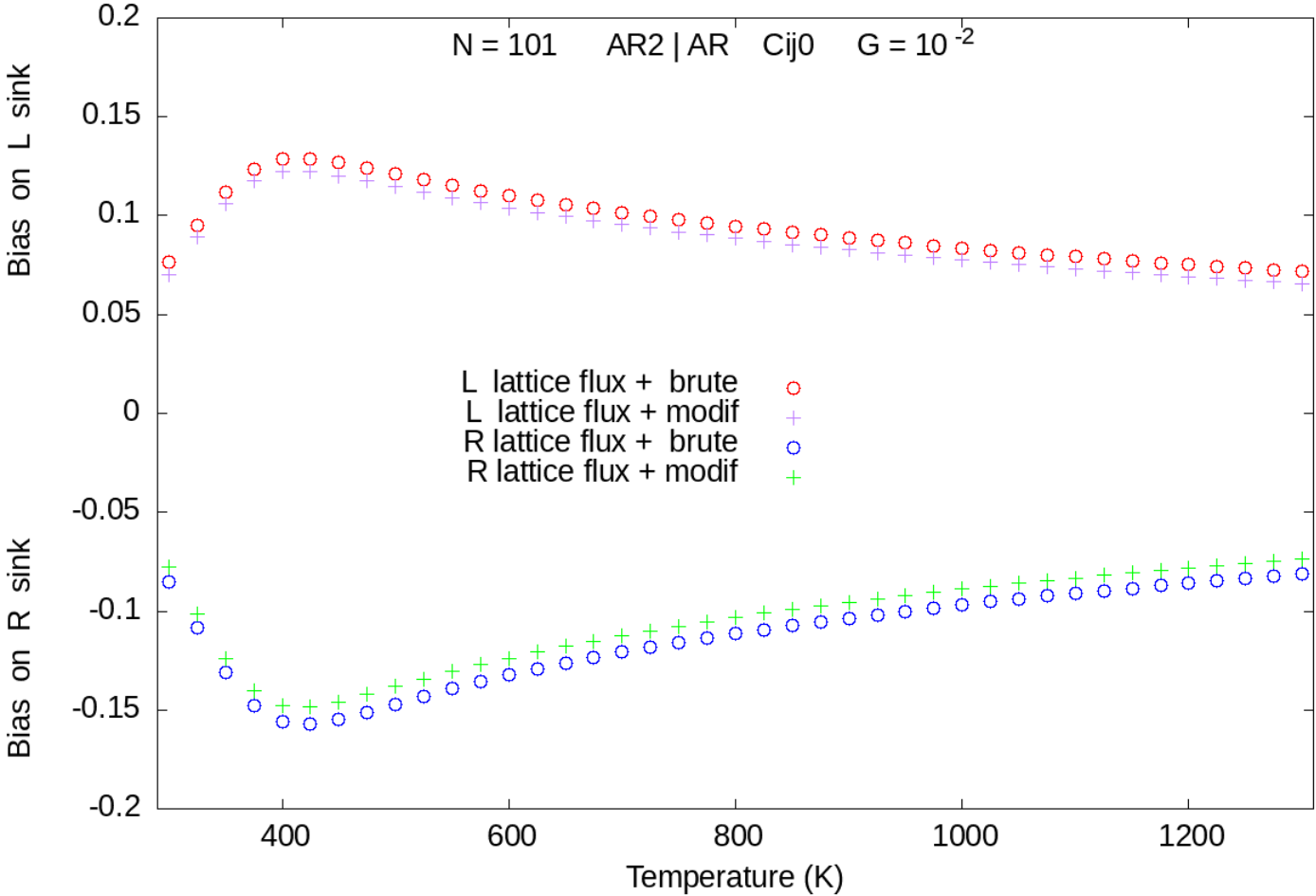

*Figure 21.*

*Absorption bias on the L and R sink with the brute and modified interactions and the lattice expression of the fluxes. Recombination fully included Rc1d1.*

### VI-5 Configuration RA|AR (N=101)

For this last case, the L sink is supposed to repel interstitials and attract vacancies. The interstitial is repelled by L sink as hard as it is attracted by the L sink (the reverse holds for the vacancy); therefore, there are neither ZforceS nor ZfluxS.

The relative weights of the elimination channels for the two defects are displayed below on Fig. 23. In the temperature range above 600K where the recombination plays no longer a noticeable role, the interstitials are mainly absorbed at the R sink (empty blue circles) at the expense of the absorption at the L sink (filled blue squares).

At the L sink, the I flux is minimum while the V flux is maximum; due to the arbitrarily non symmetric definition of the absorption bias AB, its value at the L sink hits skyrocketed values (up to several hundreds) which are displayed on Fig. 24.

The AB on the R sink is displayed on Fig. 25.

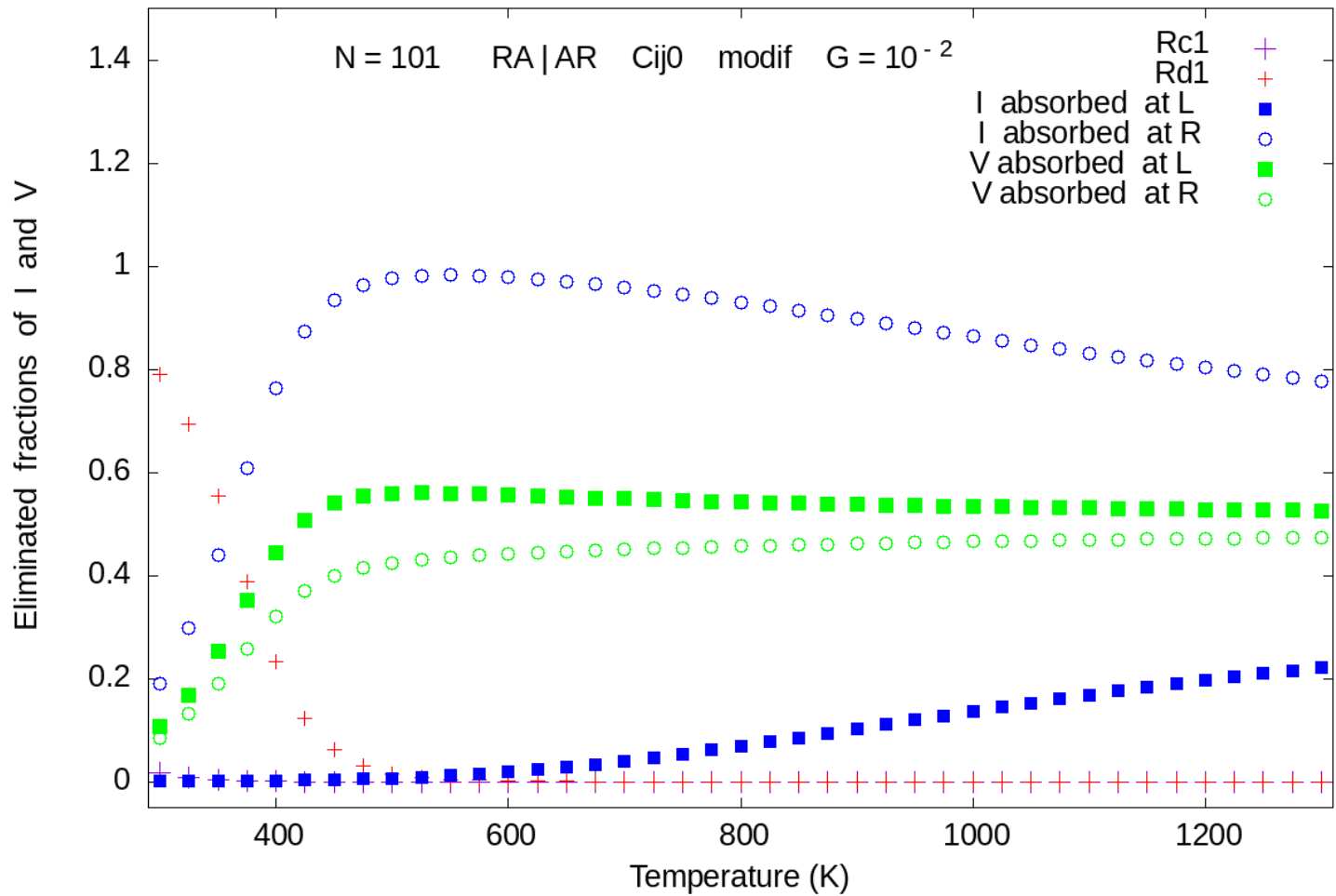

*Figure 23.*

*Eliminated fractions of I and V species as a function of temperature with the brute interaction and taking full account of the recombination Rc1d1.*

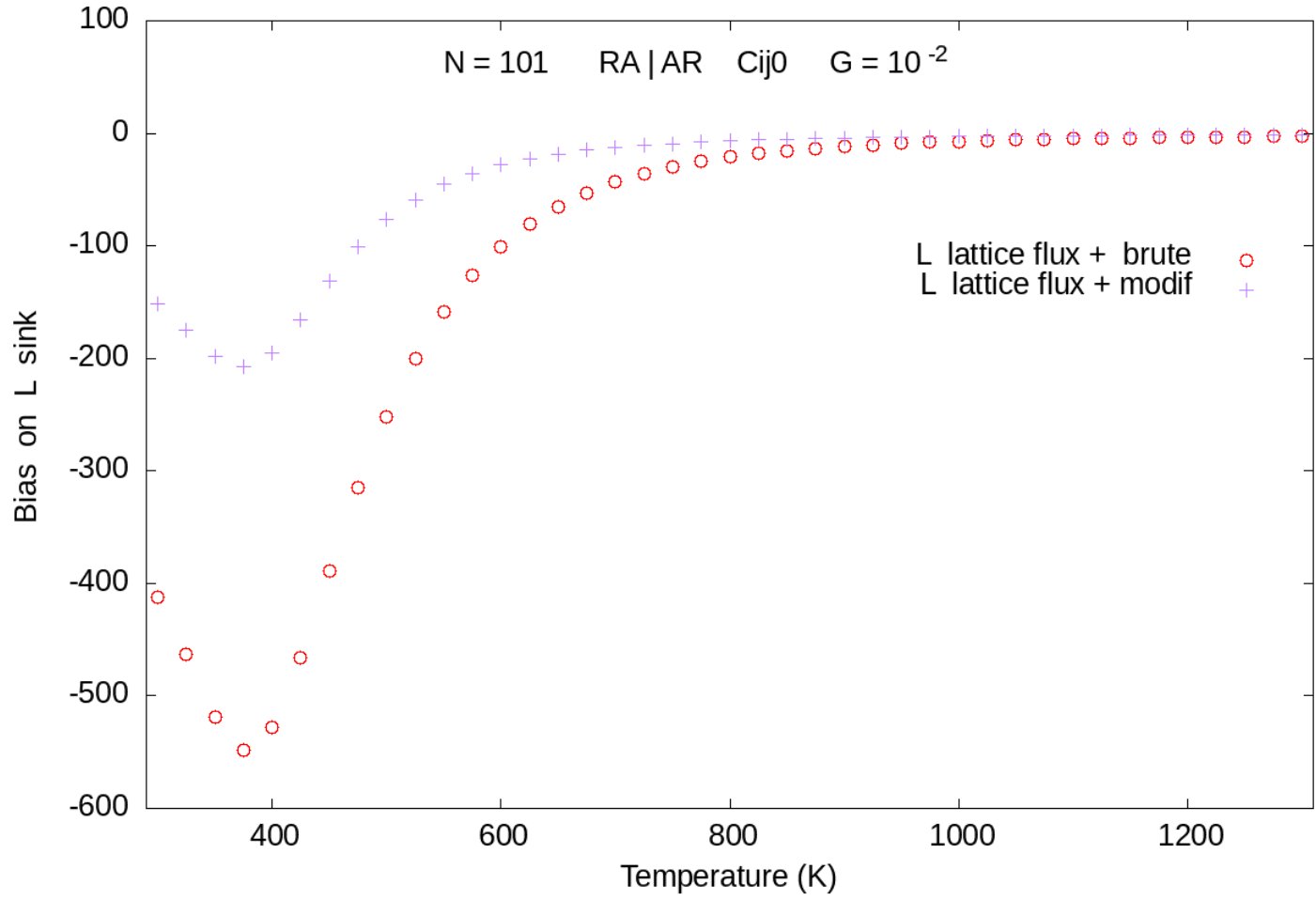

Figure 24.

*Absorption bias at the L sink with the brute and modified interaction and inclusion of full recombination Rc1d1*

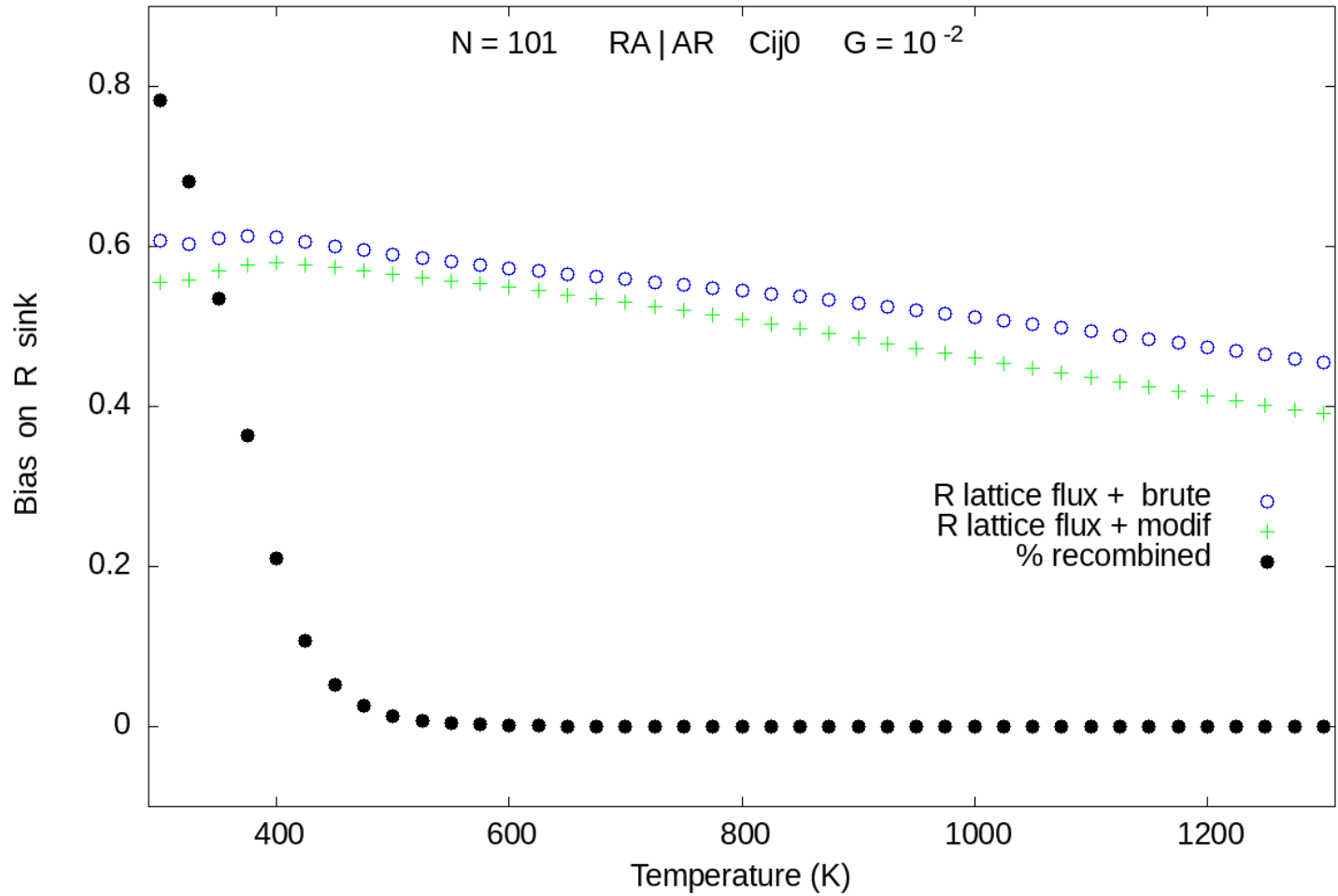

*Figure 25.*

*Absorption bias at the R sink with the brute and modified interaction and inclusion of full recombination Rc1d1*

## VI-6 Comparison of the obtained absorption bias with the standard evaluation by the Laplace method

The method has been extensively explained by previous authors. The flux of each defect 'd' at the R sink, when taken separately, is evaluated twice with a set of two Dirichlet boundary conditions:

*a first evaluation is done without any elastic interaction and with $C_{d,L}^{max}$ at the L sink, $C_{d,R}^{therm}$ at the R sink. The L and R sinks are assumed to be neutral. While the concentration at the R sink is set equal to the thermal one $C_{d,R}^{therm}$, $C_{d,L}^{max}$ is set to a value much larger than $C_{d,R}^{therm}$ (typically 10$^{-8}$ for V and 10$^{-10}$ for I). Because of the large formation energies for the two defects I and V, $C_{d,R}^{therm}$ can be numerically set equal to zero. The defect flux is constant throughout the slab and it is given by:

$$J_{d,R}^{zero} = \frac{C_{d,L}^{max} - C_{d,R}^{therm}}{(N+1)a/2} D_d^{bulk} \approx \frac{C_{d,L}^{max}}{(N+1)a/2} D_d^{bulk} \quad (15)$$

It was checked that the numerical evaluation of our code reproduces this result exactly with no source term.

*a second evaluation is done with an elastic field produced by the R sink only. The L sink is still assumed to be neutral with the same concentration $C_{d,L}^{max}$. The concentration at the R sink takes now into account the interaction of the defect 'd' with the R sink. However this interaction remains weak when compared with the formation energies of the defects and the corresponding value can be once more set equal to zero. The flux obtained while taking into account the interaction of the PD with the R sink is denoted by $J_{d,R}^{with}$. Analytical solutions are available only if the diffusion coefficient is independent of the spatial coordinates [24]. It is not the case in the present problem for the modified interaction since the local and average diffusivities $D_{d,i}^{loc}$ and $D_{d,i}^{av}$ are no

longer equal to the constant $D_d^{bulk}$ (refer to Appendix B). The evaluation of the fluxes is done numerically. With the usual notation, the two elastic bias (EB) are then defined by $X_{d,R} = J_{d,R}^{with} / J_{d,R}^{zero}$ and the proposed absorption bias AB at the R sink is calculated according to:

$$AB_R = 1 - X_{V,R} / X_{I,R}. \qquad (16)$$

This corresponding AB (purple line) is displayed together with all the AB collected above for the preceding cases for the smaller system (N=101) for the brute (Fig. 26) and modified interaction (Fig. 27). The horizontal black line stands for the AB of the symmetric configuration AR | AR in which the AB vanishes by construction. The modification brought by the change of the interaction is of small extension.

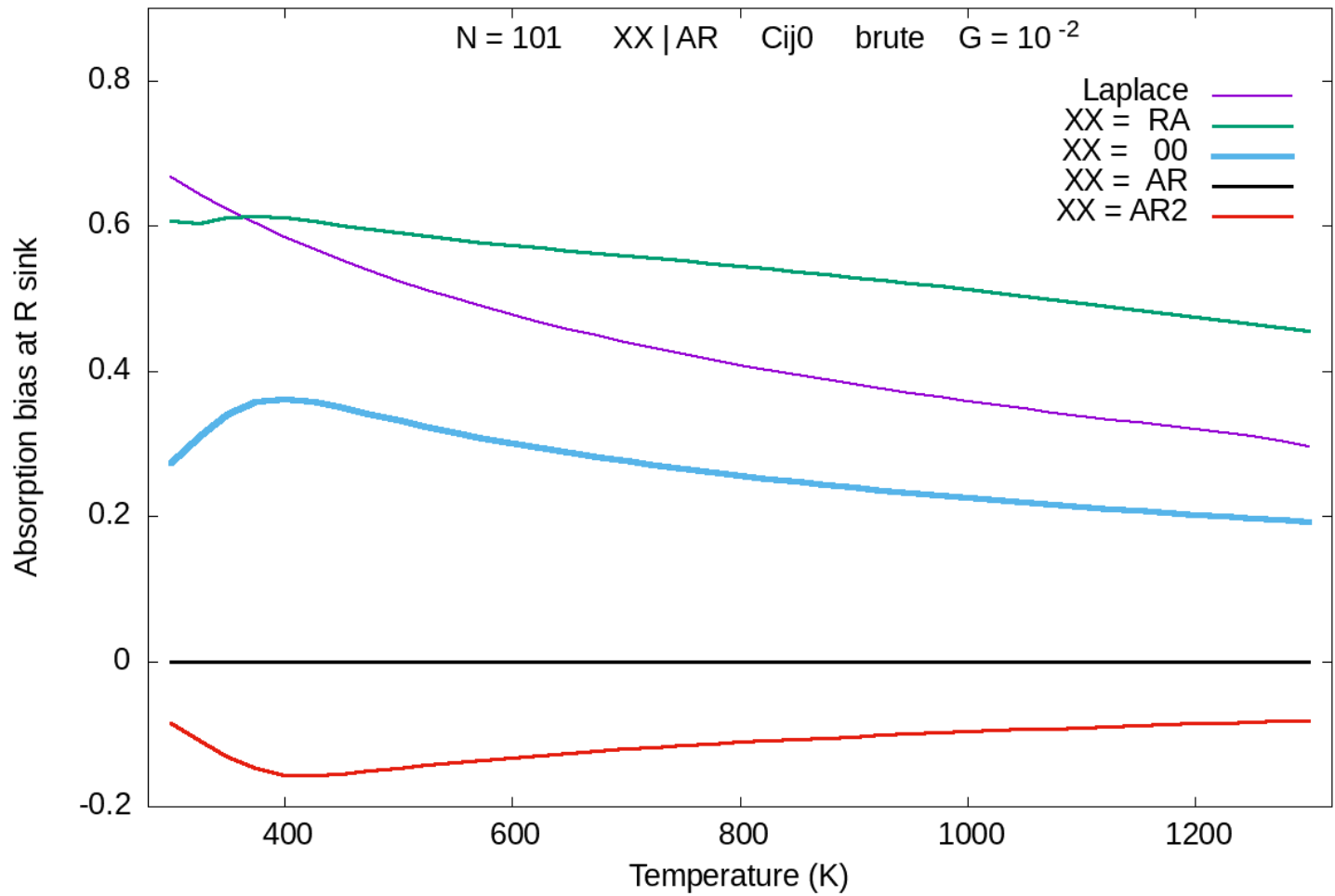

Figure 26.

*Comparison of the AB obtained through our calculations in the small system with the Laplace method for the brute interaction. The recombination is fully taken into account (Rc1d1).*

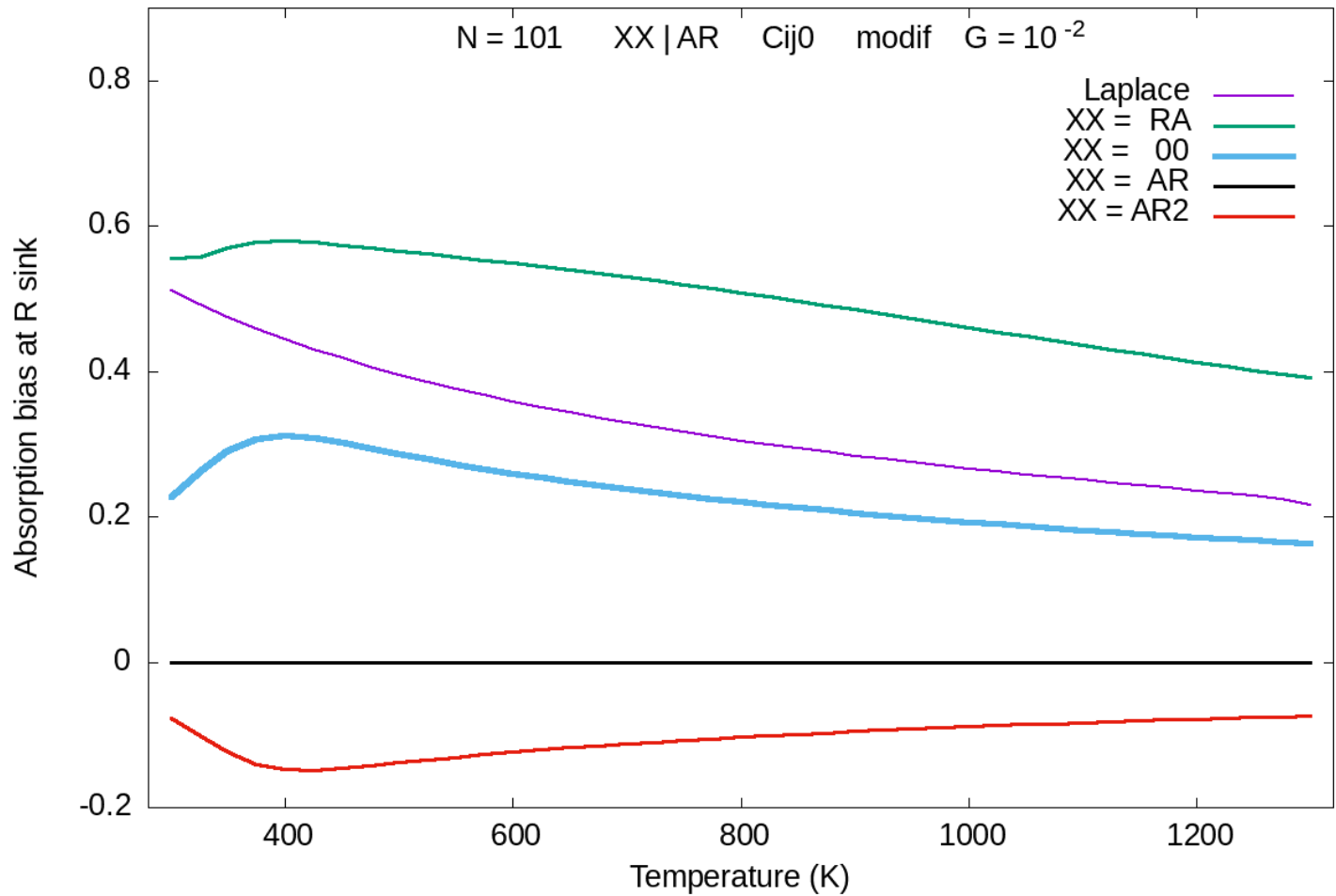

*Figure 27.*

*Comparison of the AB obtained through our calculations in the small system with the Laplace method for the modified interaction. The recombination is fully taken into account (Rc1d1).*

The same conclusion is drawn for the larger system (N=1001) and the corresponding AB are displayed on Fig. 28-29.

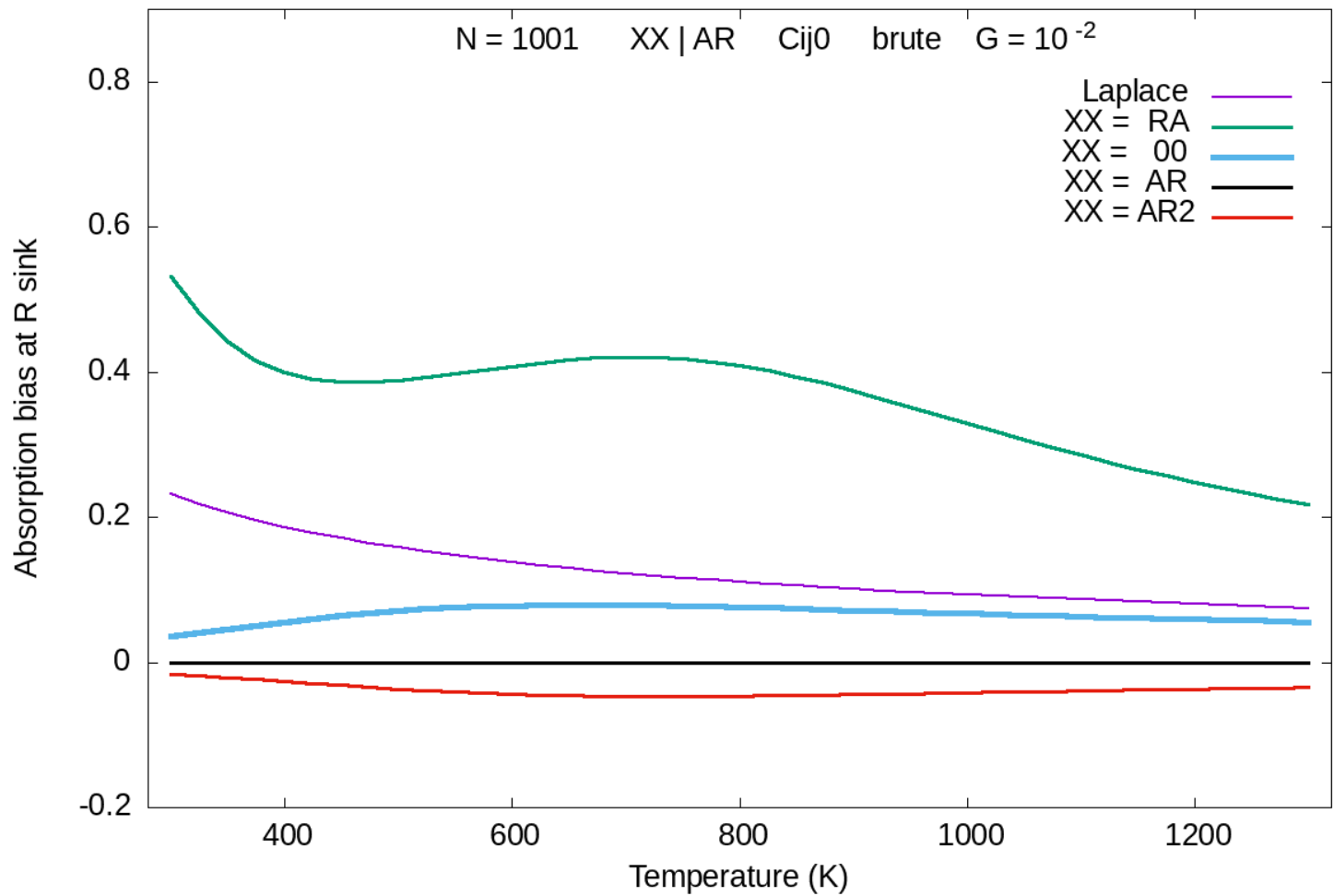

*Figure 28.*

*Comparison of the AB obtained through our calculations in the large system with the Laplace method for the brute interaction. The recombination is fully taken into account (Rc1d1).*

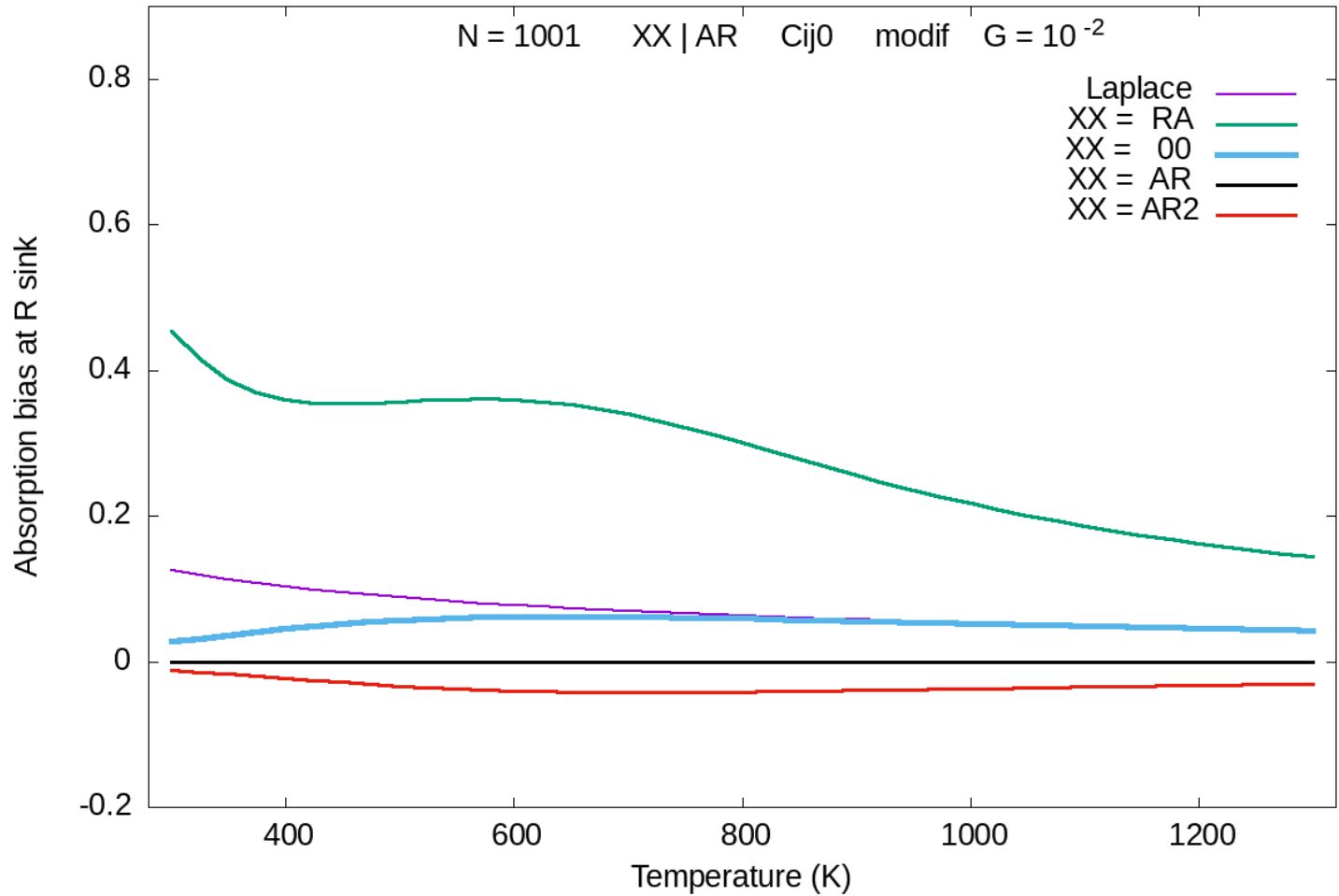

*Figure 29.*

*Comparison of the AB obtained through our calculations in the large system with the Laplace method for the modified interaction. The recombination is fully taken into account (Rc1d1).*

**VI-7 Discussion**

The stress is not to be laid on the absolute value of the bias, because the model used here is too rough for describing realistic situations. Moreover, because of the geometry chosen for the system, the overlapping of the diffusion fields for two sinks facing each other exacerbates their mutual influence: a defect which is not absorbed on a given sink is automatically absorbed by the other. This implies an imbalance between the I and V fluxes equal in magnitude but opposite in sign on the two sinks according to Eq. 2.

The important conclusion points toward the following fact: if the details of the elastic interaction (dipolar tensor against volume approximation for instance) do bring a detectable change to the AB, the boundary conditions involved in the examined cases (RA | AR, 00 | AR, AR | AR, AR2 | AR) play an overwhelming role on the result as illustrated on the AB reversal detected in the last configuration at the R sink. This should not be a surprise for a diffusion problem where the boundary conditions have always a leading contribution.

The interesting case of AB reversal obtained above for one situation was recently enlightened on a much more realistic case. A recent and sophisticated calculation took a full account of the elastic fields generated by the loops and the free surfaces in a thin foil film [31]. During a sustained irradiation, the loops in the foil were generally found to grow since they elastic bias EB tends to attract more interstitials than vacancies. It was however not the case for the loops located close to the free surfaces; the latter are viewed under a large solid angle by the migrating interstitial and become the dominant sink. As a result, the interstitials are locally absorbed mainly by the surfaces, which explains the corresponding shrinking of the nearby loops which was experimentally observed. This shading effect produced by a neighboring sink is also at work in our AR2 | AR case.

The partition of the space between the different diffusion fields was also questioned recently, suggesting that a first improvement could be the use of a Voronoi polyhedron around each type of sink [32]. This is a purely geometrical criterion and the choice of zero flux surfaces would probably yield a more appropriate partition of the volume. Unfortunately, as displayed above, the ZfluxS are not necessarily identical for the vacancies and the interstitials and, worse, their positions change with temperature; they cannot be determined without a thorough calculation of concentration profiles throughout the medium. The good point is the fact that at the higher temperatures the ZfluxS tend to merge together and to approach the faces of Voronoi polyhedron.

## VII Conclusions

The purpose of the contribution was focused on the link between the elastic bias (EB) and the absorption bias (AB) at the root of swelling and on the stiffness of this link with respect to a change of boundary conditions. The R sink was kept unmodified throughout the calculations of all configurations (attracting I and repelling V with an unmodified interaction) whereas the elastic properties of the L sink were changed in order to mimic different boundary conditions. After inspection of the results the following observations about the absorption bias (AB) can be made.

* the use of the atomistic expression of the flux is mandatory: ignoring this point may even lead to a sign change of the AB due to an incorrect evaluation of the gradient and drift terms;

* the AB is not very sensitive to the modification of the barrier profiles; the sophisticated evaluations yield slightly different values of each flux but their imbalance is only modestly affected;

* the AB is sensitive to the recombination of defects only below 500K; at higher temperatures, between 600K-1000K where swelling is expected to take place, it becomes independent of recombination;

* the AB is much more sensitive to a change of the elastic properties of the neighboring L sink than to the changes brought by the modifications listed above; we could evidence a reversal of the AB at the R sink which dragged more I than V.

This crude modelling suggests that the interaction of the diffusion fields of neighboring sinks can become dominant and exceed the effect of the elastic bias. The final consequence implies that pairing a priori one value of the AB with a sink of known EB is not possible if the surroundings of this sink are not known.

## Data availability

The author confirms that the data supporting the findings reported in this study are available within the article. Raw data that support the results are available from the corresponding author upon reasonable request.

## Declaration of competing interest

The author is freely and graciously hosted by the Centre Borelli ENS Paris-Saclay and declares to have no competing financial interest or personal relationship that could influence the data reported here.

**Aknowledgements**

I thank my former colleagues at CEA, namely L. Wang, T. Schuler, M. Nastar, F. Soisson and T. Jourdan, for their interesting remarks and stimulating discussions.

# Appendix A

## Necessary quantities to enumerate recombination events

The chosen recombination volume Vrec contains 113 crystalline sites of a BCC lattice, which belong to the first eight neighbor shells. Its center is the origin of the coordinates, which is counted as a site belonging to Vrec. Below are mentioned the coordinates of the representative site of the shell which belongs to the first octant: i, j, k all odd or even integers together with (i >=j >=k >=0) with a unit length equal to a/2 (a = lattice parameter). The shells are numbered according to their distance from origin.

| Shell number | Coordinates of the representative site | Number of sites in the shell |
|---|---|---|
| 0 | 0 0 0 | 1 |
| 1 | 1 1 1 | 8 |
| 2 | 2 0 0 | 6 |
| 3 | 2 2 0 | 12 |
| 4 | 3 1 1 | 24 |
| 5 | 2 2 2 | 8 |
| 6 | 4 0 0 | 6 |
| 7 | 3 3 1 | 24 |
| 8 | 4 2 0 | 24 |
| | | Σ = 113 |

*Table A1. Neighbor shells included in the recombination volume Vrec.*

In what follows, the number of sites in each shell follows the coordinates and is preceded by an underscore '_'.

The 4$^{th}$, 5$^{th}$, 6$^{th}$, 7$^{th}$ and 8$^{th}$ shells of Vrec are the only ones to have first neighbors outside Vrec :

* the 4$^{th}$ shell (311) has first neighbors belonging to shells 9 (422)_24

* the 5$^{th}$ shell (222) has first neighbors belonging to shells 10 (333)_8

* the 6$^{th}$ shell (400) has first neighbors belonging to shells 10 (511)_24

* the 7$^{th}$ shell (331) has first neighbors belonging to shells 9 (422)_24, 11 (440)_12 and shell 13 {(442)_24 + (600)_6}

* the 8$^{th}$ shell (420) has first neighbors belonging to shells 10 {(511)_24+(333)_8} and 12 (531)_48.

Sites (600) belong to a subshell of the 13th shell like (442) and are at the same distance from origin; but lattice geometry prevents them from reaching a site of Vrec in one jump. Thus 140 sites (sum of numbers in red) not belonging to Vrec can reach at least one site inside Vrec through one jump.

**Definition of quantities used to count the recombination events**

We need a partitioning of crystal sites according to the plane 'i' they belong to, since the elastic force depends on the 'x' coordinate and is sensitive to the + or - direction of the jump.

On Fig. A1 are depicted the quantities nint_Vrec(i) and next_Vrec(i), where 'i' can refer to plane 'i' as well as to any site belonging to plane 'i'. The recombination volume centered on a site of plane 'i' is denoted by Vrec(i). It is surrounded by another sphere containing sites which do not belong to Vrec(i) but from which a defect can reach, in one jump, one site at least belonging to Vrec(i).

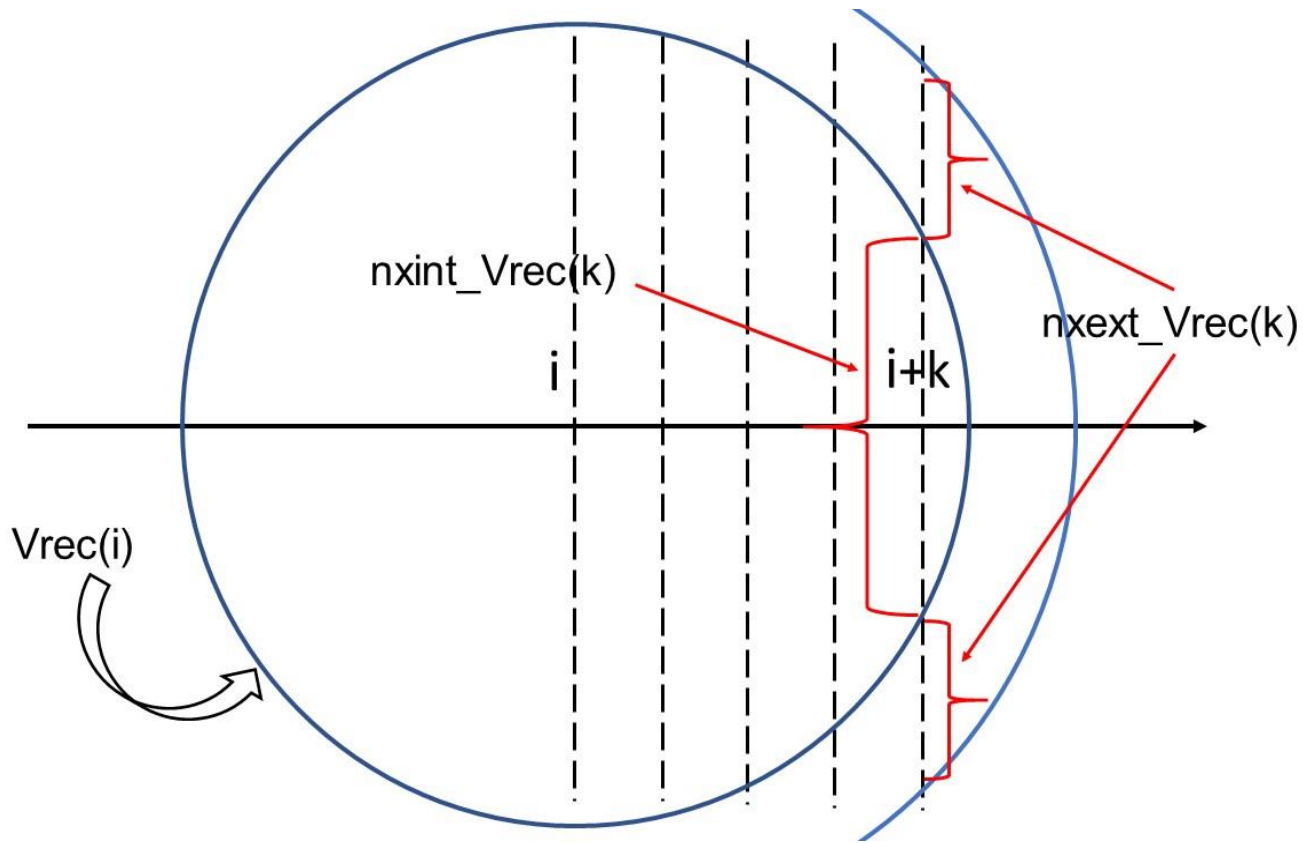

*Figure A1.*

*Definition of nint_Vrec(k) and next_Vrec(k) as a function of 'k' which measures the distance between plane 'k' and the center 'i' of the recombination volume.*

For a recombination volume Vrec(0) with a center on plane '0' , containing 113 BCC sites, we define the following quantities :

* nint_Vrec(i) : the number of sites inside Vrec(0) located on plane 'i'  equal to:

| i | -4 | -3 | -2 | -1 | 0 | 1 | 2 | 3 | 4 | |
|---|---|---|---|---|---|---|---|---|---|---|
| nint_Vrec(i) | 5 | 12 | 13 | 16 | 21 | 16 | 13 | 12 | 5 | Σ=113 |

*Table A2. Partition of sites inside the recombination volume Vrec.*

* next_Vrec(k) : the number of sites located on plane 'k', which do not belong to Vrec(0) but which are first neighbors of (at least) one site of Vrec(0),

* next_Vrec_jumm(k) and next_Vrec_jump(k): the numbers next_Vrec(k) multiplied by the number of '-' and '+' jumps (along Ox- and Ox+, respectively) which hit the surface of Vrec(0)

| k | -5 | -4 | -3 | -2 | -1 | 0 | 1 | 2 | 3 | 4 | 5 | |
|---|---|---|---|---|---|---|---|---|---|---|---|---|
| next_Vrec(k) | 12 | 16 | 12 | 12 | 16 | 4 | 16 | 12 | 12 | 16 | 12 | Σ=140 |
| next_Vrec_jumpm(k) | 0 | 0 | 0 | 8 | 8 | 4 | 24 | 20 | 12 | 28 | 20 | |
| next_Vrec_jumpp(k) | 20 | 28 | 12 | 20 | 24 | 4 | 8 | 8 | 0 | 0 | 0 | |

*Table A3. Partition of sites outside Vrec from which a site of Vrec can be reached with one – or + jump.*

On Fig. A2 are displayed some of the I jumps entering the quantities next_Vrec_jumpp( ) and next_Vrec_jumpm( ) towards the recombination volume Vrec(i) of a vacancy sitting on a site of plane 'i'.

The neighborhood relationships in the BCC lattice are such that sites on the left of the center (up to -2) can however reach the surface of Vrec via a jump with a negative x component.

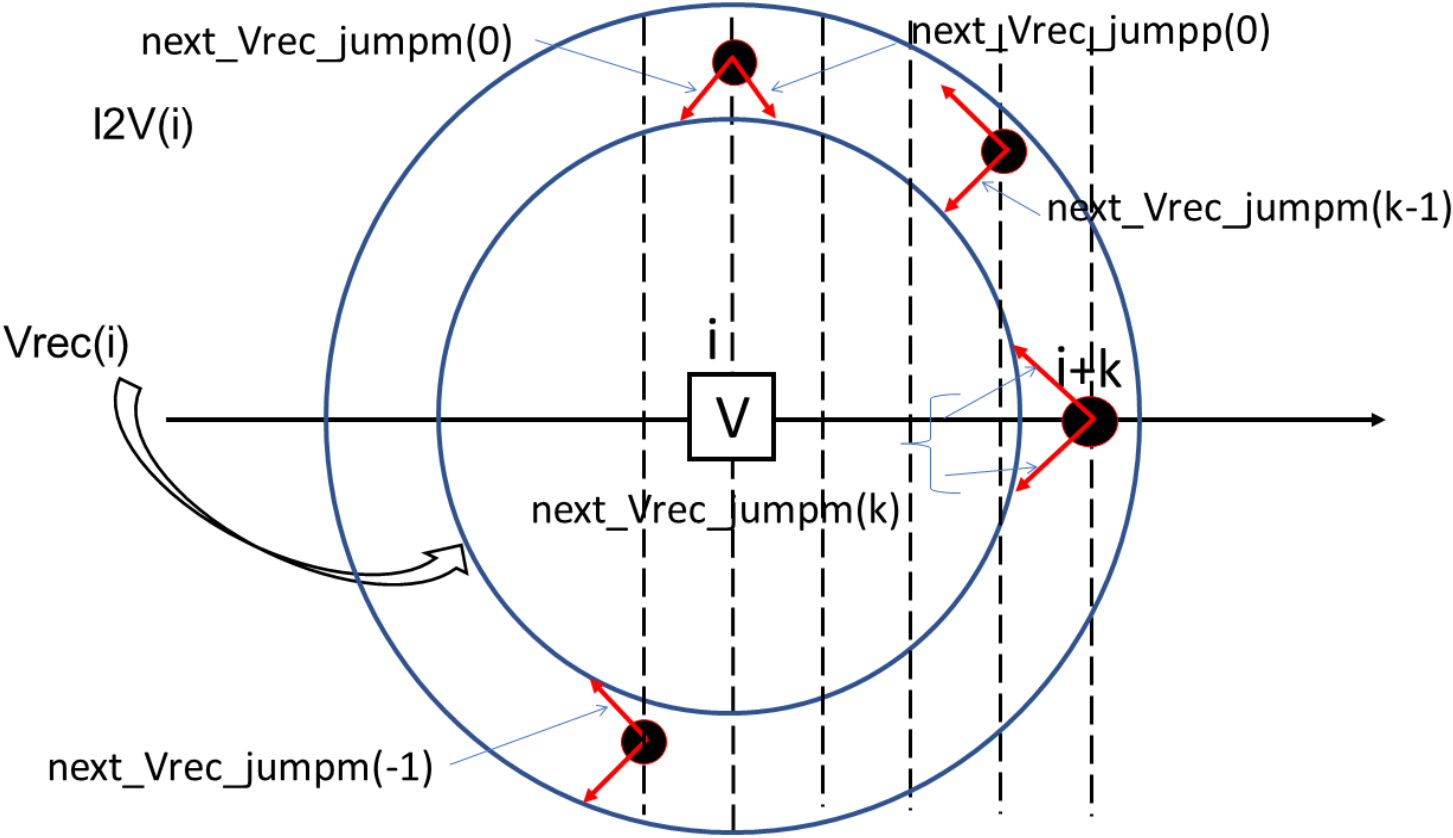

Figure A2.

Some of the jumps entering the quantities next_Vrec_jumpp( ) and next_Vrec_jumpm( ) The black filled circle and the empty white square represent the interstitial and the vacancy, respectively.

\* the numbers of recombination volumes, centered on plane 'i' and which can be reached with one defect jump (a '-' jump from k to k-1 or a '+' jump from k-1 to k.

|  | k = -4 | k = -3 | k = -2 | k = -1 | k = 0 | k = +1 | k = +2 | k = +3 | k = +4 |
|---|---|---|---|---|---|---|---|---|---|
| next_Vrec_sitem(i) (jump '-') | 5 | 7 | 3 | 5 | 6 | 1 | 2 | 2 | 0 |
| next_Vrec_sitep(i) (jump '+') | 0 | 2 | 2 | 1 | 6 | 5 | 3 | 7 | 5 |

*Table A4. Sites outside Vrec from which a site of Vrec can be reached in one jump*

On Fig. A3 are displayed some of the recombination volumes Vrec(k) which can be reached in one jump along Ox+ by a vacancy starting from a site of plane 'i'.

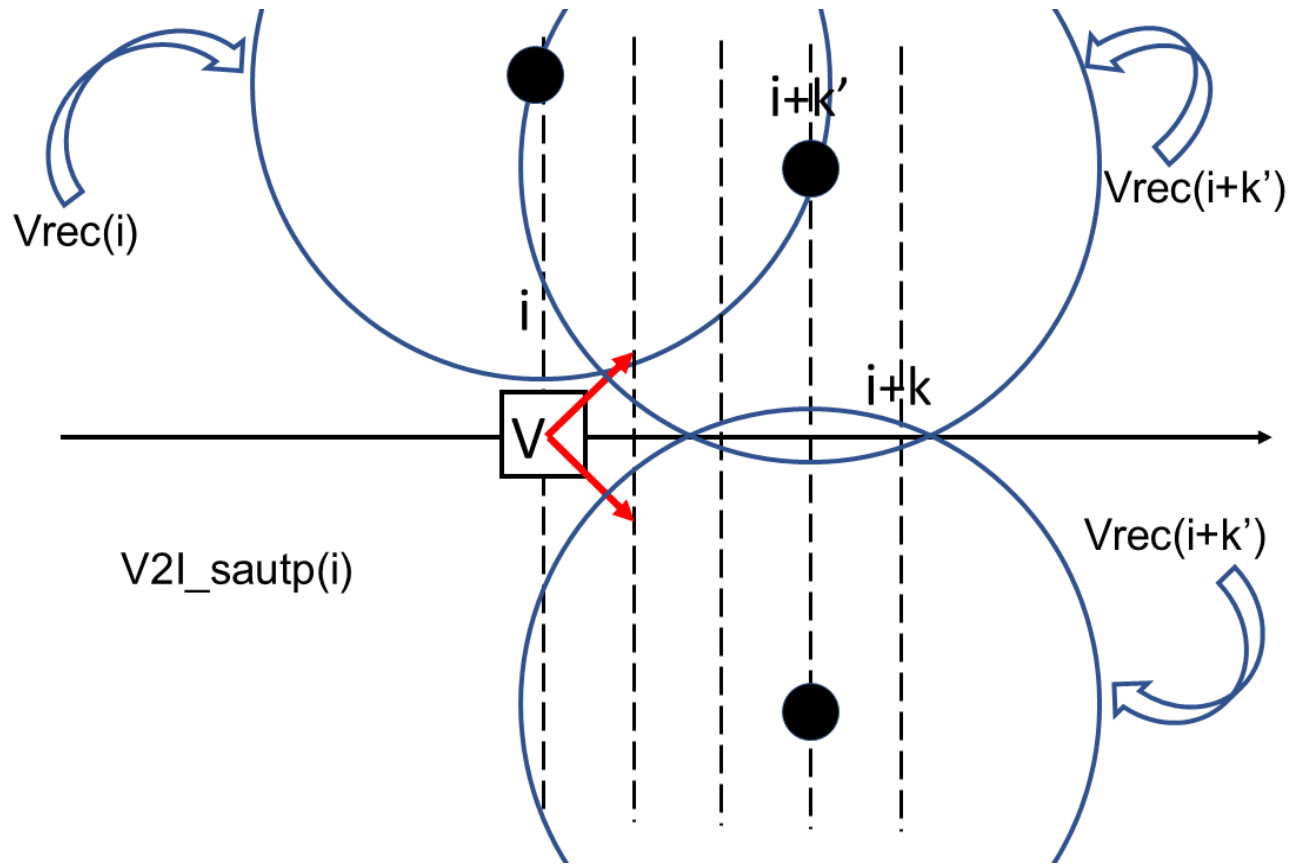

*Figure A3.*

*Some of the recombination volumes Vrec(k) centered on interstitials which can be reached in one jump, with a positive component along Ox, by the vacancy starting from 'i'.*

Hence the definition of the following quantities which will appear in the balance equations:

*summation of all first neighbor jumps of V starting from a site on plane 'j' (not belonging to Vrec(i)) to a site 'k' (belonging to Vrec(i)) and leading to a recombination event for the defect I which is waiting on site 'i':

$$V2I(i) = \sum_{j,k} C_V(j)\, W_{V, j \to k} \quad \text{where} \quad j \notin V_{rec}(i) \text{ and } k \in V_{rec}(i) \cap Vois1(j)$$

Vois1(j) is the set of first neighbors of a site 'j'; for a BCC lattice, site 'j' has 4 first neighbors on the plane 'j+1' and 4 first neighbors on the plane 'j - 1'. The '2' in the name of the variable V2I stands for '_to_' and indicates that V is performing a jump while I is waiting for its arrival;

*summation over sites 'k' occupied by V which will lead to the recombination with I when the latter jumps from 'i-1' to 'i' where 'i' belong to Vrec(k) but not 'i-1' :

$$I2V^+(i) = next\_Vrec\_jumpp(i) = \sum_{\substack{k \notin Vrec(i-1) \\ k \in Vrec(i)}} C_V(k)$$

The multiplicity of jump directions is implicitly included (a factor of 4 for the BCC and FCC lattices).

*summation over sites 'k' occupied by V which will lead to the recombination of I when the latter jumps from 'i+1' to 'i' where 'i' belong to Vrec(k) but not 'i+1' :

$$I2V^-(i) = next\_Vrec\_jumpm(i) = \sum_{\substack{k \notin Vrec(i+1) \\ k \in Vrec(i)}} C_V(k)$$

Similar quantities are defined after exchanging the symbols for the 'I' and 'V' species.

## Appendix B

### Expression of the fluxes at the atomic scale

The starting ingredient entering the building up of the driving force is the elastic interaction energy between the PD and the sink, as a function of their mutual distance measured by the plane index 'i'. The energy for defect 'd' ( I or V) is defined as follows :

$$E_{d,i} = E_d^{elast}(i=\infty) - E_d^{elast}(i)$$

where the symbol $i = \infty$ stands for the energy reference. In a medium of infinite size the corresponding energy vanishes. The energy is < 0 for an attraction, > 0 for a repulsion. In our system of finite extension, the two interactions of the two sinks sum up according to Eq. 3 in the main section.

The driving force stems from the asymmetry between the migration barriers for the jumps toward the positive $\rightarrow$ or negative $\leftarrow$ x-side between two neighbor planes. The saddle point energy is the same for the two jumps and the inequality of the jump barriers comes from the inequality of the interaction energy between the two starting sites. The driving force is $F_d(i) = -\nabla E_{d,i}$.

The jump frequencies between planes 'i' and 'i+1' are defined below:

$$W_{d,i \rightarrow i+1} = \nu_d \ \exp\left[-\beta(E_{d,i+1/2}^{sadd} - E_{d,i})\right]$$

$$W_{d,i+1 \rightarrow i} = \nu_d \ \exp\left[-\beta(E_{d,i+1/2}^{sadd} - E_{d,i+1})\right]$$

where $E_{d,i+1/2}^{sadd}$, $E_{d,i}$, $E_{d,i+1}$ stand for the energy of the system with the defect at the saddle point, on a stable site of plane 'i' and plane 'i+1' respectively; $\nu_d$ is a pre-exponential factor and $\beta = 1/kT$.

The concentrations, originally expressed as a number of defects / $A^3$, are converted to atomic volume fractions belonging to the interval [0 : 1] after multiplication by the atomic volume $V_{at}$ expressed in $A^3$. The flux is evaluated through the geometric plane at mid-distance from two neighbor atomic planes and through a facet denoted hereafter by $\Sigma$ and having a surface equal to $a^2$ or $a^2/2$ for the BCC or FCC lattices respectively. Hence the starting expression of the flux:

$$J_{d,i \to i+1} = J_{d,i+1/2} = \frac{a}{2}(4C_{d,i}W_{d,i \to i+1} - 4C_{d,i+1}W_{d,i+1 \to i})$$

The net flux at the left and right sink corresponds to $J_{d,1/2}$, $J_{d,N+1/2}$.

The energy of the system when the defect sits on its saddle point is described by the expression:

$$E_{d,i+1/2}^{sadd} = (E_{d,i+1} + E_{d,i})/2 + \Delta_{d,i+1/2}$$

where $\Delta_{d,i+1/2}$ is an increment of activation barrier introduced in the main section.

The frequencies between planes 'i' and 'i+1' are associated with the migration energies of back and forth jumps defined below:

$$E^m_{d,i\to i+1} = \Delta^{sadd}_{d,i+1/2} + (E_{d,i+1} - E_{d,i})/2 = \Delta^{sadd}_{d,i+1/2} + \delta E_{d,i+1/2}$$

$$E^m_{d,i+1\to i} = \Delta^{sadd}_{d,i+1/2} + (E_{d,i} - E_{d,i+1})/2 = \Delta^{sadd}_{d,i+1/2} - \delta E_{d,i+1/2}$$

Hence the final expressions of the jump frequencies

$$W_{d,i\to i+1} = W_{d,i+1/2}\, e^{-\beta \delta E_{d,i+1/2}} \qquad W_{d,i+1\to i} = W_{d,i+1/2}\, e^{+\beta \delta E_{d,i+1/2}}$$

with $W^{av}_{d,i+1/2} = \nu_d\, e^{-\beta \Delta^{sadd}_{d,i+1/2}} = \left( W_{d,i\to i+1}\, W_{d,i+1\to i} \right)^{1/2}$ is the geometric mean of the back and forth frequencies.

The driving force is expressed as a discrete derivative of the elastic interaction energy:

$$F_{d,i+1/2} = -(E_{d,i+1} - E_{d,i})/(a/2) \to F_{d,i+1/2} = -\frac{4}{a}\delta E_{d,i+1/2}$$

(B1)

Stated otherwise $-\delta E_{d,i+1/2}$ is nothing but the work done by the force over half the jump distance (equal to a/4 presently).

The general expression of the fluxes becomes:

$$J_{d,i+1/2} = (\frac{a}{2})4C_{d,i}W^{av}_{d,i+1/2} e^{-\beta\delta E_{d,i+1/2}} - (\frac{a}{2})4C_{d,i+1}W^{av}_{d,i+1/2} e^{+\beta\delta E_{d,i+1/2}}$$

$$= -4W^{av}_{d,i+1/2}(\frac{a^2}{4})\frac{(C_{d,i+1}-C_{d,i})}{a/2} ch(\beta\delta E_{d,i+1/2}) - 4W^{av}_{d,i+1/2}(\frac{a}{2})^2 \frac{4}{a} C_{d,i+1/2} sh(\beta\delta E_{d,i+1/2})$$

Hence the definition of a new diffusivity $D^{av}_{d,i+1/2} = W^{av}_{d,i+1/2} a^2$ which plays the role of an average diffusivity in a medium with $\Delta^{sadd}_{d,i+1/2}$ as a migration barrier. It is worth noticing that this definition is different from the standard one used for the defect diffusivity in a medium without any sink. The standard formulation rests on all the frequencies which depart from a unique site, whereas, in the present formulation, it is made of all the back and forth frequencies between two neighbor planes. If the difference between the two expressions vanishes in the absence of a driving force, it will not in general.

Hence the final formulation of the flux calculated between neighboring lattice planes with the superscript 'lattice':

$$J^{lattice}_{d,i+1/2} = -D^{av}_{d,i+1/2} ch(-\beta\delta E_{d,i+1/2})\nabla C_{d,i+1/2} - C_{d,i+1/2} D^{av}_{d,i+1/2} \frac{4}{a} sh(\beta\delta E_{d,i+1/2}) \ . \tag{B2}$$

Further, we introduce of a new local diffusivity $D^{loc}_{d,i+1/2}$ which embodies the gradient of the elastic energy:

$$D^{loc}_{d,i+1/2} = D^{av}_{d,i+1/2} ch(-\beta\delta E_{d,i+1/2}). \tag{B3}$$

An alternate formulation of Eq. 2 now reads:

$$J^{lattice}_{d,i+1/2} = -D^{loc}_{d,i+1/2} \nabla C_{d,i+1/2} - C_{d,i+1/2} D^{loc}_{d,i+1/2} \frac{4}{a} th(\beta \delta E_{d,i+1/2}) \tag{B4}$$

The first term of Eq. B2 (or B4) is the contribution of the concentration gradient to the flux; it is now intricately merged with the elastic driving force; the second term exhibits a non-linear contribution of the elastic driving force.

In the case where the driving force is weak, i.e. if $-\beta\, \delta E_{d,i+1/2}$ is sufficiently small, a first order development of the drift term with the help of Eq. B1 yields back the Nernst-Einstein expression of the flux, denoted hereafter with the superscript 'NE':

$$J^{NE}_{d,i+1/2} \approx -D^{av}_{d,i+1/2} \nabla C_{d,i+1/2} + \beta C_{d,i+1/2} D^{av}_{d,i+1/2} F_{d,i+1/2} \tag{B5}$$

In the case where $\Delta^{sadd}_{d,i+1/2}$ is chosen as independent of spatial coordinates and equal to the migration energy in the bulk $E^{mig}_d$, then $D^{av}_{d,i+1/2}$ reduces to the usual bulk diffusivity in the absence of elastic interaction $D^{bulk}_d$.

The consequences on the resulting absorption bias stemming from the adoption of Eq. B2 and Eq. B4 will be illustrated and discussed afterwards in the main section.

The modified form of the Nernst-Einstein expression is not anecdotal and plays an important role in two cases at least:

- for systems of small spatial extension in which the regions with large elastic energy gradients occupy a noticeable portion of the volume;
- for any system, whatever its size, because the absorption fluxes are, by definition, always calculated through the absorbing surface of the sink in order to get a correct value of the absorption bias (AB), that is, in a region where the elastic energy gradient reaches its maximum.

With the input parameters used for the two defects in α Fe, the portion of the system where the driving force can be considered as "not weak" (arbitrarily, whenever $\beta \, \delta E_{d,i+1/2} > 0.1$ ) encompasses atomic planes with an index ranging from 1 to 3 for V and 1 to 6 for I at T=600K. Numerical evaluations show that the correction is important for the drift term and the gradient term. As a result, the relative weights of the two contributions can be changed in such a way that even the sign of the resultant flux can be reversed. This is the main reason for keeping the atomic description in what follows.

The simplest choice for $\Delta_{d,i+1/2}^{sadd}$ consists in taking a constant value equal to the migration energy of the defect in the bulk far from any sink $\Delta_{d,i+1/2}^{sadd} = E_d^m$, but another choice will be explored (see Appendix C).

# Appendix C

**Comparison of various energy barriers evaluations in the strain field of a straight edge dislocation.**

This Appendix explores the energetic landscape through which a DP migrates when approaching a single dislocation and situates the brute evaluation with respect to more sophisticated energy calculations which take account of the dumbbell orientations. The final output is a (totally) empirical adjustment of the brute expression to bring its results closer to the best ones available today.

The coordinate axis x/y/z are along the direction [100] / [010] / [001] of the BCC lattice, respectively. The origin of the coordinates (000) is located on a site of the lattice; the sites have integer coordinates (i,j,k) in units of a/2, where 'a' stands for the lattice parameter.

The dislocation line (i.e. the edge of the extra plane perpendicular to the x-axis) is merged with the z-axis; the sites belonging to the extra plane, on the compressed side, have positive y-coordinates; the sites with negative y-coordinate are located on the extended side. The Bürgers vector is $\frac{a}{2}[100]$.

The radius of the absorption surface is arbitrarily fixed to 4 Bürgers vectors. Sites (0, 4, 0) and (0, -4, 0) sit on the absorption surface on the sides of maximal interaction with the DP: the first attracts / repulses the vacancy / dumbbell, respectively, while the reverse holds for the second.

The 'y' coordinate, measures the projected distance between the defect moving away from the dislocation line located at x=y=0. The barriers are evaluated along realistic paths located in regions of maximum attractive interaction energy for each defect. The word 'realistic' means that the path complies with the geometrical constraints of the elementary jump: a dumbbell sitting on the origin (0, 0, 0) with a dissociation axis <110> can reach only 4 neighbors, namely (1,1,1), (1,1, -1), (-1, -1,1) and (-1, -1, -1); and each neighbor is visited with two orientations. The paths are chosen somewhat arbitrarily among several possible alternatives and are contained within a square-section tube aligned along Oy axis together with a minimum lateral extension along x or z coordinates. They are described below:

\* for a dumbbell defect on the side in extension, the extra plane x=0 does not exist and is replaced by a pair of planes located at x = - 0.5 and x = + 0.5; the starting site is (0.5, -4, 0) and the chosen path encompasses sites which belong to planes  x= -0.5 and x= + 1.5 along a repeated sequence of 4 jumps (-1 -1 1), (1 -1 -1), (1 -1 1) and (-1 -1 -1) displayed on Figure C1. The brackets < > and parenthesis ( ) enclose the orientation of the dipolar tensor for the dumbbell and the coordinates of the stable sites, respectively. The orientation of the dipolar tensor at saddle position for the dumbbell is parallel to the jump vector.

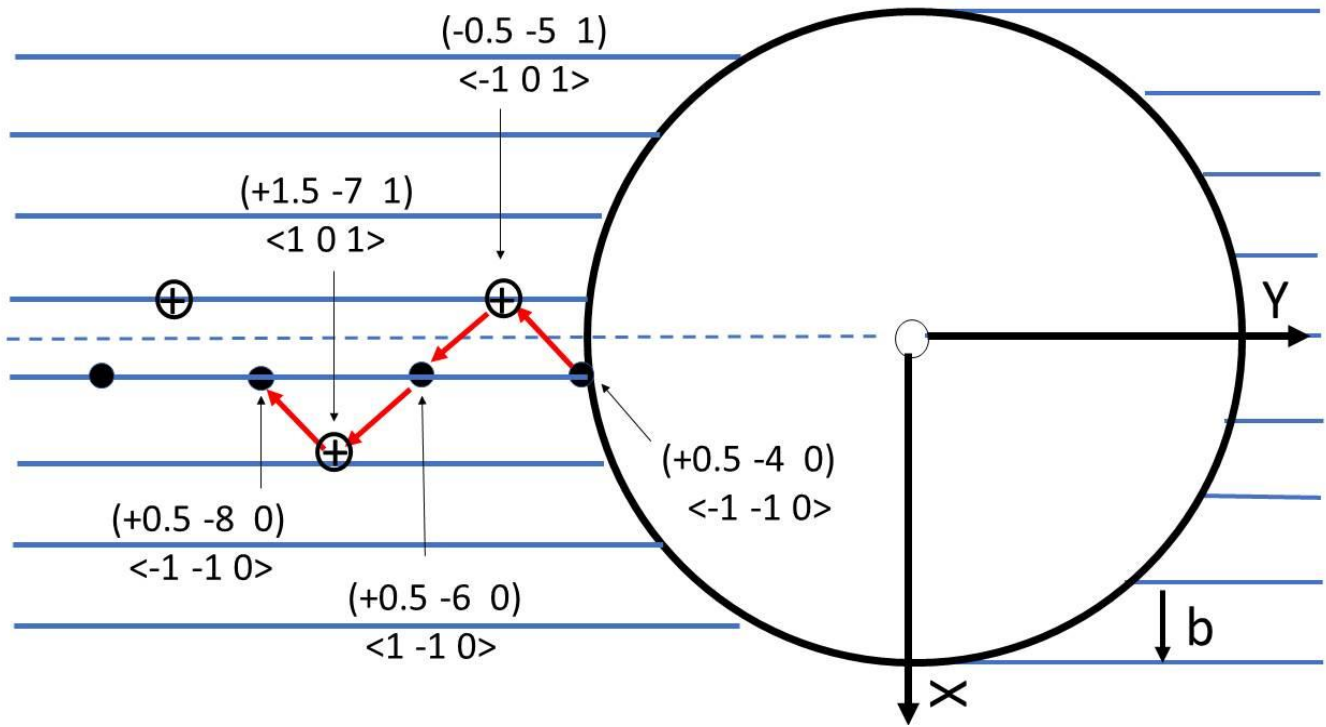

Figure C1.

*One possible path of a dumbbell defect moving away from dislocation in the extended region, along which the interaction energy at stable and saddle positions are calculated. Black disk: site in z=0 plane; circle enclosing '+': site in z= +1 plane.*

* for a vacancy on the side in compression, the path is made of sites located in the extra plane (x = 0), and sites in the adjacent one x = +1; the starting site is (0, 4, 0) and the chosen path is made of a repeated sequence of two jumps (1 1 1) and (-1 1 -1), as displayed on Figure C2.

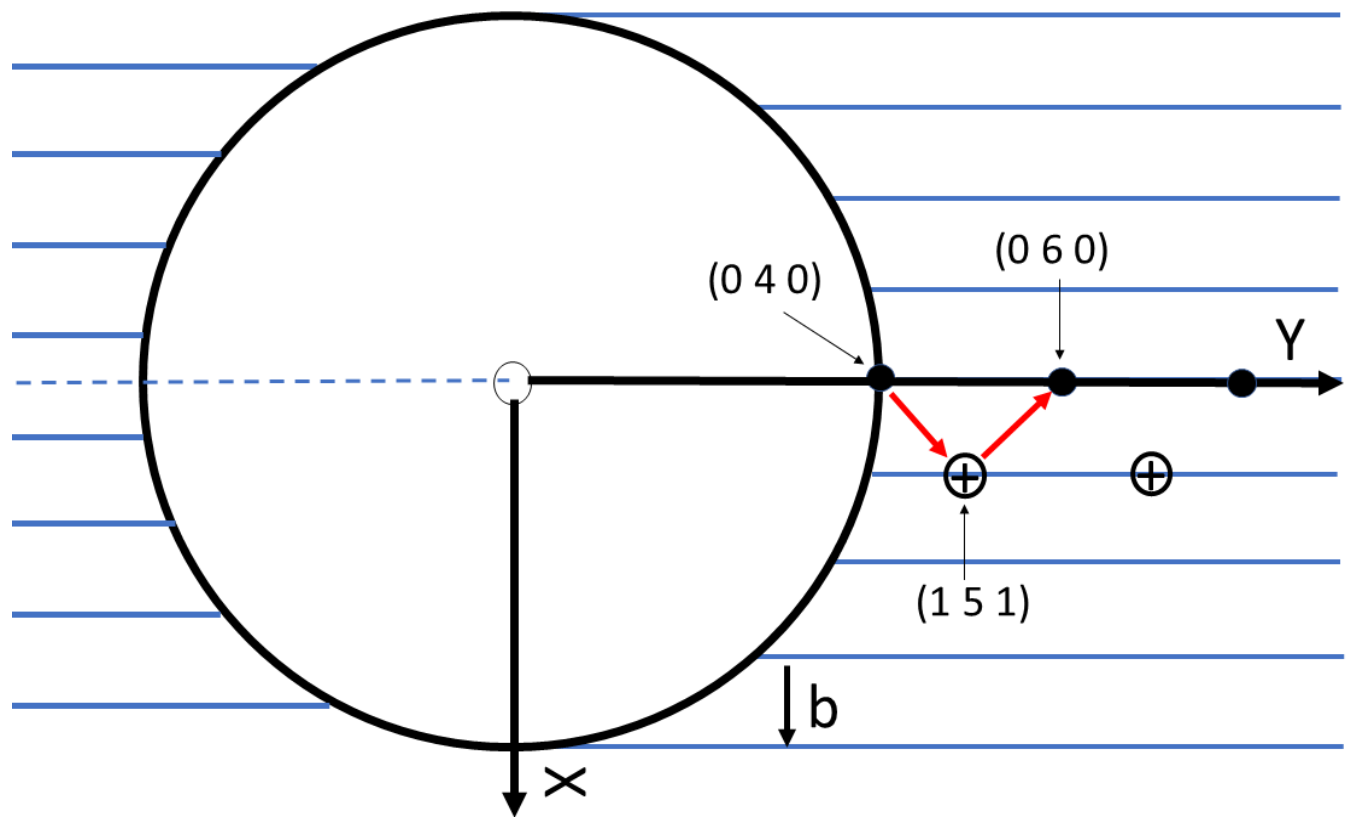

Figure C2.

One possible path of a vacancy moving away from dislocation in the compressed region, along which the interaction energy at stable and saddle positions are evaluated. Black disk: site in z=0 plane; circle enclosing '+': site in z= +1 plane.

**Expressions of stresses for a straight edge dislocation**

For an isotropic cubic crystal $C_{11} - C_{12} = 2C_{44}$ and the stresses are given in [21]:

$$\sigma_{xx} = -\frac{\mu b}{2\pi(1-\nu)} \frac{y(3x^2 + y^2)}{(x^2 + y^2)^2} \qquad \sigma_{yy} = \frac{\mu b}{2\pi(1-\nu)} \frac{y(x^2 - y^2)}{(x^2 + y^2)^2} \qquad \sigma_{xy} = \frac{\mu b}{2\pi(1-\nu)} \frac{x(x^2 - y^2)}{(x^2 + y^2)^2}$$

$$\sigma_{zz} = \nu(\sigma_{xx} + \sigma_{yy}) \qquad \mu^{isotr} = C_{44} = (C_{11} - C_{12})/2 \qquad \nu^{isotr} = C_{12}/(C_{11} + C_{12})$$

(C1)

For an anisotropic cubic crystal $C_{11} - C_{12} \neq 2C_{44}$ and the stresses are given in [26]:

$$\sigma_{xx} = -\frac{\alpha y\left[(3+H)x^2 + y^2\right]}{\left[(x^2+y^2)^2 + Hx^2 y^2\right]} \qquad \sigma_{yy} = \frac{\alpha y\left[x^2 - y^2\right]}{\left[(x^2+y^2)^2 + Hx^2 y^2\right]} \qquad \sigma_{xy} = \frac{\alpha x\left[x^2 - y^2\right]}{\left[(x^2+y^2)^2 + Hx^2 y^2\right]}$$

(C2)

with

$$\alpha = \frac{b\,I}{2\pi} \qquad I = (C_{11} + C_{12})\left[\frac{C_{44}(C_{11} - C_{12})}{C_{11}(2C_{44} + C_{11} + C_{12})}\right]^{1/2} \qquad H = \frac{(C_{11} + C_{12})(C_{11} - C_{12} - 2C_{44})}{C_{11} C_{44}}$$

(C3)

The expression of $\sigma_{zz}$ was not reported in reference [27]; it is assumed to be given by the same expression as in Eq. C1 which requires the use of the Poisson ratio.

It might be the start for some inconsistency: indeed, in an anisotropic crystal, the value of the Poisson ratio and the shear modulus are orientation dependent, and no expressions are available for them. They must be approximated by some combination of elastic constants, which were primarily designed to apply to polycrystals. Several averaged expressions are available for the shear modulus:

Voigt's approximation [28]:

$$\mu^{Voigt} = (C_{11} - C_{12} + 3C_{44})/5 \tag{C4}$$

Reuss' approximation [29]:

$$\mu^{Reuss} = \frac{5(C_{11} - C_{12})C_{44}}{4C_{11} - 4C_{12} + 3C_{44}} \tag{C5}$$

Hill's proposal [30] is a simple average of the two preceding expressions after observing that this combination is always closer to experiments performed on polycrystals:

$$\mu^{Hill} = \frac{1}{2}\left(\mu^{Reuss} + \mu^{Voigt}\right) \tag{C6}$$

Chang et al [16] made a more recent proposal:

$$\mu^{Chang} = (C_{11} - C_{12} + C_{44})/3 \tag{C7}$$

For all of them, the Poisson ratio is then deduced as a function of bulk and shear modulus, $B, \mu$ respectively. Recalling that $B = (C_{11} + 2C_{12})/3$, Hill shows that a consistent expression of the Poisson ratio is:

$$\nu^{Aver} = \frac{1}{2} \frac{3B - 2\mu^{Aver}}{3B + \mu^{Aver}} \tag{C8}$$

The final values obtained with these approximations at 300K are given in Table C1 below.

| Approximation | µ (eV/A³) | ν |
|---|---|---|
| Voigt (1928) | 0.557 | 0.285 |
| Reuss (1929) | 0.468 | 0.315 |
| Hill (1952) | 0.512 | 0.300 |
| Chang et al (2013) | 0.445 | 0.373 |

*Table C1. Values of the shear modulus and Poisson ratio for various averages.*

Notice that the choice of the Poisson ratio made by Chang et al ( $\nu^{Chang} = \nu^{isotr}$ ) is not consistent with Hill's proposal: it yields values which are noticeably different from the others, but it will not have much influence on the final results (see below).

**Calculations of migration barriers**

The values of elastic constants and of the dipolar tensors in stable $td_I^{stab}$, $td_V^{stab}$ and saddle position $td_I^{sadd}$ $td_V^{sadd}$ for the two defects are taken from [19]. The elastic constants are converted into eV/Angs³. The numerical values used in the energy calculations are gathered together in Table C2 below.

| Fe α | | unit |
|---|---|---|
| Lattice parameter | 2.831 | A |
| Atomic volume  Ω | 11.35 | $A^3$ |
| Burgers vector "1.0" : a/2 <100> | 1.415 | A |
| Burgers vector "2.0" : a <100> | 2.83 | A |
| Capture radius | 4 | Burgers vector |
| Elastic constants $C_{11}$  $C_{12}$  $C_{44}$ | 1.516   0.905   0.724 | $eV / A^3$ |
| Elastic compliances $S_{11}$  $S_{12}$  $S_{44}$ | 1.190   -0.445   1.381 | $A^3 / eV$ |
| Formation enthalpy V / I | 2.18 / 4.08 | eV |
| Formation entropy  V / I | 4.10 / 0.05 | $k_B$ |
| Migration energy V / I | 0.70 / 0.34 | eV |
| Burgers vector "1.0" : a/2 <100> | 1.415 | A |
| Burgers vector "2.0" : a <100> | 2.83 | A |

| Elastic dipole tensors | | |
|---|---|---|
| Vacancy stable point | $\begin{bmatrix} -2.38 & 0 & 0 \\ 0 & -2.38 & 0 \\ 0 & 0 & -2.38 \end{bmatrix}$ | eV / Ω |
| Vacancy saddle point <111> | $\begin{bmatrix} -2.217 & -1.641 & -1.641 \\ -1.641 & -2.217 & -1.641 \\ -1.641 & -1.641 & -2.217 \end{bmatrix}$ | eV / Ω |
| Dumbbell stable point <110> | $\begin{bmatrix} 23.752 & 4.728 & 0 \\ 4.728 & 23.752 & 0 \\ 0 & 0 & 27.906 \end{bmatrix}$ | eV / Ω |
| Dumbbell saddle point <111> from <110> to <101> | $\begin{bmatrix} 23.838 & 2.845 & -0.696 \\ 2.845 & 22.529 & 2.845 \\ -0.696 & 2.845 & 23.838 \end{bmatrix}$ | eV / Ω |
| Formation volumes  V  /  I | +0.75  /  +1.17 | Ω |
| Relaxation volumes  V  /  I | -0.22  /  +1.99 | Ω |

*Table C2. Numerical values used in the calculations of energy barriers.*

The unit eV/ Ω displayed for the dipolar tensors comes from the fact that their components in [19] are reported in eV for one defect. The point defect occupies one atomic volume Ω: this implies that the reported components hold implicitly for a volume Ω.

They must then be divided by Ω (measured in Å³) during any evaluation of energies which requires a cubic angstrom as the unit volume.

Using the dipolar tensors in stable position, the relaxation volumes upon defect formation are found equal to $\Delta V_I^{rel} = +1.99 \, \Omega$ and $\Delta V_V^{rel} = -0.22 \, \Omega$, where $\Omega$ stands for the atomic volume.

The compliances are deduced straightforwardly from the elastic constants. When multiplied by the local stresses expressed above, they yield the local strains generated by the dislocation line. These strains applied to the dipolar tensors of each defect in a given position yield finally the interaction energies. All the orientations of the dipolar tensors which are needed in the evaluation of the energies in stable and saddle positions are deduced form the value given for one single orientation in Table C2, after convenient coordinate and sign permutations.

For the stable site 'i' and a first neighbor stable site 'j', the strain tensors are denoted by $\varepsilon_{kl}(i)$ and $\varepsilon_{kl}(j)$; they are multiplied by the dipolar tensor for the dumbbell in a stable position. The latter does not depend on the location 'i' or 'j' (since polarizability effects are ignored), but depends on the orientation $h_i$ and $h_j$ of the dumbbell dissociation axis on the two sites (hence the argument below). The obtained energies depend on 'i', 'j' as well as on $h_i$ and $h_j$:

$$E_I^{stab}(i, h_i) = -\sum_{k,l} td_{I,kl}^{stab}(h_i) \, \varepsilon_{kl}(i) \qquad (C9)$$

$$E_I^{stab}(j, h_j) = -\sum_{k,l} td_{I,kl}^{stab}(h_j) \, \varepsilon_{kl}(j) \qquad (C10)$$

For the saddle positions, the strains generated by the dislocation are evaluated at the mid-point $\frac{i+j}{2}$ between the starting and the arrival site. The absolute height of the energy barrier is given by:

$$E_I^{sadd}(\tfrac{i+j}{2}) = E_I^{mig} - \sum_{k,l} td_{I,kl}^{sadd}(\tfrac{i+j}{2})\, \varepsilon_{kl}(\tfrac{i+j}{2}) \tag{C11}$$

where $E_I^m$ is the migration energy of the dumbbell far from the dislocation. The argument $\frac{i+j}{2}$ is kept for the dipolar tensor at the saddle position since it depends on the direction of the jump (which is unambiguously deduced from the knowledge of the starting and arrival sites 'i' and 'j' respectively).

In what follows, as above, the orientation of the dumbbell axis is indicated between brackets < >; the coordinates of the sites are indicated between parenthesis ( ).

**Brute force evaluation**

The PD are approximated by pure contraction or dilatation centers. The only quantity associated with each PD is its relaxation volume $\Delta V_I^{rel}$ and $\Delta V_V^{rel}$ expressed in units of atomic volume. In a solid of finite size, including the contribution of image forces, the interaction energy of the PD (denoted by the subscript 'd') located on site 'i' at a distance $r_i$ from the dislocation, is given by:

$$E_d(i) = \pm kT\, \frac{\mu b(1+\nu)\left|\Delta V_d^{rel}\right|}{3\pi(1-\nu)}\, \frac{1}{r_i}, \tag{C12}$$

where the '+' or '-' sign is taken for a repulsion or an attraction, respectively.

The absolute height of the energy barrier to be crossed between two adjacent sites 'i' and 'j' is expressed as

$$E_d^{sadd}(\tfrac{i+j}{2}) = (E_d(i) + E_d(j))/2 + \Delta_d^{sadd}(\tfrac{i+j}{2}) \quad \text{(C13)}$$

This is the most general formulation which fits all situations. The physical quantity $\Delta_d^{sadd}(\tfrac{i+j}{2})$ is not necessarily constant. If initially the simplest choice is to set $\Delta_d^{sadd}(\tfrac{i+j}{2}) = E_d^m$, it will however be fitted conveniently to reproduce as well as possible the desired energy barriers obtained through various evaluations (see below).

**Hydrostatic approximation on the dipolar tensor in isotropic crystal**

The stresses and strains used in the calculation are taken from expressions given in Eq. C1. But the PD dipolar tensors in stable and saddle positions are replaced by their hydrostatic approximation, where off-diagonals terms are equated to zero and diagonal ones are equated to one third of the trace.

**Full dipolar tensor in isotropic crystal**

The stresses and strains used in the calculation of the interaction energies are taken from expressions given in Eq. C1.

**Full dipolar tensor in anisotropic crystal**

The stresses and strains used in the calculation of the interaction energies are taken from expressions given in Eq. C2.

**Preliminary comparison between Voigt's, Reuss' and Chang's approximation**

The three approximations for the shear modulus and Poisson ratio are first compared in Figure C3. Only the coordinates of the stable positions are reported; the coordinates of the saddle points are the arithmetic average of those for the two adjacent sites. The resulting interaction energies obtained through the brute evaluation and the most refined one differ by about 18% at the shortest distance.

Reuss' expressions [29] yield slightly smaller interaction energies than Voigt's ones [28] in stable and saddle positions in the brute evaluation (the blue void circle is above the red circle); but the reverse is observed when using the full dipolar tensor in the anisotropic calculation (the blue triangle is below the red one). The approximation by Chang et al [16] yields larger interaction energies everywhere, but the gap with the preceding ones is negligible.

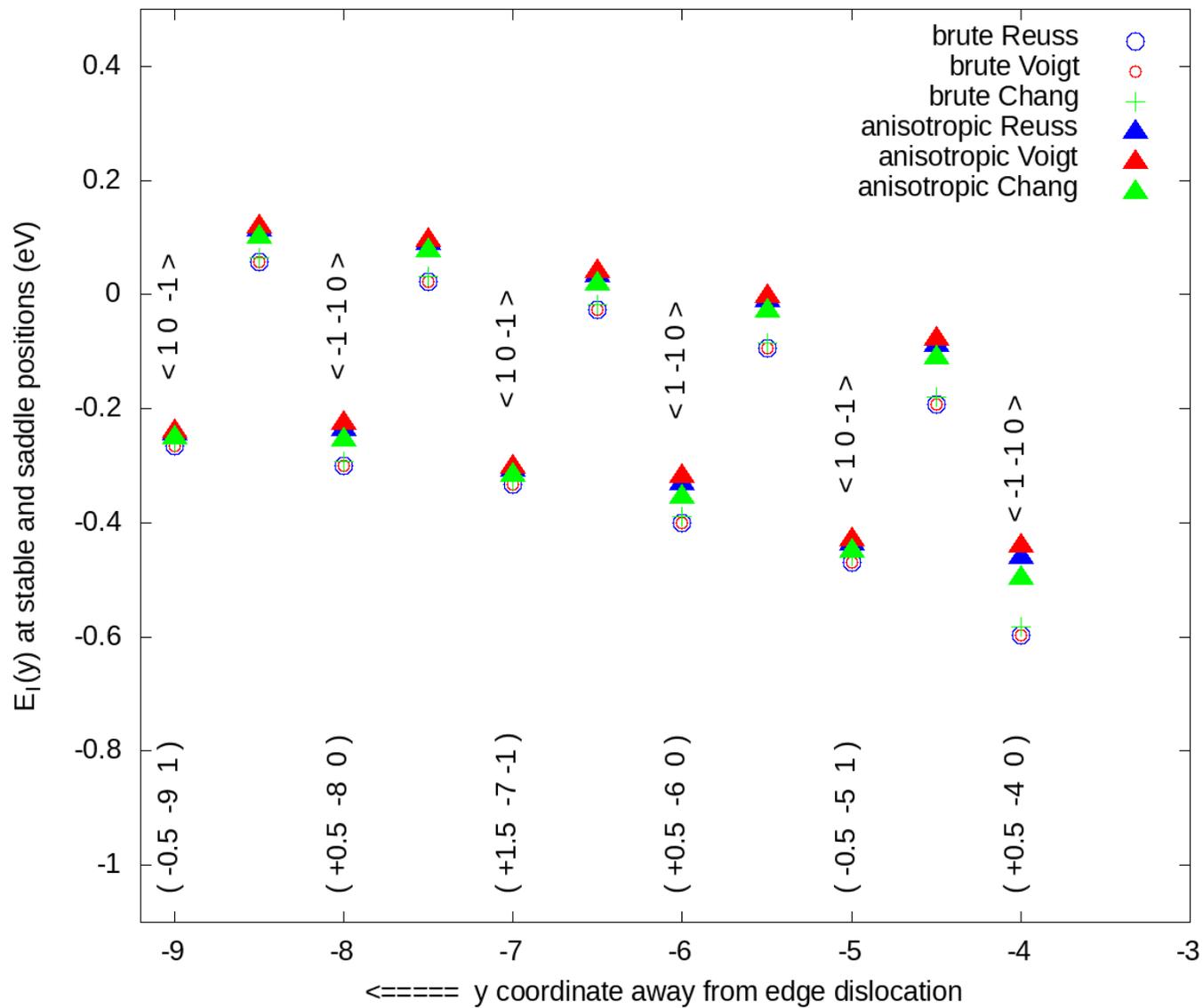

*Figure C3.*

*Diagram of interaction energies of a dumbbell with the dislocation when sitting at stable and saddle positions on the side in tension and along the chosen path: anisotropic evaluations against the brute ones, together with Reuss' [29], Voigt's [28] and Chang's [16] expressions of μ and ν.*

In all what will follow for the calculation of flux and absorption biases, we will use the average of Voigt's and Reuss' approximations, as proposed by Hill and denoted by 'V+R'.

Fig. C4 and Fig. C5 display the separate diagrams for the interaction energies of the dumbbell when sitting on stable and saddle positions located on the attractive side with V+R approximation. The energies are calculated with the four procedures enumerated above ("brute", "hydrostatic", "isotropic", "anisotropic"); the procedure denoted "brute modified" will be explained below.

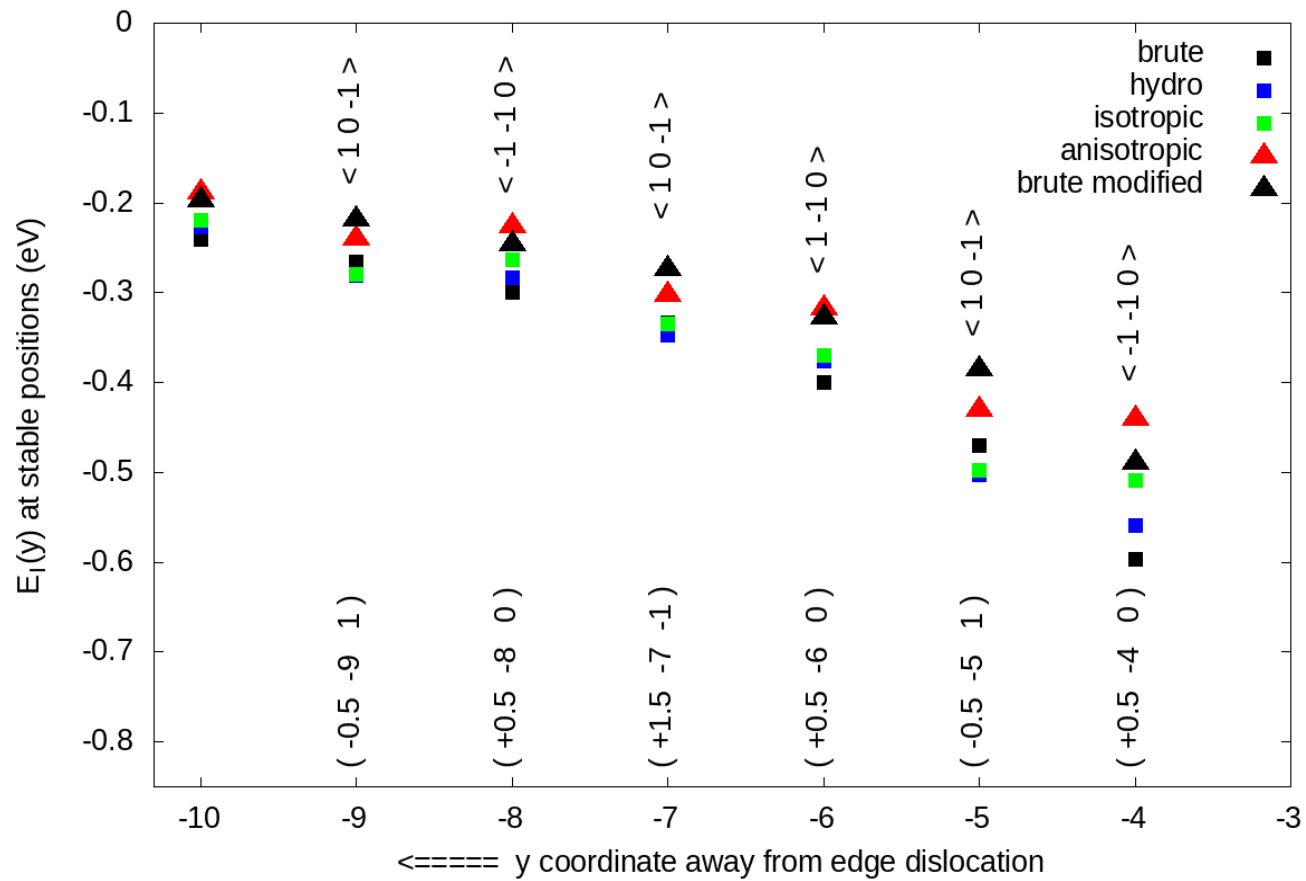

*Figure C4.*

*Diagram of interaction energies at stable positions with V+R expressions of μ and ν, close to the dislocation line for a dumbbell sitting on stable positions on the side in tension. The orientation of the dumbbell (and of its dipolar tensor) is given between brackets.*

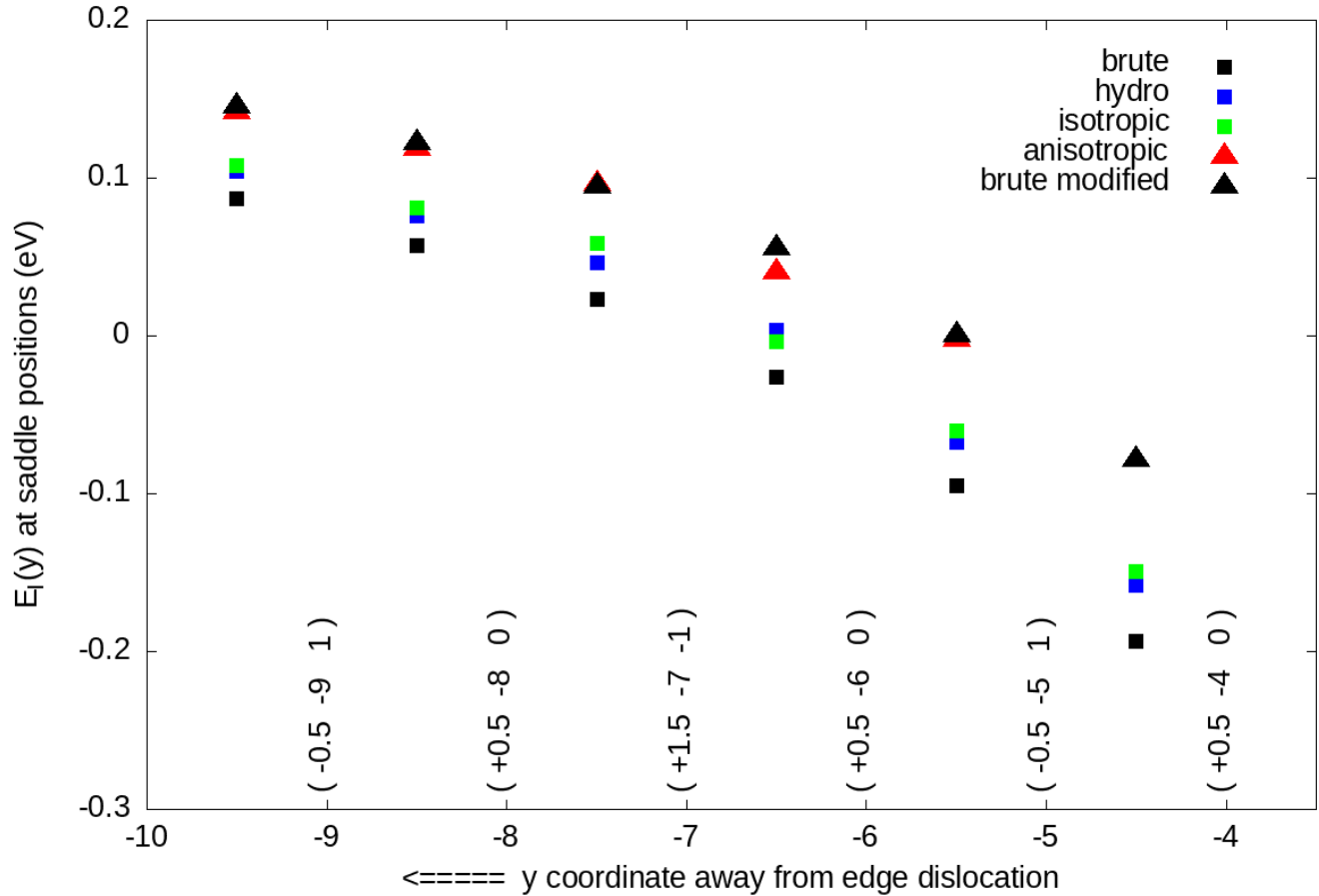

*Figure C5.*

*Diagram of interaction energies at saddle positions with V+R expressions of µ and ν, close to the dislocation line for a dumbbell on the side in tension. The coordinates of the adjacent stable sites are recalled. The orientation of the dipolar tensor at the saddle position is parallel to the jump vector which joins the two adjacent stable sites.*

**General remarks for the dumbbell migration**

The comparison of the energy profiles displayed above along the chosen path shows the following:

*i*) the brute force evaluation yields larger interaction energies (and larger energy gradients) than those obtained with the full dipolar tensor (roughly by 25% at the shortest distances);

*ii*) the use of the full dipolar tensor in the anisotropic treatment tends always to soften the interaction in stable position (for instance, - 0.59 eV for the brute evaluation against -0.44 eV for the refined one at y = - 4 in Fig. C4). The isotropic evaluation is close to, and even smaller than, the brute one, when the two ends of the dumbbell belong to the same plane perpendicular to the y-axis; this is due to the orientation change of the dumbbell;

*iii*) together with this overall softening, it is observed that the energy decrease is more like a step-down descent, and in some places, is no longer monotonous: a more distant DP can have a larger attractive energy than a closer one (see for instance Fig. C4 for y = - 8 and - 9). This is associated with the different orientations of the dissociation axis of the dumbbell along the chosen path. This feature is of course never observed with the brute force evaluation but always observed with the calculations using the dipolar tensor, even in its hydrostatic version;

*iv*) a dumbbell with a given starting orientation of the dissociation axis <1 0 -1> on site (-0.5, -5, 1) can land on the neighbor site at (+0.5, -4, 0) with two different orientations, namely < -1 -1 0 > and < 0 -1 1 >; these two orientations on the arrival site give two different interaction energies (- 0.42 eV for the first and – 0.56 eV for the second one) although the barrier to overcome was the same (Fig. C6).

This last point raises a more fundamental question about the boundary conditions to be used when solving the diffusion equation. If the dumbbell is reduced to an isotropic point defect of spherical symmetry (ignoring its dissociation axis), a single diffusion equation is sufficient to describe its migration; but there is no rule for choosing the interaction energy on the boundary between the two results quoted above. For the present problem, this difference between the two orientations of the dumbbell on the boundary has fortunately no real consequence: the formation energy of the dumbbell defect is much larger than the difference quoted above and the thermal concentrations at the boundary remain always very close to zero for any temperature of practical interest.

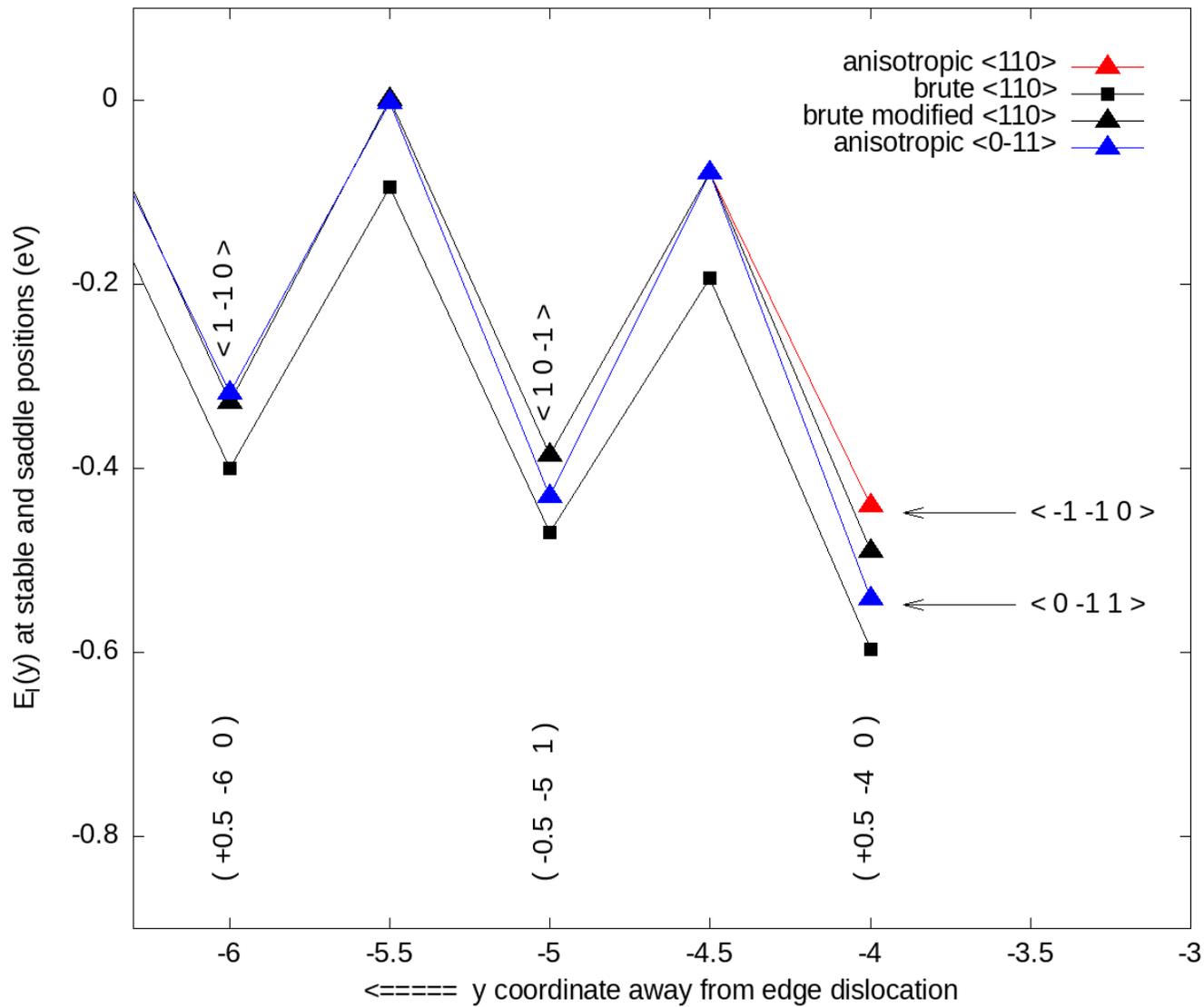

*Figure C6.*

*Diagram of energy barriers with V+R expressions of μ and ν, close to the dislocation line for a dumbbell on the side in tension. The starting point at (-0.5, -5, 1) gives access to site (+0.5, -4, 0) with two different orientations denoted with arrows.*

**Modified evaluation**

Without being able to reproduce the non-monotonous decrease of the interaction energy quoted above, a totally empirical but simple modification of the brute evaluation can however bring it closer to the results obtained with the full dipolar tensors. The modification consists in using a multiplicative correction for the energy at stable positions:

$$E_d^{modif}(i) = \pm f_d^{stable}\, kT \frac{\mu b(1+\nu)\left|\Delta V_d^{rel}\right|}{3\pi(1-\nu)\, r_i} \tag{C14}$$

In the same way, the barriers are also modified according to the empirical rule:

$$\Delta_d^{sadd}(\tfrac{i+j}{2}) = E_d^m\left(1 \mp f_d^{sadd}\left[a/(2r_{\frac{i+j}{2}})**0.5\right]\right) \tag{C15}$$

The sign to be chosen in Eq. C15 is the opposite of the sign in Eq. C14. No physical explanation is to be looked for, because it is nothing but an empirical recipe.

These corrections for the dumbbell case with $f_I^{stable} = 0.82$ and $f_I^{sadd} = 0.12$ yield the black triangle symbols in Fig. C7 for the energies at stable positions and in Fig. C8 for the energies at saddle positions denoted by the key "brute modified". The modification cannot reproduce the subtleties at short distances obtained by the dipolar tensors which were seen on Fig.C4 and C5 but it catches correctly the overall trend in the two cases.

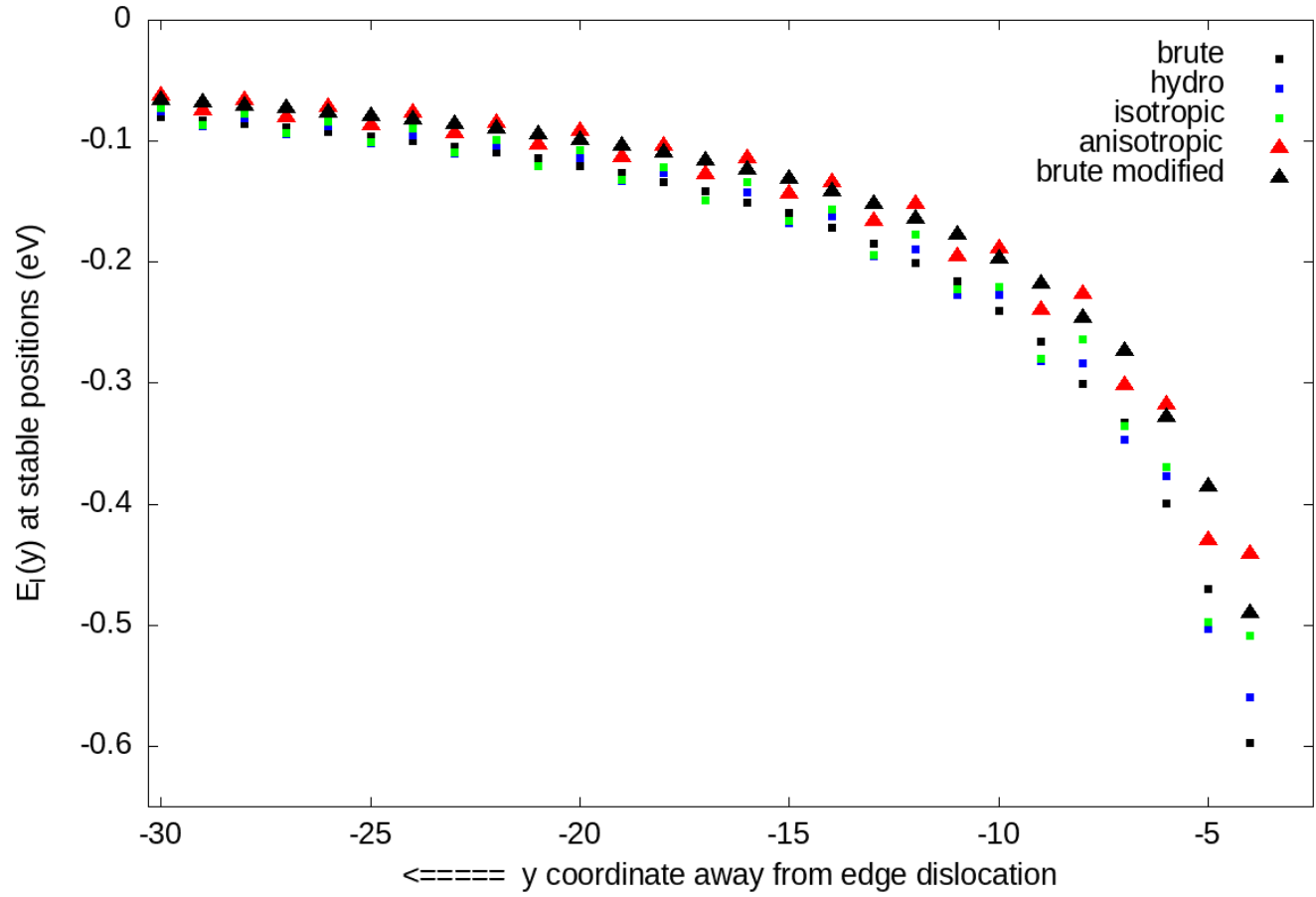

*Figure C7.*

*Energy plot at stable positions when calculated with the brute modified approximation.*

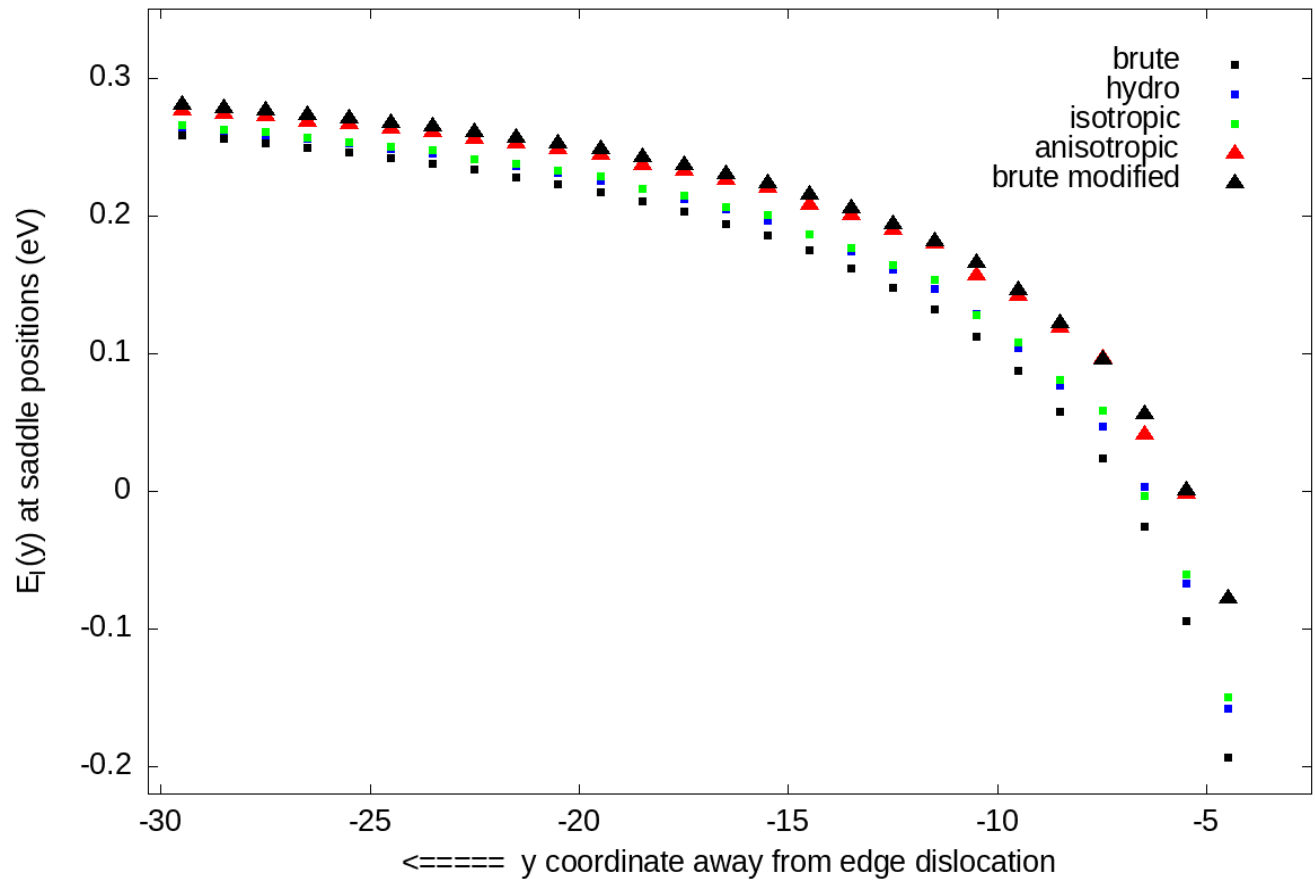

Figure C8.

Energy plot at saddle positions when calculated with the brute modified approximation.

**General remarks for the vacancy migration**

The interaction energies are roughly smaller by one order of magnitude but the migration barriers are twice higher. Displaying the interaction energies at stable and saddle positions on the same diagram like Fig. C6 would make the differences between the different approaches hardly detectable.

Fig. C9 and C10 display on separate diagrams the interaction energies of a vacancy on stable and saddle positions located on the attractive side in compression with V+R approximations. The general trends observed for the dumbbell hold also for the vacancy case. The energies at stable and saddle positions, when calculated with the brute modified procedure, become closer to the anisotropic ones with $f_V^{stab} = 0.85$ and $f_V^{sadd} = 0.015$; this yields the black triangle symbols denoted by the key "brute modified".

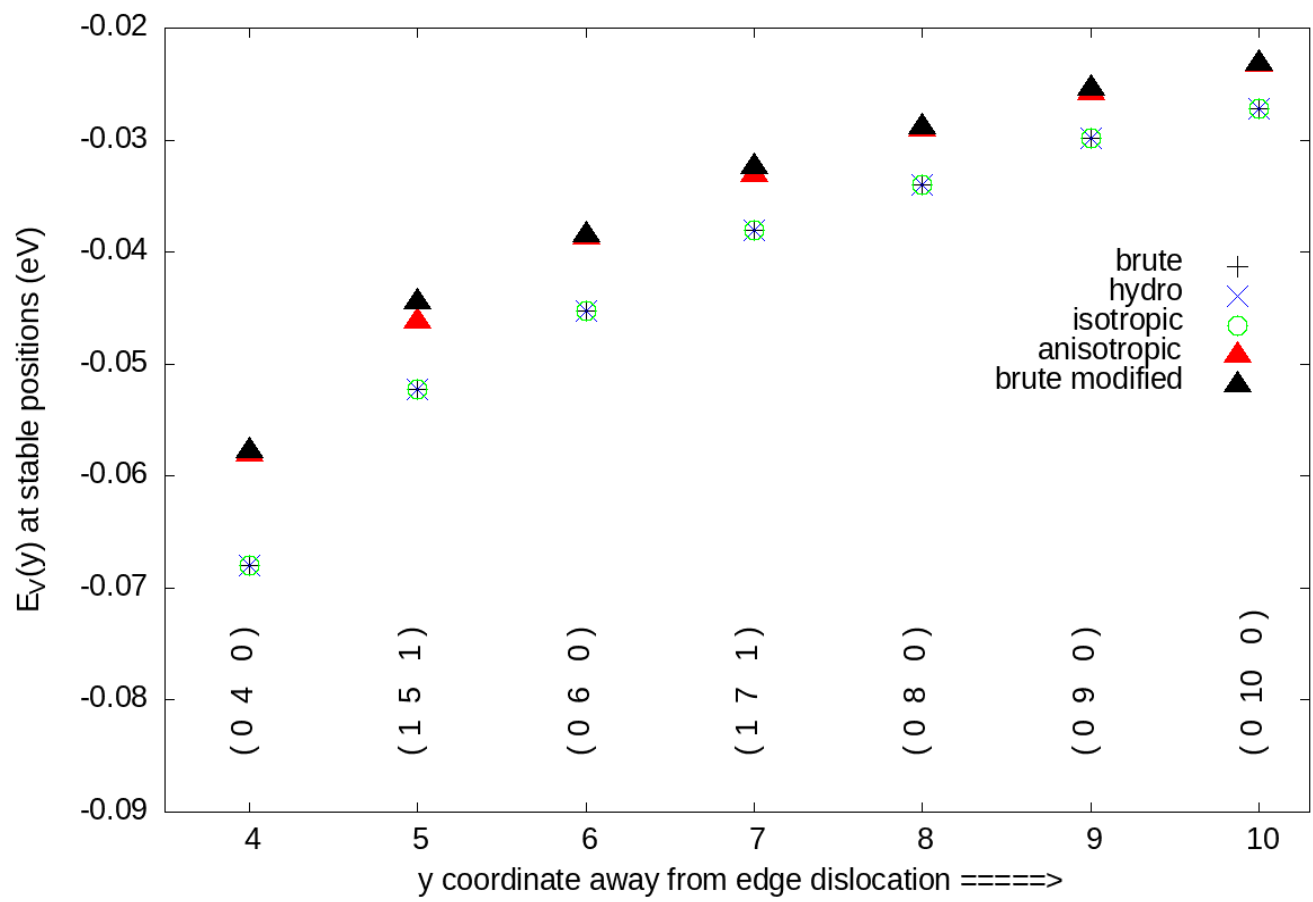

*Figure C9.*

*Diagram of interaction energies with V+R expressions of µ and v, close to the dislocation line for a vacancy sitting on stable positions on the side in compression. The coordinates of stable positions are displayed for clarity.*

The interaction energies in stable positions (Fig. C9) are superposed for the brute, the hydrostatic and the isotropic evaluations because the dipolar tensor of the vacancy at rest is fully hydrostatic. Only the anisotropic evaluation exhibits some difference and softens the interaction a bit.

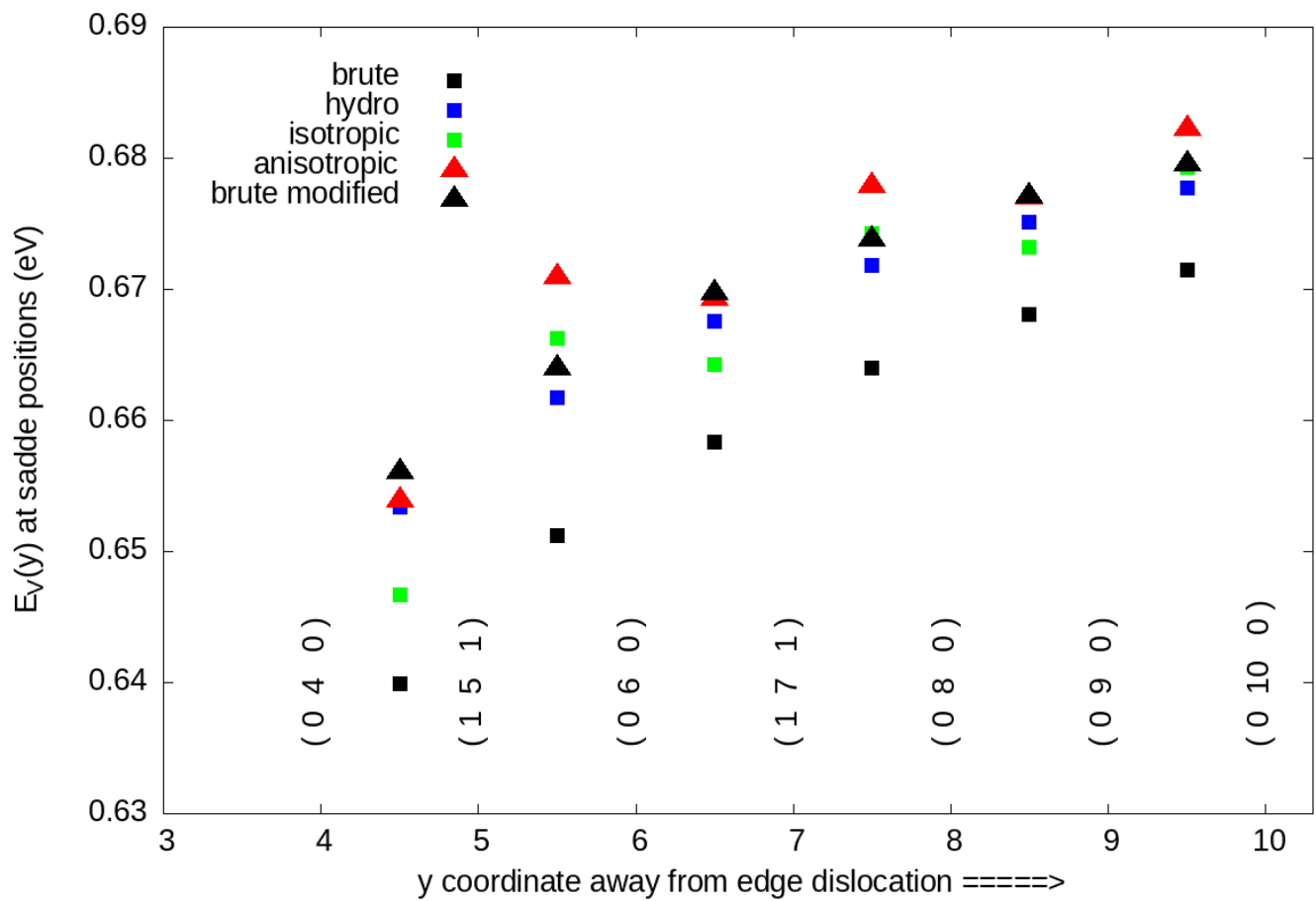

*Figure C10.*

*Diagram of interaction energies with V+R expressions of µ and v, close to the dislocation line for a vacancy sitting at saddle positions on the side in compression. The coordinates of stable positions are recalled for clarity. The orientation of the dipolar tensor at the saddle position is parallel to the jump vector which joins the two adjacent stable sites.*

For the saddle positions, the differences between all the approaches show up again (Fig. C10) and the step-up increase is again observed. The modified interaction brings only a fair agreement with the best evaluations.

The diagrams for the repulsive side are not displayed here for sake of space; they are very similar to those for the attracting side (apart from the sign change of the interaction). All the general remarks made for the two defects are still valid.

As a final remark, the full advantage of the refinements commented above can be included only if the diffusion problem is solved with one transport equation for each orientation of the dumbbell, or, alternatively, with an MC approach which uses explicitly the six orientations of the defect on a lattice site.

This is not the case of the present contribution; but the proposed empirical modification brings the brute modelling closer to the best evaluations. The empirically modified formula will be used, in parallel with the brute one, in order to detect whether this modification of the barrier profiles brings a significant change to the absorption bias or not.